\documentclass[reprint,aps,prb
]{revtex4-2}
\usepackage[utf8]{inputenc}
\usepackage{textgreek}
\usepackage{dcolumn}
\usepackage{bm}
\usepackage{amsmath, amssymb, graphicx, physics}
\usepackage{newunicodechar}

\usepackage{geometry}
\usepackage[utf8]{inputenc}
\usepackage{bbold} 
\usepackage{dsfont} 
\usepackage{xcolor}
\usepackage{natbib}
\usepackage{hyperref}
\usepackage{ulem}
\usepackage{cancel}
\usepackage[mathlines]{lineno}
\begin{document}
\preprint{APS/123-QED}
\title{Sliding phasons in Moiré Ladders}
\author{Paula Mellado}
 \affiliation{Facultad de Ingeniería y Ciencias,\\ Universidad Adolfo Ibáñez, Santiago, Chile.}
\author{Francisco Muñoz }%
 \author{Javiera Cabezas-Escares}
\affiliation{
Departamento de Física, Facultad de Ciencias,\\ Universidad de Chile and CEDENNA\\
Santiago, Chile
}%
\date{\today}
\begin{abstract}
An incommensurate charge density wave is a periodic modulation of charge that breaks translational symmetry at a momentum that does not coincide with the primitive lattice vectors. Its Goldstone excitation, the phason, comprises collective gapless phase fluctuations. Aiming to unveil the mechanism behind the onset of incommensurate charge order in layered materials, we study a half-filled, four-band tight-binding model on a ladder with a relative shift \(\delta=p/q\) between the legs, induced by the dimerization of one of them. The shift results in a moiré supercell comprising \(q\) composite cells and a modulated inter-leg tunneling. The moiré potential compresses the leg bands into flat minibands near the Fermi level, resulting in additional low-energy peaks in the density of states. Including Coulomb interactions, we find an incommensurate charge-density-wave phase in which the charge modulation is out of phase between the legs. The collective excitations of this state are long-lived neutral, acoustic phasons whose speed is controlled by the moiré parameter \(\delta\) and the inter-leg tunneling amplitude. This model sheds light on the role of interlayer incongruities in the formation of excitonic charge-ordered phases in van der Waals and heterostructured materials.
\end{abstract}
\maketitle
\section{\label{sec:level1} Introduction}
Two-dimensional layers of atoms that exhibit quantum-mechanical phenomena often comprise three-dimensional materials, which are stacked with weak interlayer coupling via van der Waals forces \cite{novoselov20162d,guo2021stacking,sung2019stacking,liu20192d}.In these layered systems, symmetry-breaking correlated phases either coexist or compete with one or more exotic phases \cite{keimer2017physics,mellado2024quantum}. Examples of natural van der Waals or layered materials are the dichalcogenides ($\rm NbSe_2$, $\rm TaS_2$, and $\rm TiSe_2$),  which can exhibit complex charge density wave patterns and phase transitions \cite{lian2018unveiling,tsen2015structure,goli2012charge,chen2016dimensional}. Transition-metal thiophosphates are another class of materials with negligible interlayer coupling, where the reduced dimensionality promotes intriguing electronic and optical quantum effects while allowing for easy tunability of magnetic exchange and anisotropy through ligand substitution \cite{chen2025introduction}. In bilayer and trilayer graphene \cite{ding2009localized} and other layered semiconductors of the type \cite{yu2023observation,xu2025correlated}, electronic, magnetic, and topological phases can coexist due to the combined effect of low dimensionality and the incommensurability between layers \cite{pantaleon2023superconductivity}. Such incommensurability arises when two atomic layers are overlaid with a slight lattice misalignment or a small rotation angle. This results in a moiré superlattice \cite{mcgilly2020visualization}, with altered properties compared to the parent materials \cite{andrei2021marvels,shi2021exotic,adak2024tunable} and flat energy bands. When bands are flat, the kinetic energy is suppressed, making the electron-electron interaction energy the dominant contribution. This is why moiré systems become playgrounds for strongly correlated electron phenomena. 

Incongruity among layers in real materials can arise from slight discrepancies in bonds within a layer or between adjacent layers \cite{abbas2020recent}. It can be by twisting homobilayers \cite{chen2019tunable} or by lattice deformation by applying external pressure and strain \cite{gao2020band,escudero2024designing,rakib2022moire,gao2022symmetry}. While uniform strain is a powerful tool to tune electronic properties, introducing spatially nonuniform strain profiles offers far greater functionality, including the generation of pseudomagnetic fields and the modulation of intra- and interlayer couplings \cite{yuan2024flat}.  A common effect in low-dimensional systems is susceptibility to charge-ordering instabilities, such as charge density waves (CDWs) \cite{mcmillan1975landau,takada1985damping,gruner1988dynamics,tucker1988theory}. It consists of a periodic modulation of the conduction-electron charge density in a crystal \cite{gor2012charge},  and it occurs primarily in low-dimensional conductors \cite{rossnagel2011origin,zhu2015classification,imada1986competition}.  In the past few decades, there have been significant efforts to understand the driving mechanisms of several CDW-hosting systems, for example, quasi-two-dimensional transition metal dichalcogenides and rare-earth tritellurides ($\rm RTe_3s$), as well as in quasi-one-dimensional (quasi-1D) systems \cite{banerjee2013charge,melikyan2005model}. 

The formation of CDWs is often attributed to an electronic instability driven by Fermi-surface nesting \cite{urban2007scaling}, combined with electron-phonon coupling, known as the Peierls mechanism \cite{hirsch1983effect,boriack1977dynamic}. However, it has been shown that the observed charge ordering phase transitions are far from being analogs of the Peierls mechanism \cite{johannes2008fermi} because electronic instabilities are easily destroyed by even slight deviations from the perfect nesting conditions of the Peierls scenario \cite{hofmann2019strong}. This is the case in the CDW materials $\rm NbSe_2$, \cite{lian2018unveiling}, $\rm TaSe_2$, \cite{ryu2018persistent}, and $\rm CeTe_3$ \cite{ralevic2016charge}, where the CDW phases result from structural phase transitions driven by the concerted action of electronic and ionic degrees of freedom \cite{johannes2008fermi}.

Electron-electron coupling can also drive CDWs in excitonic insulators or via the band Jahn-Teller effect \cite{lee1978dynamics,efetov1977charge,littlewood1982amplitude,wegner2020evidence,guo2024real}. Unlike the Peierls CDW, where the lattice distortion (mediated by phonons) is essential to lower the electronic energy, an excitonic CDW can form even without any significant electron-phonon coupling. The electronic correlations alone are sufficient \cite{gao2024observation}. In the conventional scenario of a CDW phase, an energy gap is created at the Fermi energy (E$_F$) by the modulation of the lattice at 2k$_F$, and the charge density assumes a configuration akin to a standing wave pattern. This modulation can be characterized by the complex order parameter $\rho(x)=|\rho_0|e^{i\theta}$ \cite{gor2012charge}, which comprises two components, the amplitude $|\rho_0|$, which quantifies the magnitude or intensity of the charge modulation, and the phase $\theta$, that characterizes its spatial translation relative to the underlying crystal lattice. Upon the formation of a CDW, these components become ordered \cite{tucker1988theory}; nevertheless, they are also subject to low-energy fluctuations, which give rise to two collective excitations known as amplitudons and phasons \cite{gruner1988dynamics}. The amplitudon, oscillations in the magnitude of the CDW order parameter, is a gapped excitation \cite{kwon2024dual}. The phason represents the collective phase fluctuations of the order parameter and can be gapped when pinned or gapless when unpinned. The distinction between pinned and unpinned phasons separates the static, insulating-like behavior from the dynamic, collective conduction.  An unpinned phason is the ideal, gapless (massless) Nambu-Goldstone mode associated with the spontaneous breaking of continuous translational symmetry in an incommensurate CDW.
The incommensurate nature of the CDW state implies that the CDW  wavelength is not an integer multiple of the lattice constant. In this case, the phase becomes its translational degree of freedom. In an ideal, perfectly ordered, incommensurate CDW system, the phason would be a massless mode that moves without any energy cost \cite{fisher1985sliding,brazovskii2004pinning}. 

A pinned phason arises when the CDW phase is locked or pinned to the crystal lattice. The most common cause is imperfections in the crystal lattice that act as scattering centers, locally disrupting the CDW order and creating regions where the CDW prefers to reside. Translating the CDW away from these preferred positions costs energy. In complex CDW patterns, domain walls can also contribute to pinning \cite{monceau1985charge}. 

Models that realize gapless phason modes are central to guide our search for Charge Density Wave materials with improved electrical transport properties of \cite{wilson1985charge,PhysRevLett.42.1423}. In this context,  various quasi-one-dimensional and quasi-two-dimensional materials, particularly transition metal chalcogenides, Potassium molybdates ($\rm K_{0.3}MoO_{3}$, $\rm Rb_{0.3}MoO_{3}$, blue bronzes), and bronzes like Niobium triselenide ($\rm NbSe_{3}$) \cite{ravy2006disorder} and Tantalum trisulfide ($\rm TaS_{3}$) \cite{kvashnin2020coexistence}, have exhibited a new form of collective electrical current transport, denoted as sliding CDW. The sliding CDW is the displacement of the entire CDW condensate through the crystal, involving a periodic modulation of both the electron charge density and the underlying atomic lattice. This phase is characterized by non-linear conductivity and narrow-band noise above a threshold field \cite{wilson1985charge,cheng2024engineering}.  For sliding CDW transport to occur, the CDW must typically be incommensurate with the underlying lattice.   
One case where the sliding CDW occurs in a peculiar form is the quasi 1D CDW material  $\rm Ta_2NiSe_7$ \cite{yao2024electric,he2017band}, where Angle-Resolved Photoemission Spectroscopy (ARPES) identified small hole- and electron-like pockets and a total absence of nesting of states at the primary CDW wavevector k. However, they found a possible nesting at 2k, which plausibly connects low-energy states, raising the intriguing possibility of a critical role for 2k ordering \cite{watson2023spectral}. In these materials, in addition to incommensurate CDW orders, topological edge states arise, where the wave-vector-dependent electron-phonon coupling can induce partial gaps and changes in the Fermi surface topology, rather than gapping out the entire Fermi surface. Therefore, the change in the electronic structure is not limited to the Fermi energy but occurs over a broader energy range \cite{xiao2024robust}. 

While the overall CDW carries a charge current, in incommensurate CDW phases, the phason itself is often described as charge-neutral \cite{birkbeck2024measuring,ochoa2019moire}. Neutral phasons are desirable bosonic modes that can be driven or manipulated using non-charge probes, such as thermal gradients or mechanical stimuli (e.g., strain and pressure). In 1D, this allows for a detailed separation and study of the various coupled degrees of freedom—charge, spin, and lattice—in complex electronic materials. However, the presence of neutral phasons in higher dimensions does not imply spin-charge separation in the same way it does in 1D Luttinger liquids \cite{imambekov2012one}, as the strong quantum fluctuations that lead to fractionalization into spinons and holons are generally suppressed in 2D and 3D \cite{birkbeck2024measuring}.  Determining the conditions under which these bosons remain electrically neutral in two- and three-dimensional incommensurate CDW systems is crucial for identifying candidate materials that could host them and for evaluating their possible function as mediators of other correlated phases, such as superconductivity \cite{miao2019formation, shapiro2007observation,fukuyama2020theory}.

Although a number of quasi-one-dimensional compounds hosting sliding charge-density waves \cite{ravy2006disorder,yao2024electric,he2017band} can exhibit collective transport via weakly pinned phason modes, in many such materials the CDW is strongly coupled to lattice distortions and phonons, so that the microscopic nature of the sliding mode remains closely entangled with material-specific details. Here we focus instead on $\rm HfTe_3$  (Hafnium Tritelluride), a quasi-one-dimensional transition-metal trichalcogenide \cite{abdulsalam2015structural, liu2021quasi}. $\rm HfTe_3$ naturally hosts multiple weakly coupled chains with chain-dependent dimerization, allowing for relative phase shifts between neighboring chains. In reduced dimensionality, first-principles calculations reveal long-wavelength structural instabilities that generate moiré-like modulations of electronic hopping amplitudes. These features make $\rm HfTe_3$ an especially suitable platform for connecting a minimal theoretical description of incommensurate CDWs and neutral phasons to realistic microscopic physics. 

In this article, we study a toy model of a heterostructure consisting of a half-filled, four-band ladder with a relative shift \(\delta=p/q\) between its legs, which creates a moiré supercell of \(q\) composite cells. The moiré pattern generates a moiré potential that compresses the leg bands into narrow minibands near the Fermi level, resulting in additional low-energy peaks in the density of states. Including Coulomb interactions on top of the flat band Hamiltonian, we find an incommensurate CDW phase with a moiré periodicity of \(2\pi p\), characterized by a rung-odd density mode in which charge modulations are out of phase between the legs, leading to long-lived neutral phason excitations. We show that the moiré phasons are Goldstone modes of the effective interacting theory in which the CDW slips.  From the effective action, we obtained phason velocities as functions of the inter-leg tunneling and the moiré parameter \(\delta\), showing that the phason speed decreases as \(\delta \to 1\) and increases with larger tunneling amplitudes.

The paper is structured as follows. Section \ref{sec:model} presents the Hamiltonian of the system and its spectrum in the extended Brillouin zone. Section \ref{sec:mini} derives the Hamiltonian in the reduced Brillouin zone and shows the spectrum of the flat minibands. The density of states in the reduced and extended Brillouin zones and the Lindhard electronic susceptibility for the non-interacting kinetic Hamiltonian are discussed in section \ref{sec:dos}. Section \ref{sec:cdw} reviews mean field theory for including interactions and presents the quasiparticle spectrum of the mean field Hamiltonian, its spectral features, and the results of the RPA susceptibility for charge density waves of the interacting model. Section \ref{sec:phason} addresses the dynamics of phasons in the system. In Section \ref{sec:dft}, we present DFT calculations of bands and the density of states for the material $\rm HfTe_3$ and compare them with our theoretical results. We conclude with Section \ref{sec:disc}. 
\begin{figure*}
    \centering
\includegraphics[width=\linewidth]{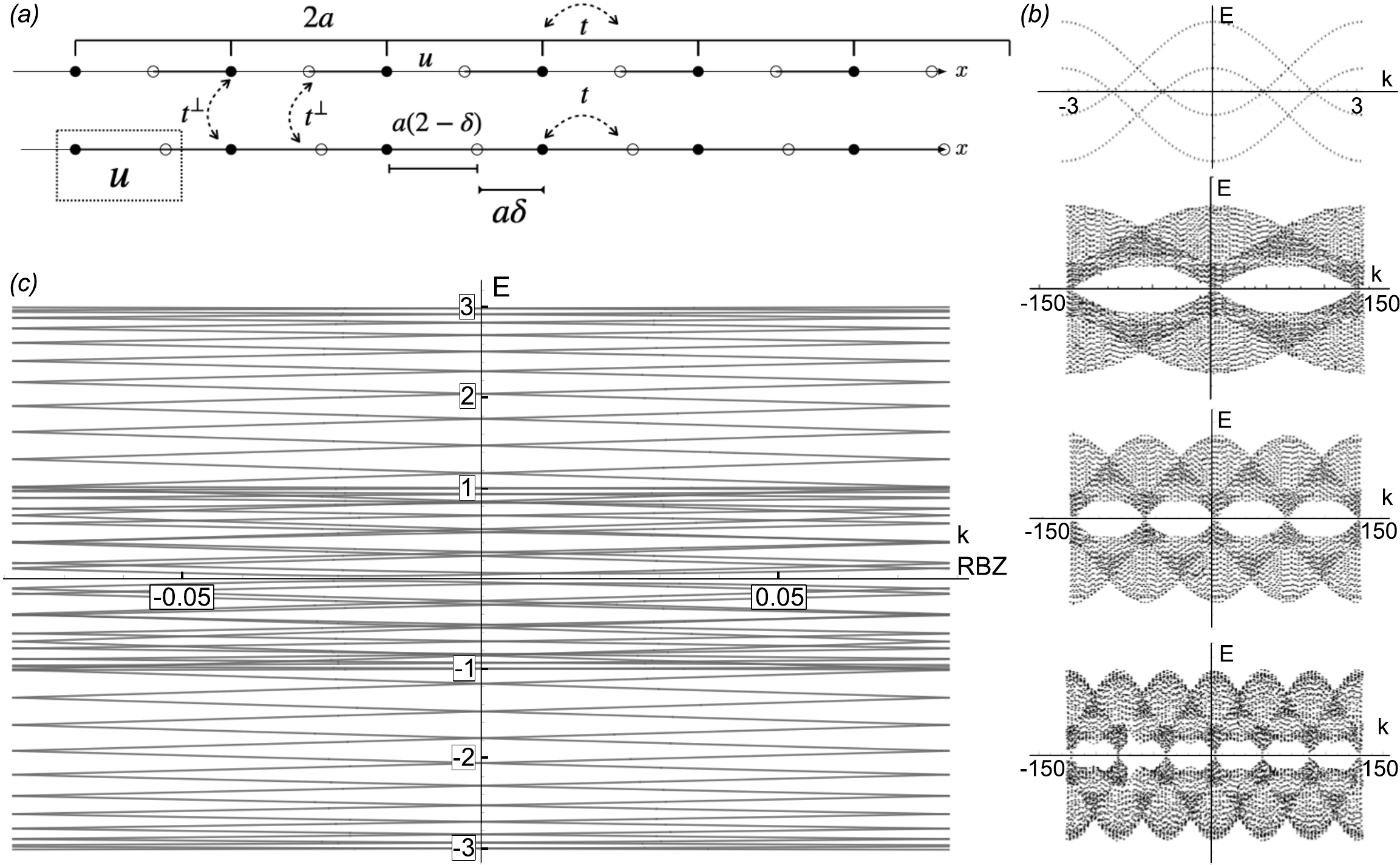}
\caption{\label{f1} (a) Schematic illustration of the effective moiré ladder model. The moiré modulation does not arise from a geometric lattice mismatch but from a long-wavelength modulation of the inter-leg tunneling, representing an emergent symmetry-broken state. (b) Band spectra of the kinetic Hamiltonian $H(k)$ in the EBZ, Eq.\ref{eq:H4eff}, at (from top to bottom), $\delta=1,\frac{19}{20},\frac{9}{10},\frac{17}{20}$. The large momentum range reflects an extended Brillouin zone representation in which all moiré replicas are unfolded. (c) Band spectra of the minibands $H_{\rm RBZ}$ in the RBZ, Eq.\ref{eq:HRBZ} at $\delta=\frac{19}{20}$.}
\end{figure*}
\begin{figure}
    \centering
\includegraphics[width=\linewidth]{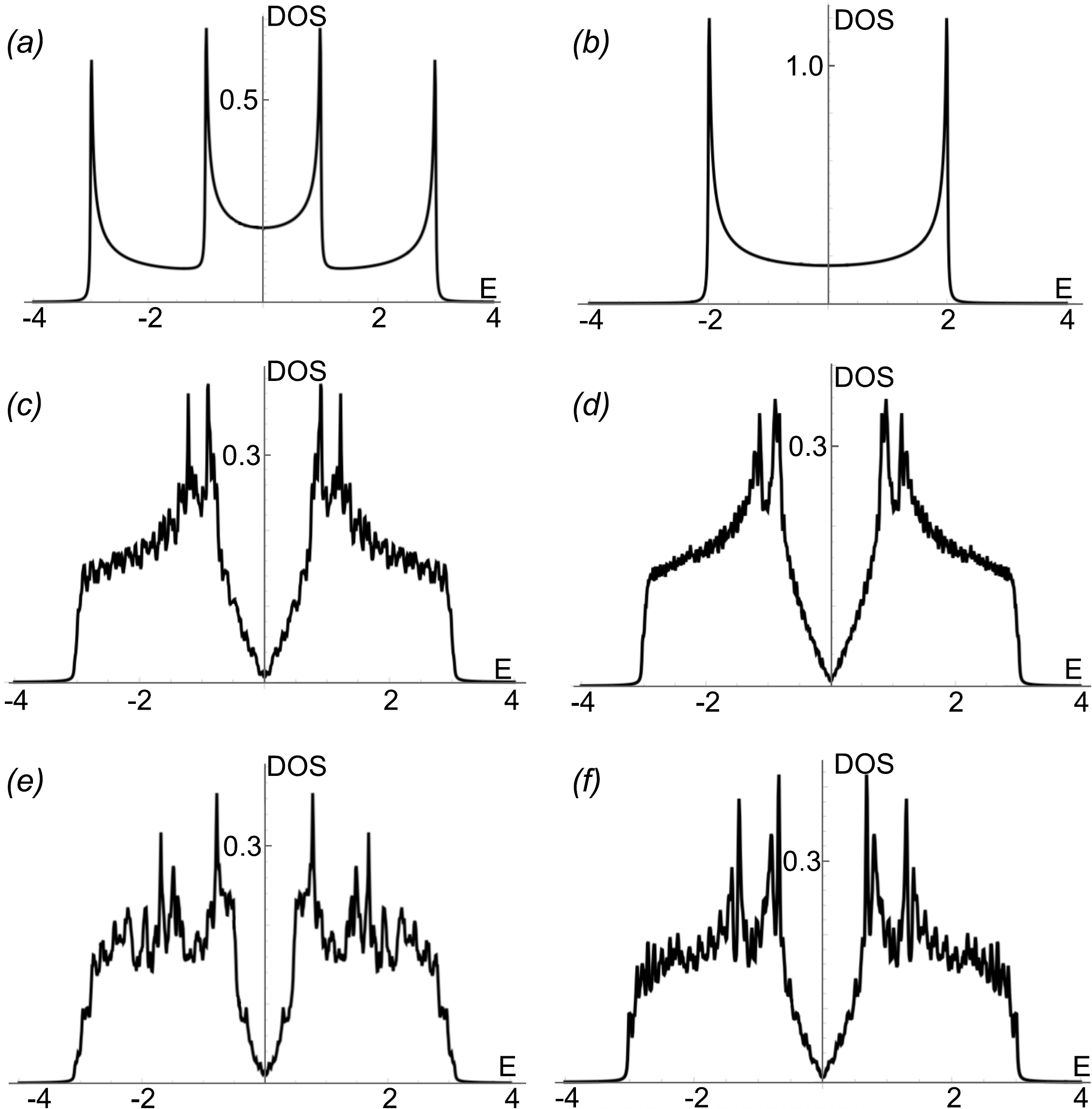}
\caption{\label{f2} DOS at T=0 and (a) $\delta=1$, (b) $\delta=1$ and no inter-leg hopping ($t_\perp =0$), (c) $\delta=\frac{19}{20}$ in the EBZ, (d) $\delta=\frac{19}{20}$ in the RBZ, (e) $\delta=\frac{17}{20}$ in the EBZ, (f) $\delta=\frac{17}{20}$ in RBZ.}
\end{figure}
\begin{figure}
    \centering
\includegraphics[width=\linewidth]{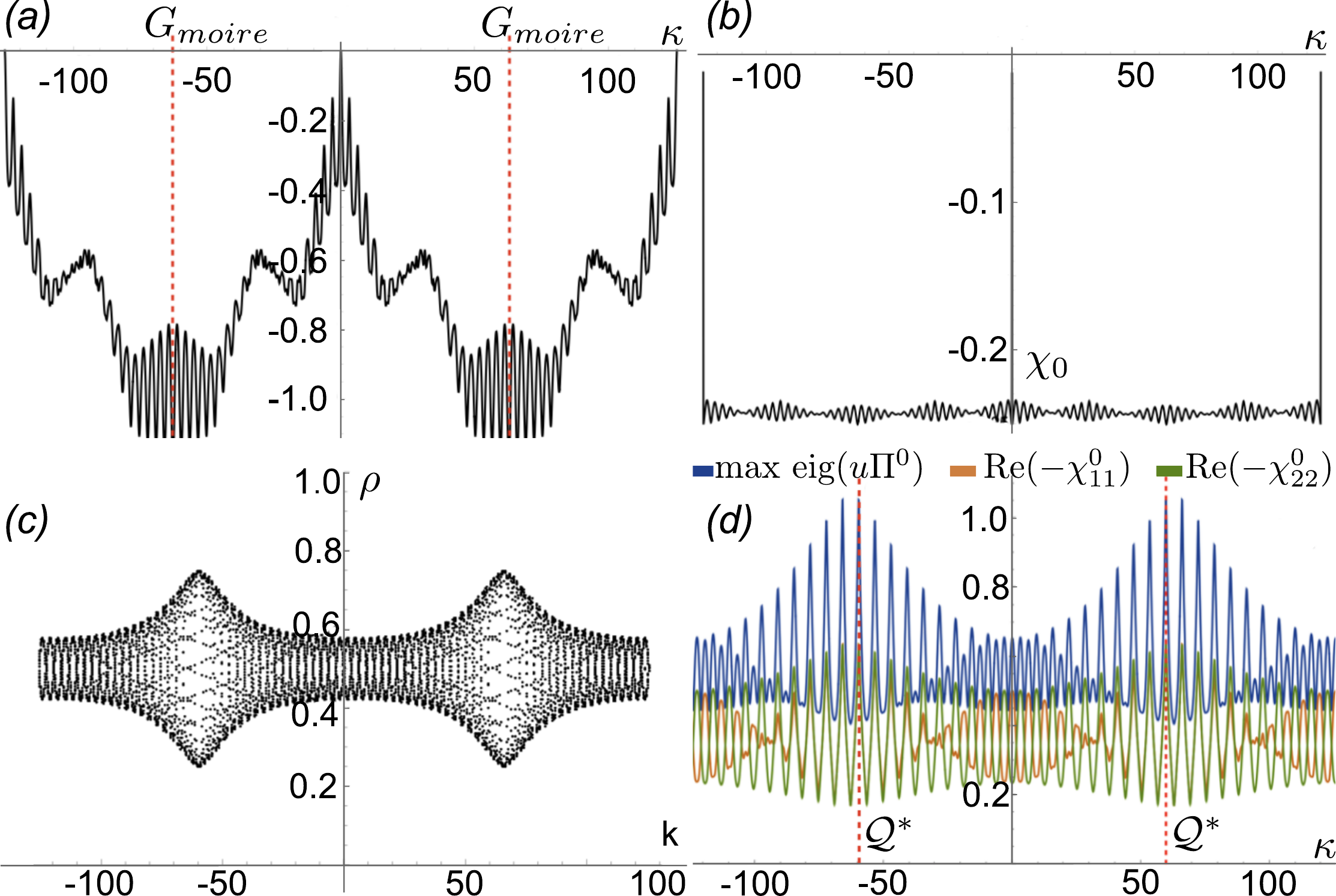}
\caption{\label{f3} Lindhard susceptibility $\chi_0(\kappa)$ at temperature (a) $T=\frac{t}{5}$ and (b) $T=5t$. (c) Site charge density associated with bands 1 and 2. (d) Largest eigenvalue $\lambda_{\max}^{(0)}(\kappa,T)$ of the bare density–density susceptibility matrix $\chi^{(0)}(\kappa,T)$ (blue), leg-1 projected (orange), leg-2 projected (green), at $T=\frac{t_1}{5}$. Q* is shown by the vertical dotted red line. In all figures $\delta=\frac{19}{20}$ and $\eta=10^{-5}$.}
\end{figure}

\begin{figure}
    \centering
\includegraphics[width=\linewidth]{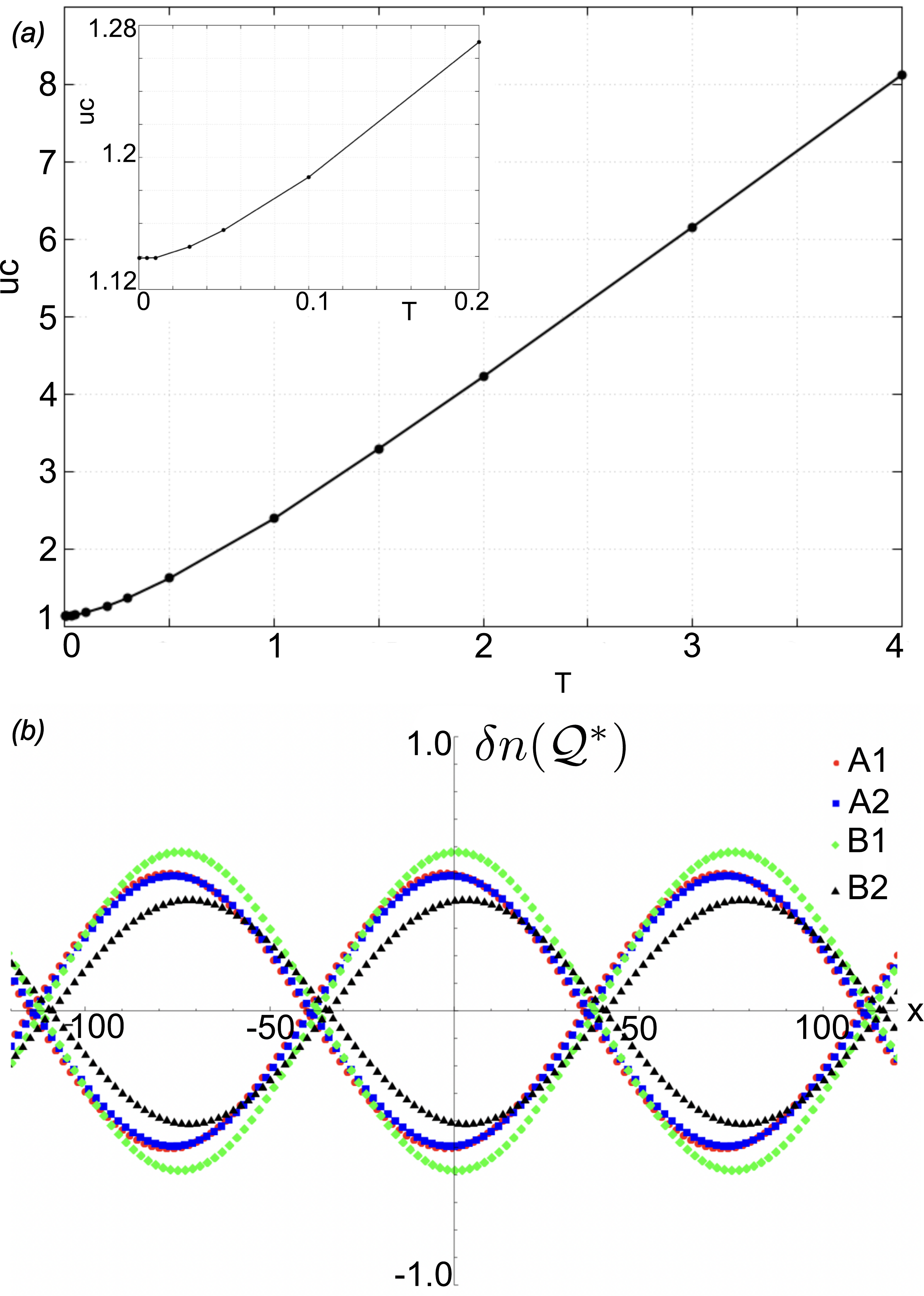}
\caption{\label{f4} (a) Critical interaction $u_c$ versus temperature T at $\delta=\frac{19}{20}$. The inset shows a zoom in of the data at low temperatures. (b) Real space charge modulation $\delta n(x)$ at the sites of the unit cell (odd mode) at $Q*$.} 
\end{figure}
\begin{figure}
    \centering
\includegraphics[width=\columnwidth]{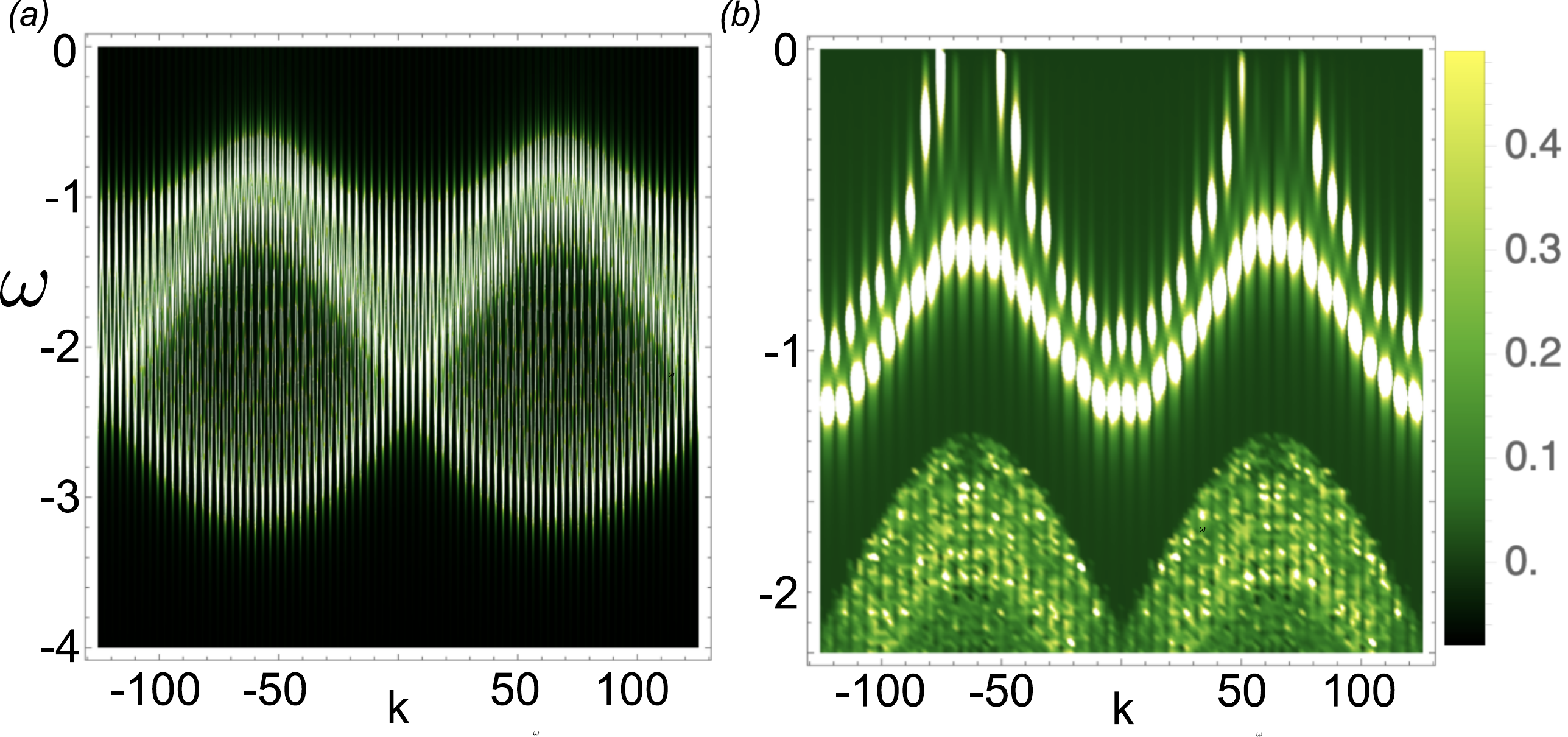}
\caption{\label{f5} Spectral function at $\delta=\frac{19}{20}$ and $T=1-\delta$ computed from (a) the single particle retarded Green function, Eq.\ref{eq:spectral_scalar} (b) the two-particle RPA susceptibility, Eq.\ref{eq:spectralphason}.}
\end{figure}
\section{\label{sec:model}Model}
Consider the ladder shown in Fig.\ref{f1}(a). It consists of two coupled chains (legs), each with two sites (A and B) per unit cell. The composite ladder unit cell contains a total of four sites 
$s: \{A_1,B_1,A_2,B_2\}$ (subscripts $1,2$ label legs). Throughout, we choose the
nearest–neighbor spacing in leg-1 as the length unit ($a=1$), so the composite cell has a length equal to $2$, and
the original reciprocal period is \(
G_0=\pi\). Leg-1 is uniform, while leg-2 exhibits phase-only dimerization due to a relative mismatch
shift \(\delta \) per composite cell, as shown in Fig.\ref{f1}(a). The shift is rationalized by a single parameter \(\delta=p/q\), with $p$ and $q$ two coprime integers. The intra-leg nearest neighbor hopping is constant and equal to $t$. This sets the intra-leg bandwidth \(W\sim 4t\).  Inter-leg (rung) hopping occurs within the same sublattice, and its average is \(t\).  Due to $\delta$, the inter-leg hopping strength oscillates along the ladder, with periodicity $q$, this produces a moiré supercell of \(q\) composite cells that has \(4q\) sites in total, a moiré reciprocal vector in the reduced Brillouin zone (RBZ) \(b_s=\frac{G_0}{q}=\frac{\pi}{q}\), and a real-space supercell length $L_s=2q$.

After downfolding (see Appendix~\ref{sec:app}), the dimerization strength of leg-2 generates a first-harmonic rung modulation of amplitude \(t_1\). Using Schrieffer–Wolff  perturbation theory \cite{schrieffer1966relation}, we parameterize
\(
t_1=\,(1-\delta)\,t, 
\)
such that the modulation vanishes at a perfect match between the legs \(\delta=1\) and increases linearly for small mismatches.
Thus, inter-leg (rung) tunneling is modulated with the moiré envelope,
\begin{equation}
  t_\perp(n)=t+t_1\cos\left(\frac{2\pi n}{q}\right)
\end{equation}

Let $n$ index composite unit cells. On each leg $\ell=1,2$ fermion operators are
$c_{\ell,\alpha,n}$ and $c^\dagger_{\ell,\alpha,n}$ ($\alpha=A,B$) . Throughout, we set the intra-leg tunneling amplitude \(t \) as our energy scale. 
The Hamiltonian of the system is
\begin{align}
\mathcal{H} &= H_{1}+H_{2}+H_{\text{rung}},\\[3pt]
H_{1}
&=\sum_n \Big[
  c^\dagger_{1,A,n}\,c^{}_{1,B,n}
 +c^\dagger_{1,B,n}\,c^{}_{1,A,n+1}
 +\text{h.c.}\Big],\nonumber
\\
H_{2}
&=\sum_n \Big[
c^\dagger_{2,A,n}\,c^{}_{2,B,n}
 + \nonumber
 c^\dagger_{2,B,n}\,c^{}_{2,A,n+1}\\
 +\text{h.c.}\Big],\nonumber
\\
H_{\text{rung}}
&=\sum_n t_\perp(n)\,\Big[
  c^\dagger_{1,A,n}c^{}_{2,A,n}
 +c^\dagger_{1,B,n}c^{}_{2,B,n}
 +\text{h.c.}\Big]\nonumber
 \label{eq:H}
\end{align}
In the moiré ladder studied here, both legs have identical microscopic lattice periodicity, so no geometric moiré pattern arises directly at the level of the underlying lattice. Rather, the moiré superstructure is introduced phenomenologically through a long-wavelength modulation of the inter-leg (rung) tunneling amplitude, as given in Eq.~(1). This modulation can be viewed as an emergent outcome of spontaneously broken translational symmetry, which arises, for instance, from a soft phonon mode that produces relative phase shifts or displacements between adjacent chains. In this sense, the moiré ladder serves as an effective low-energy description of a symmetry-broken phase where a long-wavelength modulation has already formed. A microscopic realization of this mechanism is discussed in Sec.~VII for $\rm HfTe_3$, where first-principles calculations reveal a chain-dependent structural instability leading to precisely such a modulation.

At this stage it is important to clarify our use of momentum space.  The long-wavelength modulation induced by the mismatch parameter $\delta=p/q$ introduces a supercell of size $2qa$ and couples momenta that differ by integer multiples of $G_0/q$. For clarity, in what follows we work in an extended Brillouin zone (EBZ), where all $q$ momentum replicas generated by the BZ folding are unfolded and displayed explicitly. This leads to the momentum range $k\in(\rm -qG_0,qG_0)$ used in Fig.~1(b). The EBZ does not represent a physical Brillouin zone; rather, it is a convenient representation that makes the moiré-induced band folding and hybridization transparent. The physically relevant description is recovered by folding the EBZ into the reduced Brillouin zone of width $G_0/q$, as done explicitly in Sec.~III. 

Fourier-transformed operators read
\(c^{}_{\ell,\alpha,k}=\frac{1}{\sqrt{N}}\sum_n e^{-2ikn}c_{\ell,\alpha,n}\),
where $N$ is the total number of unit cells. With this convention, the Fourier transform is defined with respect to the composite unit cell of length $2a$. In reciprocal space, structure factors associated with leg-1, leg-2, and the
dimerization phase in leg-2 are 
\begin{align}
c(k) &= 2\cos(k),\\
c_\delta(k) &= 2\cos\!\big(k-\pi\delta\big),\\
\phi(k) &= k(1-\delta)+\pi\delta,
\end{align}
respectively. As a result, nearest-neighbor hopping within and between composite cells produces real structure factors $2\cos k$, while nontrivial phase factors associated with the dimerized leg appear explicitly through the Peierls phase $\phi(k)$. The Hamiltonian is  written in a gauge where all trivial Bloch phases have been absorbed,
\[
  \mathcal{H}=\sum_k \Psi_k^\dagger\,H(k)\,\Psi_k,
\]
with 
$\Psi_k=(c_{1,A,k},c_{1,B,k},c_{2,A,k},c_{2,B,k})^\top$ and the $4\times4$ kinetic Hamiltonian matrix
\begin{equation}
\label{eq:H4eff}
H(k)=
\begin{pmatrix}
    0 & c(k) & t_\perp(k) & 0\\
    c(k) & 0 & 0 & t_\perp(k)\\
    t_\perp(k) & 0 & 0 & e^{i\phi(k)}c_{\delta(k)}\\
    0 & t_\perp(k) & e^{-i\phi(k)}c_\delta(k) & 0
\end{pmatrix}
\end{equation}
where rung modulation appears as a single harmonic
\begin{equation}
\label{eq:tperpK}
t_\perp(k)=1+(1-\delta)\cos\!\big(qk\big).
\end{equation} 
The phase $\phi(k)$ encodes the fact that
the two $\ell=2$ bonds attached to $A_{2}$ have real space displacements
$\Delta x=\delta$ and $\Delta x=-(2-\delta)$. Combining the corresponding Peierls factors
$e^{ik\delta}$ and $e^{-ik(2-\delta)}$ yields
\begin{equation}
\begin{aligned}
e^{ik\delta}+e^{-ik(2-\delta)}
=2\cos(k-\pi\delta)\,e^{\,i\big(k(1-\delta)+\pi\delta\big)}\\
= c_\delta(k)\,e^{\,i\phi(k)},
\end{aligned}
\end{equation}
The relative dimerization phase between the two legs advances by \(
\Delta\Phi=2\pi\delta\)
from one composite cell to the next, with \(\delta=p/q\in(0,1)\) written in lowest
terms. Then, after $q$ composite cells, the relative phase cancels.
As the reciprocal lattice of the supercell has a vector
\(
b_s=\frac{\pi}{q}\), the reduced first Brillouin zone (RBZ) of the moiré lattice spans the interval $[-b_s/2,b_s/2)$
, and the reciprocal superlattice vector is
\(
G_s = \{ m\,b_s\ |\ m\in\mathbb{Z}\}
\). Any moiré modulation  has Fourier components on
$G_s$ and couples states whose momenta differ by $m b_s$. 

Because the physical lattice periodicity is \(2\) but the dimerization phase accumulates in steps of \(2\pi\delta\) per composite cell, upon unfolding the RBZ bands to the EBZ, the spectrum becomes periodic under a finite momentum shift that returns all Peierls phases to themselves. The smallest shift $\Delta$ for which the whole band structure is invariant implies that there exists a $k$-independent unitary $\mathcal{U}$ with
\begin{equation}
\label{eq:similarity}
H(k+\Delta)=\mathcal{U}\,H(k)\,\mathcal{U}^\dagger
\end{equation}
Such that $H(k+\Delta)$ and $H(k)$ have the same spectrum. Using
\(c(k),\;c_\delta(k),\;\phi(k)\) we have for $\Delta=\pi q$ that the off-diagonal factors in \eqref{eq:H4eff} transform as
\begin{equation}
c_\delta(k+\pi q)\,e^{+i\phi(k+\pi q)} \;=\; (-1)^{p}\, c_\delta(k)\,e^{+i\phi(k)},
\end{equation}
\begin{equation}
c(k+\pi q) \;=\; (-1)^{q}\, c(k).
\end{equation}
Let $\mathcal{U}=\mathrm{diag}(u_1,u_2,u_3,u_4)$ with $u_j\in\{\pm1\}$ acting on
$(A_1,B_1,A_2,B_2)$. The similarity transformation
\eqref{eq:similarity} at $\Delta=\pi q$ imposes constraints on p and q that determine the generic global period $2\pi q$. This gives rise to the extended first Brillouin zone, \(k\in (-qG_0,qG_{0})\), as depicted in Fig.\ref{f1}(b).

To conclude this section, we comment on the symmetries of the Hamiltonian. With real hopping and zero onsite terms, \eqref{eq:H4eff} is time-reversal symmetric \(H(-k)=H(k)^*\) and has a chiral (sublattice)
symmetry \(\{H,\Gamma\}=0\) with \(\Gamma=\mathrm{diag}(1,-1,-1,1)\), the spectrum is particle–hole symmetric. Site parity and rung parity operators do not commute with $H(k)$. 
\section{\label{sec:mini} Minibands in the RBZ}
To obtain moiré minibands, we build the Hamiltonian matrix in the RBZ, \(|\bar{k}|\le b_s/2=\pi/(2q)\) and explicitly keep the \(q\) replica momenta
\(\{\bar{k}+m G_s\}_{m=0}^{q-1}\). Thus, the block‐diagonal piece of the Hamiltonian matrix becomes
\begin{equation}
\big[H_{\mathrm{RBZ}}^{(0)}(\bar{k})\big]_{m m'}=
\delta_{m m'}\,H\!\big(\bar{k}+m G_s\big),\label{eq:RBZdiag}\end{equation}\( m,m'=0,\dots,q-1
\). 
The rung modulation \(\cos(qk)\) implies nearest-neighbor replica coupling in the RBZ and gives rise to the off-diagonal part with cyclic boundary conditions in replica space (indices modulo \(q\))
\begin{equation}
\label{eq:RBZoffdiag}
\big[H_{\mathrm{RBZ}}^{(1)}(\bar{k})\big]_{m,m\pm 1}
=\frac{t_1}{2}\,R,\quad
R = \sigma_1 \otimes \mathbb{I}_2
\end{equation}
where $\sigma_j$ is the j-$th$ Pauli matrix. All other off-diagonal replica blocks vanish within the single-harmonic truncation.
The full \(4q\times4q\) Hamiltonian is
\begin{equation}
H_{\mathrm{RBZ}}(\bar{k})=H^{(0)}_{\mathrm{RBZ}}(\bar{k})+H^{(1)}_{\mathrm{RBZ}}(\bar{k}).
\label{eq:HRBZ}
\end{equation}
If \(t_1=0\) the \(4q\) bands split into \(q\) independent copies of the four
parent bands evaluated at $\bar{k}+m G_s$.
Finite inter-leg tunneling hybridizes neighboring replicas and opens minigaps at avoided crossings, where the minibands group into quartets inherited from the four parent branches.
The symmetry content of \(H\) \eqref{eq:H4eff} is preserved by \eqref{eq:HRBZ}, in particular, chiral symmetry remains, so a narrow band near \(E=0\) survives
for a broad window of parameters, as shown in Fig.\ref{f1}c.
\section{\label{sec:dos}Density of states and Lindhard Susceptibility} The moiré model of Eq.\ref{eq:H4eff} retains the two-leg and two-sublattice structure, reproduces the dimerization phase of leg-2, and captures the moiré-periodic
inter-leg tunneling amplitude. The spectrum, exactly solvable and particle–hole symmetric, is shown in Fig.\ref{f1}(b) for the system with a slight inter-leg mismatch ($\delta$ close to unity). In all plots where $\delta\neq 1$, a small gap occurs around the Fermi energy, E=0. This is a direct consequence of the dimerization of leg-2, which breaks the translational symmetry and opens up a gap in the energy spectrum. The mismatch \(\delta\) creates a new periodic potential that acts as a coupling term in the tight-binding Hamiltonian. 
Consider the characteristic polynomial of $H(k)$,
\begin{equation}
\begin{split}
E^4-E^2\!\Big(c_\delta^2+c^2+2t_{\perp}^2\Big)
+c_\delta^2c^2+t_{\perp}^4-\\2t_{\perp}^2\,c_\delta c \cos\phi
=0
\label{eq:charpoly}
\end{split}
\end{equation}
where, for brevity, $c_\delta\equiv c_\delta(k)$ and $c\equiv c(k)$. When \ref{eq:charpoly} is evaluated at $E=0$ it yields
$c_\delta^2c^2+t_{\perp}^4-2t_{\perp}^2\,c_\delta c \cos\phi=0$. 
The root occurs when the two conditions \(
\sin\phi(k)=0\) and \(
c_\delta(k)\,c(k)=t_{\perp}^2\)
are simultaneously satisfied. The first condition determines $\phi(k)=n\pi$; the second is a fine-tuned relationship between the two leg dispersions and the rung $t_{\perp}$. It does not hold for $k$ in the EBZ. Therefore, the spectrum of \eqref{eq:H4eff} is gapped.
This is also the case with the spectrum of $H_{\rm RBZ}$. In the  RBZ interval, \(\phi(k)\) varies slowly, and for many $k$, $\phi(k)\sim 0$ or $\pi$ and $c_\delta(k)c(k)\sim t_{\perp}^2$. Then the term, \(
c_\delta^2c^2+t_{\perp}^4-2t_{\perp}^2\,c_\delta c\cos\phi\) becomes small but nonzero, pinning eigenvalues close to $E=0$ with very narrow dispersion or flat bands. Minigaps at folded-band crossings appear and scale linearly with $t_1$ at small modulation. The position and width of these flat bands, as well as the position of the peaks in the density of states (DOS), can be precisely tuned by $\delta$. The moiré potential causes the energy bands to repel one another, leading to avoided crossings and additional band splitting. It is the source of smaller peaks in the DOS=\(\frac{1}{N_k}\sum_{n,k}
\frac{1}{\pi}\,\frac{\eta}{\big(E - E_n(k)\big)^2+\eta^2}
\) in $H$ and $H_{\rm RBZ}$ (at temperature T=0, in a discrete k-mesh approximation with $N_k$ points and Lorentzian broadening $\eta>0$). The results are shown in Fig.\ref{f2}. The system with $\delta=1$ and $t_\perp=0$ depicts a pair of van Hove singularities (VHS) at the edges of the bands ($E=\pm 3$) and a finite density of states at all energies (Fig.\ref{f2}(a)). These VHS are characteristic of one-dimensional systems at the band edges, where the group velocity approaches zero. Once inter-leg tunneling is restored, another pair of VHS shows up at $ E=\pm t$, signaling the coupling between the two legs, as shown in Fig.\ref{f2}(b). When $\delta\neq 1$, the moiré potential gives rise to three effects: i) the peaks at the edges of the spectra are smeared out, ii) the VHS at $E=\pm t$ splits into two pronounced peaks symmetrically located about $E\sim\pm t$ due to the breaking of inversion symmetry by $\delta$, and iii) the DOS $\sim \eta$ at the Fermi energy, as the system becomes gapped (Fig.\ref{f2}(c-f)). While the plots seem symmetric in energy, the fine structure within the bands exhibits nontrivial modulation arising from the interference between the legs and the mismatch in periodicity induced by dimerization.  The degree of dimerization and the strength of inter-leg tunneling determine the overall shape and width of the energy bands, which, in turn, set the energy values of the band extrema and, consequently, the distance between the van Hove singularities.  The DOS of $H(k)$ and $H_{\rm RBZ}$ reflects quasiperiodicity arising from weakly dimerized coupling and reveals a rich structure due to the interplay between inter-leg tunneling and $\delta$.

A divergence or a sharp peak in the electronic susceptibility at a specific wave vector indicates an electronic instability,  suggesting that the system is susceptible to a periodic modulation of its electron density \cite{mermin1970lindhard}. Throughout this work, we distinguish between the primary lattice-driven modulation, which generates the moiré structure, and a secondary electronic instability that may develop within the resulting miniband spectrum. To examine this possibility, we computed the Lindhard formula, which gives the non-interacting (time-independent) charge response to a perturbation at $\kappa$ \cite{mermin1970lindhard},
\begin{equation}
\begin{split}
\chi_0(\kappa) =
\sum_{n,m} \int\frac{dk}{2\pi} \frac{f(E_n(k)) - f(E_m(k+\kappa))}{E_m(k+\kappa) - E_n(k) + i\eta} \,\\ | \langle u_{m}(k+\kappa) | u_{n}(k) \rangle |^2,
\end{split}
\end{equation}
$E_n(k)$ is the $nth$ energy band of $H(k)$,
$f(E)$ is the Fermi-Dirac distribution, and $|\langle u_m(k+q) | u_n(k) \rangle|^2$ is the overlap between the periodic part of the Bl\"och wavefunction, a form factor. 

In Figs.\ref{f3}(a-b), we show $\chi_0(\kappa)$ for $\delta=\frac{19}{20}$ at low temperatures $T=\frac{t_1}{5}=\frac{1-\delta}{5}$ and high temperatures $T=5t$. At low T, the wave vectors corresponding to the sharp (negative) peaks in the susceptibility $\chi(\kappa)$ are centered around $\kappa*=\pm qG_0$. At these special points, a significant portion of the band structure translates onto itself. Away from $\kappa*$, the figure shows a comb of satellite dips spaced by $\kappa*$ (near multiples of the moiré wavevector), with fine oscillations resulting from interband overlaps and the k-dependence of the eigenvectors. No strong feature is observed at other unrelated momenta; therefore, the static, non-interacting response is driven by the form factor tied to the rung-modulated hopping. Overall, at low temperatures,  $\chi(\kappa)$ is consistent with a finite-momentum CDW around $\kappa*$, set by the modulation period, where the satellites reflect higher harmonics of the moiré modulation and band–overlap structure. At large temperatures, $T>t$, the features at $\kappa*$ are suppressed, indicating reduced susceptibility to instabilities.

The linear response of $H(k)$ at low temperatures suggests that the system is prone to forming a CDW state. In this collective electronic phase, the charge is arranged in a periodic pattern. To gain insight into this possibility, we consider the system's site-charge density.  In the site basis, where $u_{i,n}$ is the \textit{i-th} component of the column-normalized eigenvector of band n of $H(k)$, the non-interacting charge densities at all sites $i$ of bands $n=1,2$,
\(\rho_{i,n}(k)=\! f(E_{n}(k))\, |u_{i,n}(k)|^2\)
display an oscillatory behavior around $1/2$, with dominant peaks around \(\kappa*\) and several satellite peaks of smaller amplitude, as shown in Fig.\ref{f3}(c). In contrast, the site-charge density of the bands $n=3,4$ is independent of $k$. Oscillations of the charge in bands 1 and 2 are rooted in the form factor $|u_{i,n}|^2$, which becomes dependent on $k$ and $\delta$ with finite rung hopping. In this case, $|u_{i,n}|^2$ shifts weight among sites as $k$ varies, while the occupancies $f(E_n(k))$ weigh those shifts. If CDW is present in the system, it indicates that the CDW-driving channels are a pair of bands, while the other two are symmetry-protected. The total charge density in the system is constant, $\rho(k)=\sum_{i,n}\rho_{i,n}(k)=2$, every unit cell carries exactly two particles (spinless), in four single-particle states per cell, consistent with the model at half filling. Charge modulation refers to the redistribution of charge among sites within a unit cell, without altering the net charge per cell. Consequently, the overall charge density remains unmodulated in the absence of interaction- or phonon-induced effects. Motivated by the reduced kinetic energy implied by the flattened bands, in the next section, we pursue the interaction-driven scenario and evaluate the susceptibility within the Random Phase Approximation (RPA), incorporating Coulomb interactions \cite{van2021random}. 
\section{\label{sec:cdw}Charge Density Waves}
\subsection{Short Range Coulomb interactions in  Mean Field Theory}
To study CDW, we add an interacting term to $H$ and construct the mean-field theory for the system.  Following the standard approach, we build a low-energy Lagrangian from the system's full Hamiltonian using a path-integral method \cite{mahan2013many}. In the path integral formulation, the trace over Hilbert space operators is re-expressed as an integral over a set of classical-like variables, which, for fermions, are Grassmann variables (fermionic coherent states) that anticommute \cite{mahan2013many}. After this replacement, we obtain the Euclidean Lagrangian, which becomes the integrand of the path integral. For converting the fully interacting Hamiltonian from its second-quantized form into a Lagrangian with Grassmann fields, we use the Hubbard-Stratonovich transformation (HS) \cite{altland2010condensed,tsvelik2007quantum}. This transformation replaces the four-fermion interacting term with a new term that couples the fermions to a fluctuating auxiliary bosonic field. This makes the path integral Gaussian and, therefore, solvable.

Before moving forward, we briefly justify our methodological choice. Because the ladder is effectively quasi-one-dimensional, it is natural to wonder whether the interacting problem could be solved exactly via bosonization \cite{giamarchi2003quantum,giamarchi1992resistivity}. However, although bosonization is a powerful framework for one-dimensional systems with only a few Fermi points, applying it directly to the current moiré ladder setup is not feasible in practice. The incommensurate offset between the two legs gives rise to a moiré supercell of size $q$, which in turn yields $4q$ minibands and low-energy states spread over many momenta separated by the moiré reciprocal vector. The CDW instability appears at a finite wavevector $Q^*$ determined by the moiré geometry, rather than by conventional $2k_F$ nesting, and features momentum-dependent form factors that couple several minibands. Realizing this structure within a bosonization framework would necessitate introducing many coupled bosonic fields with nonlocal interactions, which would obscure the physical origin of the instability. In contrast, the Hubbard–Stratonovich approach combined with RPA offers a transparent method for analyzing finite-$Q$ density-wave instabilities in multiband systems, and it gives direct access to the neutral phason mode and its dynamics.

In what follows, we consider intra-leg Coulomb interactions between sites within the same unit cell and neglect inter-leg interactions hereafter. The interacting Hamiltonian is built by assuming density-density interactions of equal strength $u$ in both legs. In momentum space,
\begin{equation}
\begin{split}
H_{\text{int}}
=\frac{u}{L}\sum_\kappa \big[\,\rho_1(-\kappa)\,\rho_1(\kappa)+\rho_2(-\kappa)\,\rho_2(\kappa)\,\big],\label{eq:Hint}\\
\rho_1(\kappa)=\sum_{k}\sum_{s\in\{A_1,B_1\}} c^\dagger_{k+\kappa,s}\,c_{k,s},\\
\rho_2(\kappa)=\sum_{k}\sum_{s\in\{A_2,B_2\}} c^\dagger_{k+\kappa,s}\,c_{k,s}.
\end{split}
\end{equation}
Equivalently, in the band basis 
$U_k^\dagger H(k) U_k=\mathrm{diag}(E_1,\dots,E_4)=\varepsilon_{\rm diag}$ and $c_{k,s}=\sum_n U_k(s,n)\,\gamma_{k,n}$, the leg densities and the leg-projected density vertices read (X indexes leg, $X\in (1,2)$ and n,m index the bands of $H$)
\begin{equation}
\begin{split}
\rho_X(\kappa)=\sum_{k}\sum_{n,m}\Gamma^X_{nm}(k,\kappa)\;
\gamma^\dagger_{k+\kappa,n}\,\gamma_{k,m},\\
\Gamma^{1}_{nm}=\!\!\sum_{s\in\{A_1,B_1\}}U_{k+\kappa}^{*}(s,n)U_k(s,m),\\
\Gamma^{2}_{nm}=\!\!\sum_{s\in\{A_2,B_2\}}U_{k+\kappa}^{*}(s,n)U_k(s,m).
\label{eq:vertices}
\end{split}
\end{equation}
where $\Gamma^{X}$ encodes the leg X weights of the band eigenstates. Discretizing imaginary time and taking the continuum limit $\tau\in[0,\beta)$ leads to the Euclidean action (at zero chemical potential)
\begin{equation}
\begin{split}
S[\bar\psi,\psi]=\int_0^\beta d\tau\;
\Big\{
\sum_k \bar\psi_k(\partial_\tau + H(k))\psi_k\\
+\frac{u}{L}\sum_\kappa\big[\rho_1(-\kappa)\rho_1(\kappa)+\rho_2(-\kappa)\rho_2(\kappa)\big]
\Big\}.
\label{eq:action}
\end{split}
\end{equation}
where the fields $\psi,\bar{\psi}$ are anti-periodic in imaginary time $\tau\in[0,\beta]$. Consider bosonic HS fields $\phi_X(\kappa,\tau)$. Next, we use the HS transformation, which relies on the Gaussian identity to linearize the quartic term:
\begin{equation}
\begin{split}
e^{-\frac{u}{L}\int d\tau\,\rho_X(-\kappa)\rho_X(\kappa)}=\\
\int\!\mathcal D\phi_X\,\exp\Big\{-\int d\tau\Big[\frac{L}{u}\,|\phi_X(\kappa,\tau)|^2\\
-\phi_X(\kappa,\tau)\rho_X(-\kappa)-\phi_X^*(\kappa,\tau)\rho_X(\kappa)\Big]\Big\}
\end{split}
\end{equation}
After decoupling for both legs,
\begin{equation}
\begin{split}
S_{HS}[\bar\psi,\psi,\phi]=
\int_0^\beta d\tau\;\Bigg\{
\sum_k \bar\psi_k(\partial_\tau + H(k))\psi_k\\
+\frac{L}{u}\sum_\kappa\big(|\phi_1(\kappa,\tau)|^2+|\phi_2(\kappa,\tau)|^2\big)\nonumber\\
-\sum_\kappa\Big[\phi_1(\kappa,\tau)\,\rho_1(-\kappa,\tau)+\phi_2(\kappa,\tau)\,\rho_2(-\kappa,\tau)\Big]\Bigg\}
\end{split}
\label{eq:SHS}
\end{equation} 
In the band basis, the last term reads
\begin{equation}
\begin{split}
\sum_{k,\kappa}\sum_{n,m}\Big[\,\phi_1(\kappa,\tau)\,\Gamma^1_{nm}(k,\kappa)+\\\phi_2(\kappa,\tau)\,\Gamma^2_{nm}(k,\kappa)\,\Big]\;
\bar\gamma_{k+\kappa,n}\,\gamma_{k,m}+\mathrm{h.c.}
\label{eq:Yukawa-band}
\end{split}
\end{equation}
with the vertices from \eqref{eq:vertices}.
We focus on a single ordering wavevector $Q$ and retain only $\kappa=\pm Q$ modes. The complex order parameter reads
\begin{equation}
\begin{split}
\Delta_X(\tau)\equiv \phi_X(Q,\tau),\quad \Delta_X^*(\tau)\equiv \phi_X(-Q,\tau)
\end{split}
\end{equation}
Then $S_{HS}[\bar\psi,\psi,\phi]$ reduces to
\begin{multline}
S_{HS}=\int_0^\beta d\tau\Bigg\{
\sum_k\bar\gamma_k(\partial_\tau + \varepsilon_{\text{diag}}(k))\gamma_k\\
+\frac{L}{u}\sum_{X=1,2}|\Delta_X|^2 \\
-\sum_k\sum_{n,m}\Big[\Delta_1\Gamma^1_{nm}(k,Q)+\Delta_2\Gamma^2_{nm}(k,Q)\Big]\;
\bar\gamma_{k+Q,n}\,\gamma_{k,m}\\+\mathrm{h.c.}\Bigg\},
\label{eq:SHS-Q}
\end{multline}
The HS Lagrangian density associated with \eqref{eq:SHS-Q} is
\begin{multline}
\mathcal L_{\text{HS}}=
\sum_k \bar\gamma_k(\partial_\tau + \varepsilon_{\text{diag}}(k))\gamma_k
+\frac{L}{u}\sum_X |\Delta_X|^2 \nonumber\\
-\sum_k\sum_{n,m}\Big[\Delta_1\,\Gamma^1_{nm}(k,Q)+\Delta_2\,\Gamma^2_{nm}(k,Q)\Big]\,
\bar\gamma_{k+Q,n}\gamma_{k,m}
\end{multline}
At the saddle point, the order parameter is time independent $\Delta_X(\tau)\to\Delta_X$. The fermions appear quadratically and can be integrated out using the Grassmann Gaussian identity \cite{tsvelik2007quantum}, which yields the effective action:
\begin{equation}
S_{\text{eff}}[\Delta]=\beta\frac{L}{u}\sum_X|\Delta_X|^2 - \Tr\ln\big[\mathcal G_0^{-1}-\hat\Phi(\Delta)\big],
\label{eq:Seff}
\end{equation}
with $\mathcal G_0^{-1}(i\omega,k)=i\omega\mathbb{I}_{4}-\varepsilon_{\text{diag}}(k)$, and
\begin{equation}
\begin{split}
\hat\Phi(\Delta)=\sum_{k}\sum_{n,m}\Big[\Delta_1\,\Gamma^1_{nm}(k,Q)+\Delta_2\,\Gamma^2_{nm}(k,Q)\Big]\\\,
\ket{k+Q,n}\bra{k,m}+\mathrm{h.c.}\nonumber
\end{split}
\end{equation}
Here, \textit{Tr} refers to a full trace over Matsubara frequencies, momenta, and internal indices (band, leg, site). 

Stationarity $\delta S_{\text{eff}}/\delta \Delta_X^*=0$ yields the gap equations
\begin{equation}
\frac{L}{u}\,\Delta_X
=\sum_{k}\sum_{n,m}\Gamma^X_{nm}(k,Q)\;
\big\langle \bar\gamma_{k+Q,n}\,\gamma_{k,m}\big\rangle_{\Delta_1,\Delta_2}
\label{eq:gap-general}
\end{equation}
Let $(m,n)$ be the dominant pair of bands at $Q$. Consider a separable form factor
\begin{equation}
\begin{split}
F(k)\equiv \alpha_1\,\Gamma^1_{nm}(k,Q)+\alpha_2\,\Gamma^2_{nm}(k,Q),
\end{split}
\end{equation}
and the order parameter \(\Delta_k\equiv \Delta_0\,F(k)\) with \(\Delta_0\in\mathbb C\).
After expanding the term $\rm Trln$ to quadratic order in Eq.\ref{eq:Seff}, the linear contribution vanishes, and Eq.\ref{eq:Seff} becomes the Gaussian action. 
Restricting to the dominant $(m,n)$ and choosing a gauge where $\Delta_0$ is real, the mean-field Lagrangian becomes
\begin{multline}
\mathcal L_{\text{MF}}=
\sum_{k}\bar\Psi_k\big(\partial_\sigma+E_+\sigma_0+E_-\sigma_3+\Delta_0 F(k)\sigma_1\big)\Psi_k\\
+\frac{L}{u}|\Delta_0|^2,
\label{eq:LMF}
\end{multline}
where 
\begin{equation}
\begin{split}
E_\pm(k)=\frac{E_m(k)\pm E_n(k{+}Q)}{2}.
\end{split}
\end{equation}
$\Psi_k=(c_{k},c_{k+Q})^\top$ and the form factor $F(k)$ are the k-dependent weights with which the CDW order parameter couples states at $k\pm Q$. The mean-field Hamiltonian is
\begin{multline}
\mathcal H_{\text{MF}}(k)=E_+(k)\sigma_0+E_-(k)\sigma_3+\Re\Delta_k\,\sigma_1\\-\Im\Delta_k\,\sigma_2,
\label{eq:HMF}
\end{multline}
The quasiparticle spectrum of Eq.~\ref{eq:HMF}, $\tilde{E}(k)=E_+\pm\sqrt{|\Delta_k|^2+E_-^2(k)}$ consists of two dispersion branches, and the energy separation $2|\Delta_k|$ represents the CDW gap opened by electron-hole mixing.

The single-particle retarded Green's function describes the propagation of electron-like excitations with momentum $k$ and real frequency $\omega$,
\begin{equation}
G^R(k,\omega)
=\sum_{m}
\frac{|\tilde{u}_{m}(k)|^{2}}{\omega - \tilde{E}_{m}(k) + i\eta},
\label{eq:GR_MF_scalar}
\end{equation}
where $|\tilde{u}_{m}(k)|^{2}$ is the total weight of the eigenstate $m$ of $\mathcal{H}_{\rm MF}$. The associated spectral function is the imaginary part of $G^{R}$
\begin{equation}
A(k,\omega)=
-\frac{1}{\pi}\,\mathrm{Im}\,G^{R}(k,\omega)\label{eq:spectral_scalar}
\end{equation}
and measures the density of available states at momentum $k$ and energy $\omega$. It controls the ARPES intensity via $I_{\text{ARPES}}(k,\omega) \;\propto\; f(\omega,T)\, A(k,\omega)$, with $f(\omega,T)$ the Fermi–Dirac distribution. The map of ARPES intensity, shown in Fig.~\ref{f5}(a), displays two mean field quasiparticle bands. The two bright arcs (upper and lower) correspond to the dispersions of the MF quasiparticles. The dark area between them is the gap region where no spectral weight exists. 
\subsection{RPA susceptibility and Charge Density Waves}   The matrix of leg-resolved (leg-channel) bare susceptibilities \(\chi^{(0)}(\kappa,i\omega)\) is obtained by expanding the term \textit{Trln} in the effective action to quadratic order, Eq.\ref{eq:Seff} \cite{mahan2013many}.  In the static limit,
\(\chi^{(0)}_{XY}(\kappa)\equiv \chi^{(0)}_{XY}(\kappa,i\omega{=}0)\),
\begin{widetext}
\begin{equation}
\chi^{(0)}_{XY}(\kappa)
=\frac{1}{N_k}\sum_{k}\sum_{n,m}
\frac{f\!\big(E_m(k)\big)-f\!\big(E_n(k{+}\kappa)\big)}
{E_m(k)-E_n(k{+}\kappa)+i\eta}
\Gamma^{X}_{nm}(k,\kappa)\,\Gamma^{Y}_{nm}(k,\kappa)^{*}
\label{eq:chi0-static}
\end{equation}
\end{widetext}
Next, we consider intra-leg density-density repulsive Coulomb interactions of equal strength \(u\) on both legs. In the leg channel, the interaction matrix is 
\begin{equation}
V=\begin{pmatrix} u & 0\\ 0 & u\end{pmatrix}=u\,\mathbb{I}_{2}.
\end{equation}
Because $\Delta$ couples linearly to the leg density, the functional derivative gives the density response; thus, the interacting susceptibility is
\begin{multline}
\chi(\kappa,i\omega)=\big[\,\mathbb{I}_2-V\,\chi^{(0)}(\kappa,i\omega)\,\big]^{-1}\,\chi^{(0)}(\kappa,i\omega)\\
=\big[\,V^{-1}-\chi^{(0)}(\kappa,i\omega)\,\big]^{-1}.
\label{eq:RPA}
\end{multline}
In the static limit \(i\omega\!\to\!0\), this gives \(\chi(\kappa)\).

At a given temperature T, a density-wave instability occurs when Eq.~\ref{eq:RPA} diverges, 
\begin{equation}
\det\!\big[\mathbb{I}_2-V\,\chi^{(0)}(\kappa,T)\big]=0
\end{equation}
i.e., when the largest eigenvalue of the \emph{kernel}
\(V\chi^{(0)}(\kappa,T)\) reaches \(+1\). For \(V=u\,\mathbb{I}_2\), the eigenvalues of \(V\,\chi^{(0)}(\kappa)\) are \(u\) times the eigenvalues
of \(\chi^{(0)}(\kappa)\). Let \(\lambda_{\max}^{(0)}(\kappa)\) be the largest eigenvalue of \(\chi^{(0)}(\kappa)\), the instability criterion reads
\begin{equation}
u\,\lambda_{\max}^{(0)}(\kappa,T)=1,\quad
Q*=\underset{\kappa}{\arg\max}\ \lambda_{\max}^{(0)}(\kappa,T)\ 
\label{eq:lambda-criterion}
\end{equation}
at the transition. The instability wavevector \(Q*\) is the point at which the largest eigenvalue of the leg-projected bare matrix \(\chi^{(0)}(\kappa,T)\) is maximal. 
In Fig.\ref{f3}(d), we show the largest eigenvalue $\lambda_{\max}^{(0)}(\kappa,T)$ of $\chi^{(0)}(\kappa,T)$ and the leg-1 and leg-2 projected density–density susceptibilities for $\delta=\frac{19}{20}$ and $u=1.2$. At low temperatures, the system is unstable to a CDW with wavevector $Q^*=p\pi=pG_0$ (as indicated by the vertical dotted red line). At high temperatures, the peaks are smeared out, and instability does not occur.

The critical coupling $u_c=1/\lambda_{\max}^{(0)}(Q^*,T)$ represents the smallest Coulomb interaction which, at fixed $T$, drives the system unstable toward a charge-density-wave phase. For $u>u_c$, a CDW with wavevector $Q^*$ develops in the system. Likewise, for a given u, the critical temperature is defined by \(
g(T)\equiv 1- u\;\lambda_{\max}^{(0)}(Q*,T_c)\;=\;0.
\), therefore, $u$ and $T_c$ lie on the same instability curve. Fig.\ref{f4}(a) shows that $u_c(T)=1/\lambda_{\rm max}(T)$ increases monotonically with \(T\). At low temperatures, $T<t$, $u_c$ rises quadratically, while at higher temperatures, it grows roughly linearly. 
Because \(Q^*\) follows the moiré geometry, the enhancement of \(\chi_0\)
with increasing \(\delta\) implies a lower \(u_c(\delta)\) hence a larger
\(T_c(\delta)\).

The eigenvector corresponding to $\lambda^{0}_{\rm max}$ is $v_{\rm odd}=\frac{1}{\sqrt{2}}(-1,1)$, indicating a leg-odd mode. Furthermore, the leg-resolved weights at Q* are each $\sim 0.5$, so the RPA-unstable eigenmode is evenly distributed over the two legs. The charge modulation on the two legs is opposite or out-of-phase \(\delta n_{1}(x)\sim \cos{(Q^*x)},\quad \delta n_{2}(x)\sim \cos{(Q^*x+\pi)}\).  Therefore, this is a rung-odd density wave where the net charge modulation on a rung cancels to leading order. 
\subsection{Sliding Charge Density Waves}
The moiré reciprocal periodicity is $\frac{\pi}{q}$, so the unstable mode sitting at $Q*$, which is not an integer multiple of a moiré reciprocal vector, shows that, relative to the moiré lattice, this is an incommensurate CDW whose periodicity breaks the translational symmetry of the moiré lattice.  However, $Q*=p\pi$ is an integer multiple of $G_0=\pi$, so the CDW state commensurates with the atomic scale and sits at the $\Gamma$ point in the RBZ. Therefore, when $Q*$ folds into $\Gamma$, there is an intracellular charge accumulation.
In a CDW state, its phase field $\theta(x)$ plays an analogous role to the displacement field in elasticity, describing distortions in the periodic modulation or CDW phasons \cite{coddens2006problem}. The phason velocity (\(\nu_\theta\)) indicates how easily the CDW translates as a whole in the lattice. A high phase velocity implies a very stiff CDW that resists changes in its phase, while a low speed indicates a soft CDW that can be easily displaced \cite{tucker1988theory}. In the leg channel, the CDW phason tied to the total-density modulation in our setting belongs to the odd neutral sector. This corresponds to configurations in which charge is added to both legs and to both sites with opposite phases. Because this phason is neutral under long-range Coulomb forces, the long-wavelength mode is expected to remain gapless and acoustic, and to exhibit a linear dispersion $\omega \sim \nu_{\theta}|k|$ \cite{wilson1985charge,fisher1985sliding,brazovskii2004pinning,PhysRevLett.42.1423}. 

Since the RPA two-by-two kernel is on a leg basis, to obtain the charge modulation on each microscopic site, we project the unstable leg mode back to the sites using the unitary matrix of eigenvectors $U(k)$ projected into the odd channel, $U_-(k)$. The site-resolved form factor at momentum $Q^*$ is 
\begin{widetext}
\begin{equation}
F_s(Q^*)=
\frac{1}{N_k}\sum_k\sum_{m,n}\frac{f(E_m(k))-f(E_n(k+Q^*))}{E_m(k)-E_n(k+Q^*)+i\eta}U^*_{-sn}(k+Q^*)U_{-sm}(k)
\end{equation}
\end{widetext}
The lattice charge modulation, shown in Fig.\ref{f4}(b), \(\delta n_{s}(x)\propto \Re[F_s(Q^*) e^{i Q^* x}]\) oscillates between legs.

\section{\label{sec:phason}Phason dynamics}
To compute the phason dispersion, we must go beyond the static charge-density picture and calculate the retarded RPA susceptibility, where the phason is a pole when the system is in the CDW state ($T<T_c$). The retarded leg-resolved Lindhard function 
\(\Pi^R(q,\omega)\), a \(2\times 2\) matrix in the basis of legs is
\begin{widetext}
\begin{equation}
\Pi^{R}_{XY}(\kappa,\omega)
= \frac{1}{N_k}\sum_{k}\sum_{n,m}
\frac{ f\!\big(\tilde{E}_m(k)\big) - f\!\big(\tilde{E}_n(k{+}\kappa)\big) }
     { \omega + i\eta + \tilde{E}_m(k) - \tilde{E}_n(k{+}\kappa) }\,
\tilde{\Gamma}^{X}_{nm}(k,\kappa)\,\tilde{\Gamma}^{Y}_{nm}(k,\kappa)^{*}
\label{eq:bubble}
\end{equation}
\end{widetext}
Leg projected vertices are computed using the quasiparticle spectrum.
The retarded RPA susceptibility is
\begin{equation}
\chi^{R}(\kappa,\omega)=\Big[\mathbb{I}_2 - V\,\Pi^{R}(\kappa,\omega)\Big]^{-1}\,\Pi^{R}(\kappa,\omega).
\end{equation}
 Poles (the collective modes of the CDW state) are zeros of the  kernel
\begin{equation}
\det\!\Big[\mathbb{I}_2 - V\,\Pi^{R}(\kappa,\omega)\Big]=0 .
\label{eq:pole_condition}
\end{equation}
We project the retarded Lindhard function onto the odd channel
\begin{equation}
\Pi^{R}_{-}(\kappa,\omega)= e_-^\dagger\,\Pi^{R}(\kappa,\omega)\,e_-,
\nonumber
\end{equation}
Then
\begin{equation}
\chi^{R}_{-}(\kappa,\omega)
= \frac{\Pi^{R}_{-}(\kappa,\omega)}{1-u \Pi^{R}_{-}(\kappa,\omega)} ,
\end{equation}
\begin{equation}
\text{poles:}\;\; 1- u\Pi^{R}_{-}(\kappa,\omega)=0 .
\end{equation}
The pair $(\omega,\kappa)$ that satisfies the previous equation defines the phason dispersion $\omega(\kappa)$ at the ordering wave-vector \(Q^{*}\), and a higher-energy gapped collective mode. The two-particle spectral function 
\begin{equation}
\mathcal{A}_-(\kappa,\omega)=-\frac{1}{\pi}\,\Im\,\chi^{R}_-(\kappa,\omega)
\label{eq:spectralphason}
\end{equation}
shown in Fig.\ref{f5}(b), depicts three regimes. The lowest band $(\omega\lesssim -1.3)$ corresponds to the particle–hole continuum due to the imaginary part of the retarded bubble, \(\Im \Pi^{R}(\kappa,\omega)\), which corresponds to the density of states of pair-hole excitations. It measures the energy that must be added to the system to create a particle–hole pair by removing an electron from band n with momentum k and adding an electron to band m with momentum $k+Q^*$. The middle bright dispersive band \((-1.2\lesssim\omega \lesssim -0.8)\) corresponds to the amplitudon, a Higgs-like oscillation of the magnitude of the excitonic order parameter. The upper bright, sharp dispersive branch \((-0.7\lesssim \omega \lesssim 0)\) shows the poles of the RPA susceptibility that produce the gapless phason. The CDW ordering vector is \(Q^{*}\), so the Goldstone mode softens at  \(\kappa\sim Q^{*}\).
\subsection{Moiré Phason}
To study the sliding (phason) mode we start from the mean–field Lagrangian, Eq.\ref{eq:LMF} and promote the order parameter to a slowly varying
complex field, \(\Delta_0 F(k)\,e^{i\theta(\tau,x)}\), which carries a slowly varying phase, and a fixed amplitude $\Delta_0$ (amplitude fluctuations are gapped). As above, $F(k)$ is the form factor.
The unitary rotation \(R_\theta=e^{+\frac{i}{2}\theta\tau_3}\) removes the phase
from the gap term, and, to linear order in gradients, minimally couples the fictitious gauge field \(a_\mu=\tfrac12\partial_\mu\theta\) to the
fermions \cite{chaikin1995principles}. 
Integrating out the gapped fermions and expanding the fermionic determinant to second order in $a_\mu$ gives the action for the phason in the long-wavelength, low-frequency limit.
\begin{multline}
S_{\rm ph}=\frac12 \int d\tau dx\;
\Bigl[
K\,(\partial_\tau\theta)^2+\rho\,(\partial_x\theta)^2
\Bigr],\\\omega^2(k)=\nu_{\theta}^2 k^2,\;\;
\nu_{\theta}^2=\rho/K.
\end{multline}
with \(K\) and \(\rho\) the dynamic and static stiffness of the phase. For weak coupling, at $T<T_c$ and at $k\sim Q*$, a gradient expansion retains terms $\omega^2$ and $k^2 $ with (BCS) coefficients \cite{gruner1988dynamics}
\begin{multline}
K=\frac14\sum_{k,b}\frac{|\Delta_{k,b}|^2}{E_{k,b}^3}\tanh\frac{E_{k,b}}{2T}
,\\
\rho=\frac{1}{2}\sum_{k,b}v_{-}^{2}\frac{|\Delta_{k,b}|^2}{E_{k,b}^3}\tanh\frac{E_{k,b}}{2T}
\label{eq:stif}
\end{multline}
where $v_{-}=\partial_k E_{-}(k,b)$, $|\Delta_{k,b}|^2=|\Delta_0 F_-(k)|^2$ and $F_-(k)$ is the form factor projected in the odd channel.
Instead of solving the linearized gap equation, a good estimate of $\Delta_0$ can be predicted from the ordered-state MF form factors at $k=k*$; the CDW \textit{hot spots}. Momentum $k*$ marks the points of the Fermi surface connected by $Q*$ with the largest susceptibility — these are the so-called \textit{nested} segments. $k*$ can be estimated by minimizing $|E_m(k)-E_n(k+Q*)|$ with respect to k and identifying the relevant bands. At $\delta=\frac{19}{20}$, we found $m=n=4$, $k*\sim 21.3$, and thus $\Delta_0\sim 0.7$. The intraband character of the instability is not universal; rather, it arises from the inclusion of short-range intra-leg interactions that preserve band independence. In this channel, the instability gives a neutral phason.

The mean-field gap mixes single-particle states at k and k + Q*. In addition, it creates the collective phase-oscillation mode of that condensate. Having $\Delta_0$ and the relevant band, and remembering that the energy scale is set by the hopping constant $t$, we find from Eq.\ref{eq:stif} the spatial stiffness (elastic constant) $\rho\sim 0.32$ (in units of $t$), and the dynamical stiffness (inverse inertia constant) $K\sim 6.7$ (in units of $t^{-1}$). Since $\rho/K\sim 0.05$; this is a soft mode. The phason speed $\nu_\theta\sim 0.2-0.17 i$, in units of $\frac{a t}{\hbar}=v_F$, where $v_F$ is the Fermi velocity. Therefore $\nu_\theta\ll v_F$. For typical values of the tunneling amplitude $t\sim 0.1-1$ [eV] and lattice constant $a\sim 3-5$ [A], we obtain for the phason velocity $\nu_\theta\sim 2\times 10^4-2\times 10^5$ [m/s]. $\nu_\theta$ can be compared, for instance, with the speed of acoustic phonons, which is on the order of $v_{ph}\sim 10^3-10^4$ [m/s]. The damping ratio is defined as $\operatorname{Im}[\nu_\theta]/\operatorname{Re}[\nu_\theta]\sim 0.05$; therefore, for the slight inter-leg mismatch, these soft phasons are long-lived.  

Near the hot spots, one can estimate how the phason speed scales with t and $\delta$. Minibands appear whenever a small periodic modulation couples states separated by multiples of the moiré reciprocal vector $b_s$. In our model, moiré minibands exist only because of the moiré rung-modulation. As we established above, the rung modulation comes from the term \((1−\delta)\,t\cos(q\,k)\), so the entire superlattice potential is proportional to $(1-\delta)$.  In perturbation theory, the miniband width scales like 
\begin{equation}
    W(\delta)\sim \frac{V_{\text{moire}}^{\text{eff}\,2}}{\Delta E}
\end{equation} and in the weak-coupling limit, $W(\delta)\sim V_{\text{moire}}^{\text{eff}}$. In our model, the effective moiré modulation amplitude is $V^{\mathrm{eff}}_{\mathrm{moire}} \sim (1-\delta)t$. Therefore $W(\delta)\sim (1-\delta)t$.  

In the CDW phase, the phason stiffness \(\rho\) is primarily determined by the energetic cost of spatial phase distortions, which in turn depends on the effective density of states \(N_{\mathrm{eff}}\) and the Fermi velocity of the low-energy band undergoing condensation \(\rho \sim N_{\mathrm{eff}} v_F^2 \). In the ladder, the relevant Fermi velocity in the miniband comes from the miniband width
$v_F \sim \frac{W(\delta)}{\Delta k} \propto (1 - \delta)\, t$. Plugging that in yields
\(\rho \sim  N_{\mathrm{eff}} (1 - \delta)^2 t^2\). 
As minibands flatten when $\delta\to 1$, the effective density of states scales like $N_{\mathrm{eff}}\sim const/t$. That gives
\begin{equation}
    \rho \propto (1 - \delta)^2 t
\end{equation}
The inertia $K$ of the phason comes from the $\omega$-dependence of the dynamical bubble $\Pi^R(k,\omega)$ at 
$k\to 0$ and the scaling is dominated by
\begin{equation}
K \sim 1/t
\end{equation}
Therefore,
\begin{equation}
\nu_{\theta}(\delta)\sim W(\delta)\;\propto\;t(1-\delta).
\end{equation}
In summary, increasing $t$ raises \(v_F\), and with that,
\(\rho(t)\) and \(\nu_{\theta}(t)\) increase roughly linearly. On the other hand, \(K(t)\) decreases with $t$
as \(1/v_F(t)\).
Altogether, turning up \(t\) speeds up phasons, while pushing \(\delta\to 1\) slows them down. 

As a final observation, when leg-2 becomes nearly fully dimerized, the minibands disappear because the low-energy spectrum develops an almost perfectly flat bonding–antibonding structure,
\(E_{\pm}(k) \sim \pm t \delta + \mathcal{O}(1 - \delta)\).
One of the bands thus becomes essentially flat, and the rung-induced hybridization with leg-1 is strongly reduced, since the cosine modulation can no longer generate substantial momentum-space replicas to narrow the minibands. The intra-cell splitting in the dimerized chain is \(\Delta_{\mathrm{dimer}} \sim 2 t \delta\); once this gap is large, the modulation is unable to mix the states efficiently.
\section{\label{sec:dft}DFT calculations}
\begin{figure}
\centering
\includegraphics[width=0.95\linewidth]{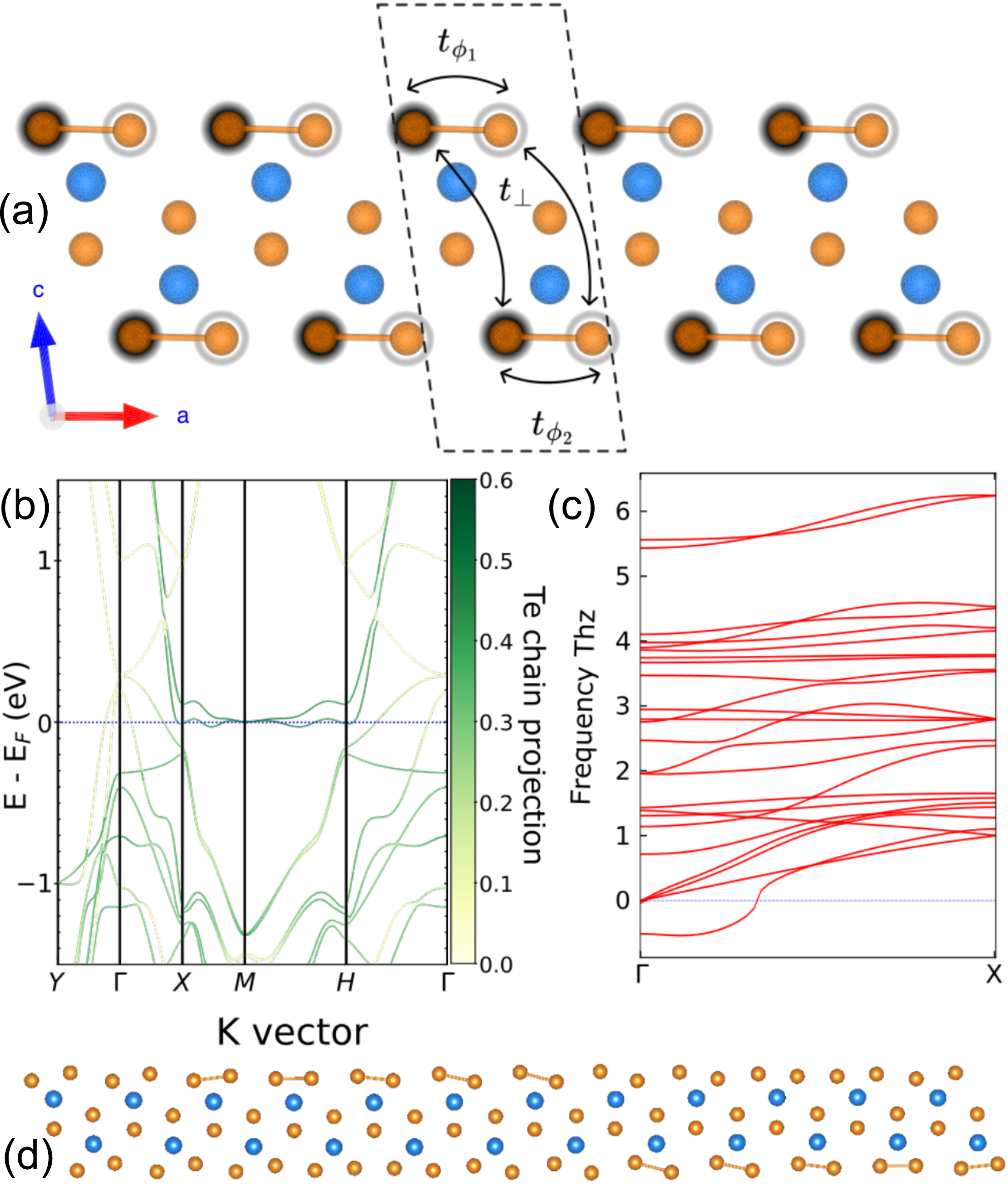}
\caption{(a) Structural representation of monolayer $\mathrm{HfTe_3}$. Hf atoms are shown in blue and Te atoms in orange. The Te atoms belonging to the dimerized chains are additionally shaded in white and black, indicating the A and B sublattice positions of the model; the corresponding intra- and inter-chain hoppings are also illustrated. In the minimum-energy configuration, both Te chains are dimerized but exhibit a relative phase shift, reflected in the condition $t_{\phi_1} \neq t_{\phi_2}$. The dashed lines indicate the unit cell, which remains open along one direction due to the monolayer geometry.
(b) DFT band structure of the corresponding surface configuration, with states projected onto Te atoms.
(c) Phonon dispersion calculated using DFPT, revealing a clear instability along the $\Gamma \to X$ path.
(d) Real-space distortion associated with the negative-frequency modes; this distortion defines the enlarged unit cell used in the moiré analysis. For visualization purposes, the distortion amplitude has been artificially enhanced.}
\label{fig:HfTe3_model}
\end{figure}
\begin{figure}
\centering
\includegraphics[width=0.9\linewidth]{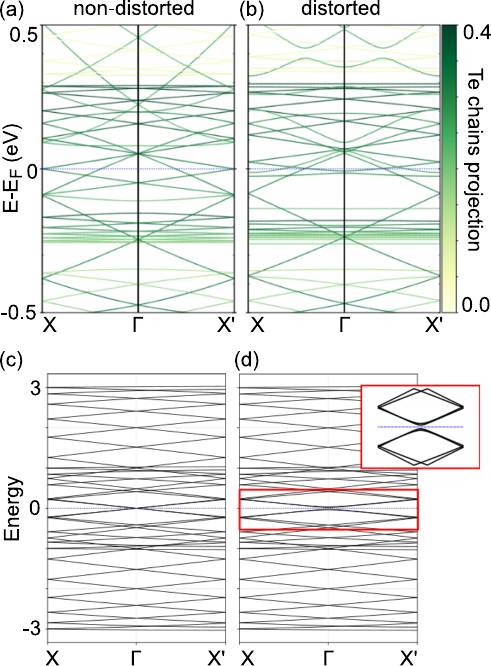}
\caption{Comparison between first-principles results and the moiré-like tight-binding model for $\mathrm{HfTe_3}$.
(a,b) DFT band structures for a $12 \times 1 \times 1$ supercell: (a) unperturbed configuration and (b) distorted structure with a phase difference between the two dimerized Te chains.
(c,d) Corresponding band structures obtained from the effective moiré-like Hamiltonian: (c) unperturbed model and (d) modulated model including the chain-dependent phase shift.}
\label{fig:bands-12}
\end{figure}
We illustrate the relevance of our moiré model by applying it to $\text{HfTe}_3$, a quasi-one-dimensional compound whose chain-dependent dimerization and weak coupling between chains make it a natural realization of the moiré-induced CDW mechanism investigated in this work. We focus on a single monolayer of $\text{HfTe}_3$. Because experimental probes access the phason dynamics in $\text{HfTe}_3$ primarily through transport and collective response, rather than direct dispersion measurements, our comparison focuses on microscopic origin, symmetry, and characteristic velocity scales. 
$\text{HfTe}_3$ crystallizes in a monoclinic structure with space group P2$_1$/m \cite{abdulsalam2015structural, liu2021quasi}. As illustrated in Fig. \ref{fig:HfTe3_model}.a, the lattice consists of a multichain structure: infinite prismatic $\text{HfTe}_3$ chains extending along the $b$-axis, and two dimerized Te–Te chains along the $a$-axis. These two-dimensional slabs stack along the $c$-axis through weak van der Waals (vdW) interactions. Due to its structural motif, $\text{HfTe}_3$ belongs to the broader family of transition-metal trichalcogenides (TMTCs), commonly denoted as $MX_3$ with
$M = \mathrm{Ti, Zr, Hf, V, Nb, Ta}$ and $X = \mathrm{S, Se, Te}$. These materials form quasi-one-dimensional (1D) van der Waals (vdW) crystals composed of strongly covalently bonded chains, weakly coupled through interchain interactions. This quasi-1D character has been shown to host a variety of emergent quantum phenomena, including charge-density waves, exotic transport behavior, and tunable correlated states \cite{zhao2018band, li2017anisotropic,island2016titanium, denholme2017coexistence}. In this context, $\text{HfTe}_3$ provides an ideal testbed for bridging our simplified theoretical construction with the more complex physics of the TMTC family, whose structural anisotropy,  chain-dependent coupling, and presence of CDW and superconductivity  \cite{li2017anisotropic, denholme2017coexistence, liu2021quasi} make them candidates for moiré-like modulation effects.

Although experimental CDWs have been found in $\text{HfTe}_3$ bulk systems \cite{denholme2017coexistence,liu2021quasi}, several recent works have explored routes to reduce their effective dimensionality \cite{meyer2019metal,chen2024thickness}, motivated by the weak vdW coupling between its chain-like units. From the perspective of the electronic structure, the surface (Fig.\ref{fig:HfTe3_model}(b)), and bulk remain remarkably similar: both exhibit nearly flat bands along the $b$-direction and significantly more dispersive bands along the $a$-direction (see Appendix \ref{sec:dft}). The crucial difference, instead, emerges from the dynamics. In the bulk system, the interlayer vdW interactions provide sufficient restoring forces to stabilize the vibrational spectrum, whereas at the surface, the absence of these couplings leads to a phonon instability (Fig.\ref{fig:HfTe3_model}(c)). The vibrational mode responsible for the structural instability (Fig.\ref{fig:HfTe3_model}(d)) corresponds to a displacement of the dimerized Te atoms primarily along the $a$–$c$ plane, forming a pattern that is symmetry-allowed in both bulk and surface geometries. The mentioned mode becomes unstable only at the surface, where the absence of interlayer vdW restoring forces fails to suppress the distortion. Phonon modes involving the external Te chains have been reported in experimental studies of related TMTCs, most notably ZrTe$_3$ \cite{gleason2015structural, hu2015charge}, where they exhibit strong coupling to the conduction electrons associated with CDW formation. The oscillatory pattern of this mode (Fig. \ref{fig:HfTe3_model}(d)) produces a periodic modulation in the relative positions of the \textit{black–white} Te pairs, with each chain oscillating with a distinct phase. This phase-shifted motion naturally leads to modulation of the hopping between sites intra-chain, $t_{\phi_i}\to t_{\phi_i}(n)$, providing the microscopic mechanism by which lattice vibrations generate the moiré-like mismatch, validating the central hypothesis of our model. Our DFPT calculations along the $\Gamma \to X$ path identify the lowest unstable phonon at $q=(0.06,0,0)$, with a minimum frequency of –0.54 THz. To remain within the dynamically unstable region while keeping the computational cost manageable, we construct a $12 \times 1 \times 1$ supercell using the real-space distortion obtained from the DFPT eigenvector. This choice is close to the $16 \times 1 \times 1$ supercell required to reach the exact energy minimum, yet still captures the essential features of the long-wavelength modulation. 
Upon structural relaxation, the distorted supercell exhibits only a marginal decrease in total energy compared to the undistorted structure. Nevertheless, the atoms relax into positions that display periodic modulations (with a maximal amplitude of 0.03 \AA) across the enlarged unit cell, consistent with a moiré-like displacement field. This macroscopic modulation underscores the lattice's softness and the robustness of the relaxed structure's moiré ground state. These results indicate that chain systems can not only exhibit an experimental random mismatch shift but, in this case, $\text{HfTe}_3$, naturally tend to a configuration characterized by long-wavelength structural oscillations. This distortion provides a small energy gain, further stabilizing the few-layered system.

To accurately describe the system within the moiré framework, we combined DFT calculations with Wannier projection and downfolding to extract the hierarchy of electronic couplings in the undistorted structure. The Wannier Hamiltonian reveals strong intra-chain hoppings within the dimerized Te chains and much weaker inter-chain tunneling mediated by the Hf–Te zigzag layer, identifying the essential channels for a minimal model. The long-wavelength structural modulation was obtained from the relaxed geometry associated with the soft phonon mode: the Te$(A_i)$–Te$(B_i)$ bond lengths exhibit smooth periodic oscillations with distinct phases on the two chains.
This observation necessitates a refinement of our theoretical Hamiltonian. While we preserve the overall structure of Eq. \ref{eq:H4eff}, we now incorporate an oscillatory intra-chain hopping on both chains, consistent with the distortion pattern obtained from the DFPT phonon mode. The modified model therefore includes a chain-dependent $c_\delta(k)\,e^{\pm i\phi_{1,2}(k)}$, as illustrated in Fig. \ref{fig:HfTe3_model}.a and in the equation \ref{eq:H4-HfTe3}. This refinement captures the essential physics introduced by the real-space modulation—namely, the chain-selective dimerization and the resulting moiré-like variation in the hopping phases $\phi_i$.\\
To compare the DFT results with the moiré tight-binding model, we focus on the perturbed but unrelaxed electronic structure obtained by displacing the atoms along the unstable mode while keeping the periodicity of the original cell. In this configuration, the two inter-chain hoppings 
$t_\perp (A_1-A_2)$ and $t_\perp (B_1-B_2)$ remain equivalent, preserving the symmetry and translational invariance required for a well-defined RBZ folding and a direct correspondence with the effective Hamiltonian. After full structural relaxation, the system develops inequivalent $t_\perp$ bonds and loses strict periodicity, which complicates the interpretation of the bands and obscures their relation to the reduced model. Nevertheless, this relaxed geometry represents a physically meaningful state that may be stabilized or tuned in realistic settings—for example, through substrate coupling, external pressure, epitaxial strain, or interface-induced mismatch—suggesting that the modulation captured by the model reflects a broader class of experimentally accessible moiré-type distortions.\\
In the high-symmetry configuration—corresponding to the unstable structure—the modified Hamiltonian with equal phases on both chains $t_{\phi_1} = t_{\phi_2} $ yields reduced-Brillouin-zone (RBZ) bands that include Dirac cones at $\Gamma$, including one highly degenerate cone located at the Fermi level (see Fig.\ref{fig:bands-12}.c). By contrast, when we incorporate the structural modulation, the RBZ bands (Fig.\ref{fig:bands-12}.d) of the resulting \textit{double-modulated} system $t_{\phi_1} \neq t_{\phi_1} $ reproduce the miniband structure characteristic of moiré models Fig. \ref{f1}(c).
Our DFT band structures, Fig. \ref{fig:bands-12}(a-b), exhibit a closely analogous evolution. Although the Dirac crossing does not lie exactly at the Fermi level, it appears only $\approx$0.05 eV above $E_F$, and the distortion opens a small gap while flattening the bands in its vicinity. Additionally, the pronounced flat band near 0.25 eV, which persists before and after distortion, mirrors the flat minibands predicted by our model at $\approx 1$~eV. The repeated zig-zag patterns across $\Gamma$ and the similarity between the computed and model densities of states further reinforce this correspondence. In general, this type of distortion produces a global shift of the band, leaving flat bands extremely close to the Fermi level—an aspect of particular interest for superconducting potential, for example—and selectively opening the band gap at $\Gamma$, see Fig.\ref{fig:bands-12}(b) and Fig.\ref{fig:bands-12}(c). This selective modification, along with the pattern of crossings and flat bands, is precisely the central feature captured in Fig. \ref{f1}.

\begin{figure}
\centering
\includegraphics[width=\columnwidth]{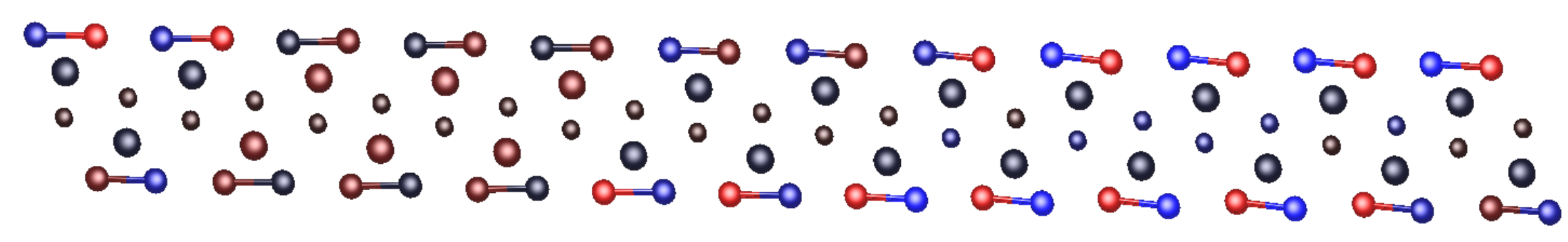}
\caption{Charge redistribution map obtained from DFT relative to the unperturbed structure. The color gradient indicates charge accumulation (Blue, $\delta\rho > 0$) and depletion (Red, $\delta\rho < 0$), with a maximum amplitude of $|\delta\rho| \approx 0.002\,e$. The pattern reveals a distinct anti-correlation between the two legs, confirming the rung-odd (dipolar) character of the instability predicted by the RPA analysis.}
    \label{fig:dft_charge}
\end{figure}

To confirm the main findings of the model, a neutral acoustic phonon, we calculated the redistribution of the perturbed geometry previously described (\textit{i.e.} 12 times the unit cell), around each atom. The results are shown in Fig.~\ref{fig:dft_charge}. Quantitatively, the computed charge modulation is small (amplitude of $\pm0.002~e$), consistent with an instability of the Fermi surface driven by a weak moiré potential rather than a chemical reconstruction of the bonding patterns. Despite the small amplitude, the dipolar pattern ($\delta\rho_1=-\delta\rho_2$) is clear along the whole supercell; the ground state spontaneously breaks symmetry into a neutral pattern (in the long-range). This neutrality protects the associated phason mode from Coulomb-induced stiffening, preserving its acoustic nature, as predicted by the model.  
\section{\label{sec:disc}Conclusions}
In this article, we propose a model that provides a mechanism for the onset of charge orders in layered materials. It consists of a half-filled, four-band ladder system comprising two shifted legs. A moiré potential arises from the slight dimerization of one leg, transforming the broad, high-kinetic-energy bands of the individual chains into narrow, flat minibands near the Fermi level. We found that the spectrum in the extended Brillouin zone is periodic under a finite momentum shift that returns all Peierls phases to their original states. This shift establishes the fundamental global period of the moiré lattice. As a result, we obtain the extended first Brillouin zone, which determines the periodicity and the locations of the hot patches near the Fermi energy, where the energy bands repeat. The density of states reflects quasiperiodicity due to weakly dimerized coupling and exhibits a complex structure stemming from modulated inter-leg tunneling and Peierls modulation.

Sharp peaks in the noninteracting Lindhard response function anticipate the incommensurate charge modulation and satellite structure associated with the moiré and base lattices. To confirm the possibility of CDW in the system, we included short-range intra-leg Coulomb interactions and computed the system's susceptibility using the Random Phase Approximation. We found that the largest eigenvalue of the leg-projected bare matrix depends on dimerization $\delta$ through $Q*=pG_0$. The moiré reciprocal periodicity is $\pi/q$, so the unstable mode sitting at $Q*$, which is not an integer multiple of a moiré reciprocal vector relative to the moiré lattice, is incommensurate. The unstable eigenmode is odd and is equally distributed across both legs. Therefore, the charge modulation on the two legs is out of phase, and the net charge modulation on a rung cancels. As the unstable mode is leg-odd, its phase mode—the phason—is neutral with respect to long-range Coulomb interactions, and the long-wavelength dispersion remains gapless and acoustic, exhibiting linear dispersion. Therefore, the lowest-energy collective excitations of our model consist of soft, long-lived collective bosons realized by neutral acoustic phasons, which are the phase modes of an incommensurate excitonic CDW phase. The dispersions of the CDW phason and of the gapped collective mode, the amplitudon, show up in the two-particle spectral function computed from the retarded susceptibility, the imaginary part of which depicts the CDW continuum at higher energies. The resulting gapless dispersion of the phason at the CDW wavevector, along with its neutrality, supports the idea that these modes do not experience a restoring force, rendering them as Goldstone modes when translational symmetry is continuously broken. 

From the mean-field theory of the interacting system, we derived the phason action in the long-wavelength, low-frequency regime. We showed that near the hot patches, where minibands tend to flatten, the miniband width, set by the inter-leg tunneling,  and the moiré pattern $\delta$ significantly affect phason velocities and the DOS.

We identify the interleg moiré-modulated potential as the dominant mechanism stabilizing the incommensurate CDW phase in this class of layered models. After flat bands emerge, electronic interactions supply the remaining ingredient that precipitates the instability.  
Subsequent work should examine the effect of weak inter-leg Coulomb couplings and disorder.

We further examined the relevance of this mechanism in a realistic setting by analyzing the surface of HfTe$_3$, a quasi-one-dimensional transition–metal trichalcogenide whose bulk exhibits a charge-density-wave transition. Our DFPT calculations revealed a phonon instability at a long wavelength, whose eigen-displacement induces a chain-dependent structural modulation. Upon relaxation, this instability produces a periodic distortion in which the two Te chains acquire distinct phases, forming a real-space moiré pattern. The miniband flattening, gap openings, and band reconstructions obtained in the distorted DFT supercell closely match those predicted by the effective moiré Hamiltonian. The density of charge shows leg-odd oscillations and an overall neutral behavior, establishing a direct correspondence between the lattice instability in HfTe$_3$ and the moiré electronic texture described in our model. More broadly, such chain-dependent phase-shifted modulations may arise in other quasi–1D systems through unstable phonon modes, energetically favorable geometric relaxations, or experimentally induced mismatch conditions, highlighting a general route by which structural incongruities can drive emergent electronic instabilities.

As a concluding point, it is important to clarify how the lattice-driven instability identified in $\rm HfTe_3$ is logically connected to the excitonic charge-density wave analyzed in the moiré ladder model. The DFT and DFPT calculations presented here reveal an intrinsic structural instability associated with a soft phonon mode, which generates a long-wavelength, chain-resolved modulation of the electronic hopping amplitudes. This modulation generates a moiré-like electronic structure that forms the foundation of our effective model. The excitonic CDW examined in Secs.~V and VI should be interpreted as a secondary, interaction-induced electronic instability that can emerge once the moiré minibands have formed and the electronic kinetic energy has been substantially suppressed. At present, there is no direct experimental or first-principles evidence that such an excitonic CDW is realized in HfTe$_3$. Rather, our results demonstrate that the moiré modulation generated by a lattice instability provides a natural and robust platform for stabilizing an incommensurate electronic CDW with a neutral phason mode. In this sense, HfTe$_3$ serves as a physically motivated example showing how lattice-driven moiré patterns can enable emergent electronic collective phases at lower energies.

Controlling CDW states and their phason dynamics could lead to materials with tunable electronic properties, such as dissipationless transport, enabling the creation of ultra-low-power memory and computing devices. The non-linear dynamics of sliding CDWs can produce harmonics, which may be valuable for high-frequency applications. 

\begin{acknowledgments}
P.M. acknowledges support from the Fondo Nacional de Desarrollo Científico y Tecnológico (Fondecyt) under Grant No. 1250122 and from the Indian Institute of Technology Madras, India, where part of this work was conducted. J.C.E gratefully acknowledges ANID for her national doctoral scholarship No. 21231429 and partial support from UAI. F.M.  acknowledges support from Fondecyt grants 1231487 and 1220715, CEDENNA CIA250002, and partial support by the supercomputing infrastructure of the NLHPC (CCSS210001). This work used Bridges-2 at Pittsburgh Supercomputing Center through allocation PHY150003P from the Advanced Cyberinfrastructure Coordination Ecosystem: Services \& Support (ACCESS) program, which is supported by National Science Foundation grants No. 2138259,  2138286, 2138307, 2137603, and 2138296.
\end{acknowledgments}
\appendix
\section{\label{sec:app}Rung harmonic $t_1$ from downfolding}
Rung modulation arises naturally from the coupling of the uniform leg with the dimerized leg. A convenient formal route is the Schrieffer–Wolff transformation \cite{schrieffer1966relation}. We split the Hilbert space into the
\(m=0\) replica (low‐energy subspace) and the \(m\neq0\) sectors (virtual replicas). To second order in the inter-replica coupling \(V\), one obtains
\[
H_{\mathrm{eff}}=H_{00}+V_{0\bar m}\frac{1}{E-H_{\bar m\bar m}}V_{\bar m 0}+\cdots,
\]
where bars denote the high-energy subspace. In our problem, \(V\) is generated by the Peierls phase of the dimerized chain, which carries momentum \(\pm G_s\) and thus connects \(m\) and \(m\pm1\). In a continuum formulation, this is equivalent
to retain the first moiré harmonic in the interlayer tunneling.
Tracing the virtual replicas renormalizes the rung matrix element as
\[
t_\perp(k)\simeq t+\sum_{m=\pm1}\frac{\abs{V_m}^2}{\Delta_m}\cos(mqk+\varphi_m)
\;\;\]\[\Rightarrow\;\; t+t_1\cos(qk+\varphi_r),
\]
with \(V_{\pm1}\propto(1-\delta)\,t\) set by the dimerization strength and \(\Delta_m\), an energy denominator controlled by the parent dispersions. This produces
\[
t_1=\alpha\,(1-\delta)\,t,\qquad \alpha=\mathcal{O}(1),
\]
and the choice \(\varphi_r=0\), a gauge that minimizes parameters without breaking time-reversal or inversion.
\section{\label{sec:dft}Details and Methods of DFT and DFPT calculations}
\subsection{DFT}
\begin{figure}[h!]
\centering
\includegraphics[width=1.0\linewidth]{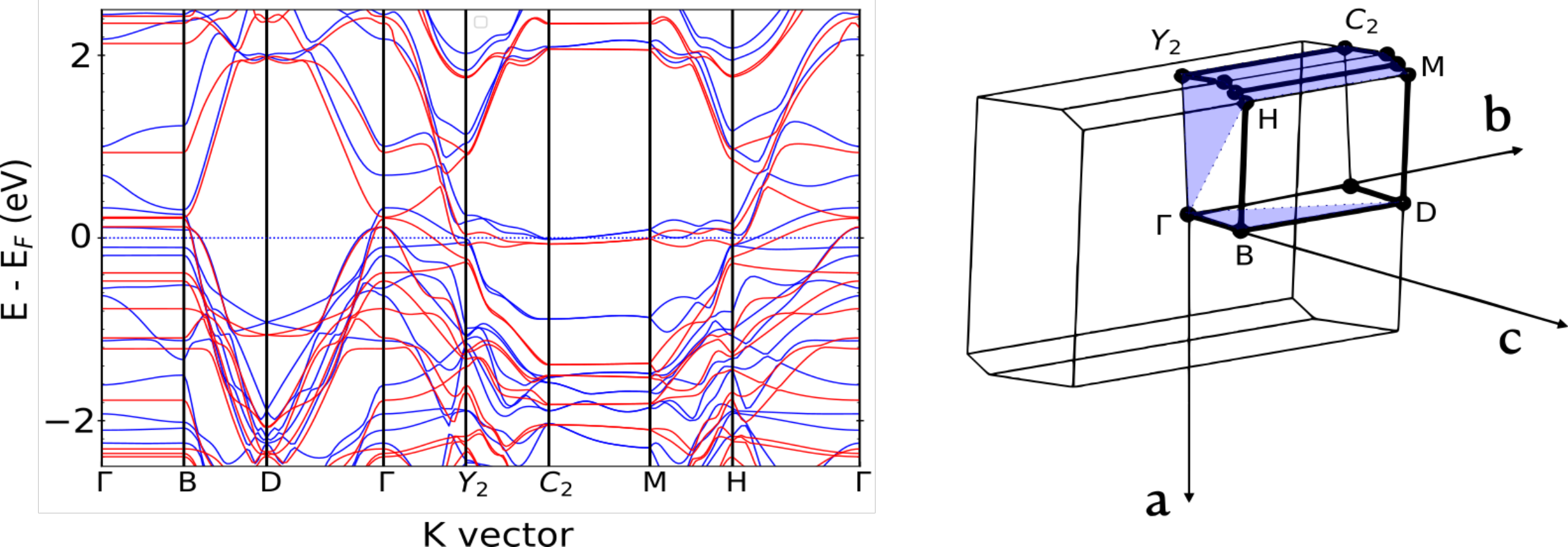}
\caption{Left: Electronic band structure of $\mathrm{HfTe_3}$ comparing the bulk (blue) and surface (red) systems. Right: Schematic representation of the bulk Brillouin zone and its relation to real-space lattice directions $a,b$ and $c$. The surface preserves the in-plane symmetry of the bulk but lacks translational symmetry along $c$, reducing the 3D Brillouin zone to a 2D surface Brillouin zone with identical in-plane reciprocal vectors. For comparison, the surface bands are plotted along the same high-symmetry path as for the bulk.}
    \label{fig:compare-bands}
\end{figure}

Density functional theory calculations were performed using the Vienna Ab initio Simulation Package (VASP) \cite{vasp1,vasp2,vasp3,vasp4}, within the projector–augmented-wave (PAW) \cite{paw} framework, using the PBE exchange–correlation functional\cite{pbe}. A plane-wave basis with an energy cutoff of 520 eV was used, which we verified to be sufficient for total-energy and force convergence for both Hf and Te PAW datasets, including semicore states in the case of Hf (Hf\_pv). Electronic self-consistency was achieved using a stringent energy threshold of $10^{-8}$ eV, and partial occupancies were treated with a small-width metallic smearing scheme suitable for quasi-metallic systems. 
Although lower-dimensional realizations of $\mathrm{HfTe_3}$ have been explored, particularly configurations reduced to isolated chains, we found that they do not reproduce the key electronic features of the bulk. For this reason, we adopt a two-dimensional slab geometry, which retains the essential interchain coupling and yields a band structure much closer to that of the experimentally characterized bulk system (see Fig. \ref{fig:compare-bands}), that hosts both charge-density-wave order and superconductivity. The band structure was plotted using the Pyprocar package \cite{pyprocar}, and the visualization of the structures was done with VESTA \cite{vesta}.\\
For surface calculations, the Brillouin zone was sampled using a $10\times16\times1$ Monkhorst–Pack mesh, dense enough to capture the anisotropic dispersion of the quasi-1D Te chains and to guarantee convergence of the surface electronic structure. Slab geometries used to model the surface retained periodicity only in the in-plane directions; the nonperiodic direction $c$ contained a vacuum spacing of 24 \AA, large enough to suppress inter-slab interactions. Structural relaxations were carried out by simultaneously optimizing atomic positions and cell shape while enforcing force convergence below $10 ^{−4}$  eV/\AA. \\
\subsection{DFPT}
Phonon dispersion relations and dynamical matrices were obtained using Density Functional Perturbation Theory (DFPT) of VASP interfaced with Phonopy \cite{togo2015first,togo2023first,togo2023implementation}. The force-constant matrices were computed within a $2\times 2 \times 1$ supercell sampling, which is sufficient to capture the quasi-1D character of the Te chains while preserving the in-plane periodicity of the slab. The perturbative self-consistent cycle was converged to an energy tolerance of $10^{-8}$eV. The resulting force-constant matrix was used to construct the phonon spectrum along high-symmetry paths of the surface Brillouin zone. We systematically verified the robustness of the soft-mode instability by varying all relevant numerical parameters. We adjusted the electronic self-consistency threshold, tightening the energy convergence, and tested increasingly dense Monkhorst–Pack meshes as well as larger supercells in the perturbative DFPT calculations. In all cases, the phonon dispersion consistently displayed an unstable branch. For the calculation described before, the minimal frequency for the unstable mode at $q = (0.06, 0, 0)$ was –0.54 THz.\\
\subsection{Parametrization of the moiré hopping modulation}
The relaxed surface structure obtained by following the unstable DFPT phonon mode ($12\times 1\times 1$ supercell) exhibits a long-wavelength modulation of the interatomic distances along the quasi-one-dimensional Te chains. These modulations are smooth, harmonic, and chain-dependent, and can be accurately described by cosine functions with distinct phases for each chain. For a given bond type $\alpha$, the distance profile takes the form
$$
d_{\alpha}(n) = d^{(0)}_{\alpha} + \delta d_{\alpha}\cos\!\left(\frac{2\pi n}{\lambda_{\alpha}} + \phi_{\alpha}\right),
$$
where $\lambda_{\alpha}$ is the modulation wavelength and $\phi_{\alpha}$ is a chain-dependent phase. Importantly, the two inequivalent Te chains show different phases, $\phi_{1}\neq\phi_{2}$, producing a relative phase shift in their structural modulation. Here, we restrict our analysis to the case in which the $t_\perp = t_\perp(A_1A2) = t_\perp(B_1B2)$. This choice preserves the periodicity and symmetry structure
required for a direct correspondence with the moiré model introduced in the
main text. Exploring the consequences of inequivalent $t_{AA}$ and $t_{BB}$, as
indeed emerges after full structural relaxation, remains an interesting direction
for future work, but lies beyond the scope of the present study.

A Wannier analysis was carried out to identify the relevant electronic channels: the dominant matrix elements correspond to the intra-chain hoppings within the dimerized Te chains, while the inter-chain and Hf-Te couplings are substantially weaker and strongly site-selective. This information establishes the structure of the effective Hamiltonian but does not determine how the hoppings vary in space.

Since the local hopping amplitude depends smoothly on the corresponding bond length, the spatial dependence of $t_{\alpha}(n)$ inherits the same periodicity and phase as the underlying distance modulation. Thus, the effective hopping
modulations enter the tight-binding model as
$$
t_{\alpha}(n)=t_{\alpha}^{(0)}+
\delta t_{\alpha}\cos\!\left(\frac{2\pi n}{\lambda_{\alpha}}+\phi_{\alpha}\right),
$$
with amplitudes $\delta t_{\alpha}$ chosen to preserve the hierarchy of hopping
strengths revealed by the Wannier analysis.

In Fourier space, a cosine modulation produces two sharp harmonics at $\pm 2\pi/\lambda_{\alpha}$, so that the real-space phase $\phi_{\alpha}$ becomes a complex phase factor multiplying the Fourier components. This provides a direct microscopic mechanism by which chain-dependent structural phase shifts generate a moiré-like phase structure in the tight-binding Hamiltonian.

In this representation, the effective four-band Hamiltonian takes the form:
\begin{equation}
\label{eq:H4-HfTe3}
H(k)=
\begin{pmatrix}
0 & t_{\phi_1}^{+} & t_\perp(k) & 0 \\
t_{\phi_1}^{-} & 0 & 0 & t_\perp(k) \\
t_\perp(k) & 0 & 0 & t_{\phi_2}^{+}\\
0 & t_\perp(k) & t_{\phi_2}^{-} & 0
\end{pmatrix},
\end{equation}
where $t_{\phi_\alpha}^{\pm}=c_\delta(k)\,e^{\pm i\phi_\alpha(k)}$, similar to Eq\ref{eq:H4eff}, with $c_\delta(k)$ the common harmonic amplitude, and $\phi_\alpha(k)$ the chain-dependent phase inherited from the relaxed structural modulation. For the upper chain, due to the periodicity of 12 cells, we use a modulation ratio $\delta_1 = 11/12 \approx 0.92$ to determine $c_\delta(k)$ and $\phi_1(k)$, while for the lower chain we use $\delta_2 = 0.95$ to obtain the corresponding phase $\phi_2(k)$. These values encode the distinct structural phases of the two Te chains and generate the chain-dependent complex phases in the Fourier components of the hopping modulation.

Using these modulated hoppings, the effective Hamiltonian resulting Eq.~\eqref{eq:H4-HfTe3} reproduces the miniband formation, band reconstruction, and gap openings observed in the DFT supercell calculations (Fig.\ref{fig:bands-12}.(a-b)), thus establishing a direct microscopic connection between the structural distortion and the moiré tight-binding description developed in the main text.\\
%


\begin{thebibliography}{99}%
\makeatletter
\providecommand \@ifxundefined [1]{%
 \@ifx{#1\undefined}
}%
\providecommand \@ifnum [1]{%
 \ifnum #1\expandafter \@firstoftwo
 \else \expandafter \@secondoftwo
 \fi
}%
\providecommand \@ifx [1]{%
 \ifx #1\expandafter \@firstoftwo
 \else \expandafter \@secondoftwo
 \fi
}%
\providecommand \natexlab [1]{#1}%
\providecommand \enquote  [1]{``#1''}%
\providecommand \bibnamefont  [1]{#1}%
\providecommand \bibfnamefont [1]{#1}%
\providecommand \citenamefont [1]{#1}%
\providecommand \href@noop [0]{\@secondoftwo}%
\providecommand \href [0]{\begingroup \@sanitize@url \@href}%
\providecommand \@href[1]{\@@startlink{#1}\@@href}%
\providecommand \@@href[1]{\endgroup#1\@@endlink}%
\providecommand \@sanitize@url [0]{\catcode `\\12\catcode `\$12\catcode `\&12\catcode `\#12\catcode `\^12\catcode `\_12\catcode `\%12\relax}%
\providecommand \@@startlink[1]{}%
\providecommand \@@endlink[0]{}%
\providecommand \url  [0]{\begingroup\@sanitize@url \@url }%
\providecommand \@url [1]{\endgroup\@href {#1}{\urlprefix }}%
\providecommand \urlprefix  [0]{URL }%
\providecommand \Eprint [0]{\href }%
\providecommand \doibase [0]{https://doi.org/}%
\providecommand \selectlanguage [0]{\@gobble}%
\providecommand \bibinfo  [0]{\@secondoftwo}%
\providecommand \bibfield  [0]{\@secondoftwo}%
\providecommand \translation [1]{[#1]}%
\providecommand \BibitemOpen [0]{}%
\providecommand \bibitemStop [0]{}%
\providecommand \bibitemNoStop [0]{.\EOS\space}%
\providecommand \EOS [0]{\spacefactor3000\relax}%
\providecommand \BibitemShut  [1]{\csname bibitem#1\endcsname}%
\let\auto@bib@innerbib\@empty
\bibitem [{\citenamefont {Novoselov}\ \emph {et~al.}(2016)\citenamefont {Novoselov}, \citenamefont {Mishchenko}, \citenamefont {Carvalho},\ and\ \citenamefont {Castro~Neto}}]{novoselov20162d}%
  \BibitemOpen
  \bibfield  {author} {\bibinfo {author} {\bibfnamefont {K.~S.}\ \bibnamefont {Novoselov}}, \bibinfo {author} {\bibfnamefont {A.}~\bibnamefont {Mishchenko}}, \bibinfo {author} {\bibfnamefont {A.}~\bibnamefont {Carvalho}},\ and\ \bibinfo {author} {\bibfnamefont {A.}~\bibnamefont {Castro~Neto}},\ }\bibfield  {title} {\bibinfo {title} {2d materials and van der waals heterostructures},\ }\href@noop {} {\bibfield  {journal} {\bibinfo  {journal} {Science}\ }\textbf {\bibinfo {volume} {353}},\ \bibinfo {pages} {aac9439} (\bibinfo {year} {2016})}\BibitemShut {NoStop}%
\bibitem [{\citenamefont {Guo}\ \emph {et~al.}(2021)\citenamefont {Guo}, \citenamefont {Hu}, \citenamefont {Liu},\ and\ \citenamefont {Tian}}]{guo2021stacking}%
  \BibitemOpen
  \bibfield  {author} {\bibinfo {author} {\bibfnamefont {H.-W.}\ \bibnamefont {Guo}}, \bibinfo {author} {\bibfnamefont {Z.}~\bibnamefont {Hu}}, \bibinfo {author} {\bibfnamefont {Z.-B.}\ \bibnamefont {Liu}},\ and\ \bibinfo {author} {\bibfnamefont {J.-G.}\ \bibnamefont {Tian}},\ }\bibfield  {title} {\bibinfo {title} {Stacking of 2d materials},\ }\href@noop {} {\bibfield  {journal} {\bibinfo  {journal} {Advanced Functional Materials}\ }\textbf {\bibinfo {volume} {31}},\ \bibinfo {pages} {2007810} (\bibinfo {year} {2021})}\BibitemShut {NoStop}%
\bibitem [{\citenamefont {Sung}\ \emph {et~al.}(2019)\citenamefont {Sung}, \citenamefont {Schnitzer}, \citenamefont {Brown}, \citenamefont {Park},\ and\ \citenamefont {Hovden}}]{sung2019stacking}%
  \BibitemOpen
  \bibfield  {author} {\bibinfo {author} {\bibfnamefont {S.~H.}\ \bibnamefont {Sung}}, \bibinfo {author} {\bibfnamefont {N.}~\bibnamefont {Schnitzer}}, \bibinfo {author} {\bibfnamefont {L.}~\bibnamefont {Brown}}, \bibinfo {author} {\bibfnamefont {J.}~\bibnamefont {Park}},\ and\ \bibinfo {author} {\bibfnamefont {R.}~\bibnamefont {Hovden}},\ }\bibfield  {title} {\bibinfo {title} {Stacking, strain, and twist in 2d materials quantified by 3d electron diffraction},\ }\href@noop {} {\bibfield  {journal} {\bibinfo  {journal} {Physical Review Materials}\ }\textbf {\bibinfo {volume} {3}},\ \bibinfo {pages} {064003} (\bibinfo {year} {2019})}\BibitemShut {NoStop}%
\bibitem [{\citenamefont {Liu}\ and\ \citenamefont {Hersam}(2019)}]{liu20192d}%
  \BibitemOpen
  \bibfield  {author} {\bibinfo {author} {\bibfnamefont {X.}~\bibnamefont {Liu}}\ and\ \bibinfo {author} {\bibfnamefont {M.~C.}\ \bibnamefont {Hersam}},\ }\bibfield  {title} {\bibinfo {title} {2d materials for quantum information science},\ }\href@noop {} {\bibfield  {journal} {\bibinfo  {journal} {Nature Reviews Materials}\ }\textbf {\bibinfo {volume} {4}},\ \bibinfo {pages} {669} (\bibinfo {year} {2019})}\BibitemShut {NoStop}%
\bibitem [{\citenamefont {Keimer}\ and\ \citenamefont {Moore}(2017)}]{keimer2017physics}%
  \BibitemOpen
  \bibfield  {author} {\bibinfo {author} {\bibfnamefont {B.}~\bibnamefont {Keimer}}\ and\ \bibinfo {author} {\bibfnamefont {J.~E.}\ \bibnamefont {Moore}},\ }\bibfield  {title} {\bibinfo {title} {The physics of quantum materials},\ }\href@noop {} {\bibfield  {journal} {\bibinfo  {journal} {Nature Physics}\ }\textbf {\bibinfo {volume} {13}},\ \bibinfo {pages} {1045} (\bibinfo {year} {2017})}\BibitemShut {NoStop}%
\bibitem [{\citenamefont {Mellado}\ and\ \citenamefont {Sturla}(2024)}]{mellado2024quantum}%
  \BibitemOpen
  \bibfield  {author} {\bibinfo {author} {\bibfnamefont {P.}~\bibnamefont {Mellado}}\ and\ \bibinfo {author} {\bibfnamefont {M.}~\bibnamefont {Sturla}},\ }\bibfield  {title} {\bibinfo {title} {Quantum fluctuations in the van der waals material nips 3},\ }\href@noop {} {\bibfield  {journal} {\bibinfo  {journal} {Physical Review B}\ }\textbf {\bibinfo {volume} {110}},\ \bibinfo {pages} {134438} (\bibinfo {year} {2024})}\BibitemShut {NoStop}%
\bibitem [{\citenamefont {Lian}\ \emph {et~al.}(2018)\citenamefont {Lian}, \citenamefont {Si},\ and\ \citenamefont {Duan}}]{lian2018unveiling}%
  \BibitemOpen
  \bibfield  {author} {\bibinfo {author} {\bibfnamefont {C.-S.}\ \bibnamefont {Lian}}, \bibinfo {author} {\bibfnamefont {C.}~\bibnamefont {Si}},\ and\ \bibinfo {author} {\bibfnamefont {W.}~\bibnamefont {Duan}},\ }\bibfield  {title} {\bibinfo {title} {Unveiling charge-density wave, superconductivity, and their competitive nature in two-dimensional nbse2},\ }\href@noop {} {\bibfield  {journal} {\bibinfo  {journal} {Nano letters}\ }\textbf {\bibinfo {volume} {18}},\ \bibinfo {pages} {2924} (\bibinfo {year} {2018})}\BibitemShut {NoStop}%
\bibitem [{\citenamefont {Tsen}\ \emph {et~al.}(2015)\citenamefont {Tsen}, \citenamefont {Hovden}, \citenamefont {Wang}, \citenamefont {Kim}, \citenamefont {Okamoto}, \citenamefont {Spoth}, \citenamefont {Liu}, \citenamefont {Lu}, \citenamefont {Sun}, \citenamefont {Hone} \emph {et~al.}}]{tsen2015structure}%
  \BibitemOpen
  \bibfield  {author} {\bibinfo {author} {\bibfnamefont {A.~W.}\ \bibnamefont {Tsen}}, \bibinfo {author} {\bibfnamefont {R.}~\bibnamefont {Hovden}}, \bibinfo {author} {\bibfnamefont {D.}~\bibnamefont {Wang}}, \bibinfo {author} {\bibfnamefont {Y.~D.}\ \bibnamefont {Kim}}, \bibinfo {author} {\bibfnamefont {J.}~\bibnamefont {Okamoto}}, \bibinfo {author} {\bibfnamefont {K.~A.}\ \bibnamefont {Spoth}}, \bibinfo {author} {\bibfnamefont {Y.}~\bibnamefont {Liu}}, \bibinfo {author} {\bibfnamefont {W.}~\bibnamefont {Lu}}, \bibinfo {author} {\bibfnamefont {Y.}~\bibnamefont {Sun}}, \bibinfo {author} {\bibfnamefont {J.~C.}\ \bibnamefont {Hone}}, \emph {et~al.},\ }\bibfield  {title} {\bibinfo {title} {Structure and control of charge density waves in two-dimensional 1t-tas2},\ }\href@noop {} {\bibfield  {journal} {\bibinfo  {journal} {Proceedings of the National Academy of Sciences}\ }\textbf {\bibinfo {volume} {112}},\ \bibinfo {pages} {15054} (\bibinfo {year} {2015})}\BibitemShut {NoStop}%
\bibitem [{\citenamefont {Goli}\ \emph {et~al.}(2012)\citenamefont {Goli}, \citenamefont {Khan}, \citenamefont {Wickramaratne}, \citenamefont {Lake},\ and\ \citenamefont {Balandin}}]{goli2012charge}%
  \BibitemOpen
  \bibfield  {author} {\bibinfo {author} {\bibfnamefont {P.}~\bibnamefont {Goli}}, \bibinfo {author} {\bibfnamefont {J.}~\bibnamefont {Khan}}, \bibinfo {author} {\bibfnamefont {D.}~\bibnamefont {Wickramaratne}}, \bibinfo {author} {\bibfnamefont {R.~K.}\ \bibnamefont {Lake}},\ and\ \bibinfo {author} {\bibfnamefont {A.~A.}\ \bibnamefont {Balandin}},\ }\bibfield  {title} {\bibinfo {title} {Charge density waves in exfoliated films of van der waals materials: evolution of raman spectrum in tise2},\ }\href@noop {} {\bibfield  {journal} {\bibinfo  {journal} {Nano letters}\ }\textbf {\bibinfo {volume} {12}},\ \bibinfo {pages} {5941} (\bibinfo {year} {2012})}\BibitemShut {NoStop}%
\bibitem [{\citenamefont {Chen}\ \emph {et~al.}(2016)\citenamefont {Chen}, \citenamefont {Chan}, \citenamefont {Wong}, \citenamefont {Fang}, \citenamefont {Chou}, \citenamefont {Mo}, \citenamefont {Hussain}, \citenamefont {Fedorov},\ and\ \citenamefont {Chiang}}]{chen2016dimensional}%
  \BibitemOpen
  \bibfield  {author} {\bibinfo {author} {\bibfnamefont {P.}~\bibnamefont {Chen}}, \bibinfo {author} {\bibfnamefont {Y.-H.}\ \bibnamefont {Chan}}, \bibinfo {author} {\bibfnamefont {M.-H.}\ \bibnamefont {Wong}}, \bibinfo {author} {\bibfnamefont {X.-Y.}\ \bibnamefont {Fang}}, \bibinfo {author} {\bibfnamefont {M.-Y.}\ \bibnamefont {Chou}}, \bibinfo {author} {\bibfnamefont {S.-K.}\ \bibnamefont {Mo}}, \bibinfo {author} {\bibfnamefont {Z.}~\bibnamefont {Hussain}}, \bibinfo {author} {\bibfnamefont {A.-V.}\ \bibnamefont {Fedorov}},\ and\ \bibinfo {author} {\bibfnamefont {T.-C.}\ \bibnamefont {Chiang}},\ }\bibfield  {title} {\bibinfo {title} {Dimensional effects on the charge density waves in ultrathin films of tise2},\ }\href@noop {} {\bibfield  {journal} {\bibinfo  {journal} {Nano letters}\ }\textbf {\bibinfo {volume} {16}},\ \bibinfo {pages} {6331} (\bibinfo {year} {2016})}\BibitemShut {NoStop}%
\bibitem [{\citenamefont {Chen}\ \emph {et~al.}(2025)\citenamefont {Chen}, \citenamefont {Xu}, \citenamefont {Yao}, \citenamefont {Li}, \citenamefont {Zhao}, \citenamefont {Liu},\ and\ \citenamefont {Guo}}]{chen2025introduction}%
  \BibitemOpen
  \bibfield  {author} {\bibinfo {author} {\bibfnamefont {Z.-R.}\ \bibnamefont {Chen}}, \bibinfo {author} {\bibfnamefont {H.-P.}\ \bibnamefont {Xu}}, \bibinfo {author} {\bibfnamefont {W.-D.}\ \bibnamefont {Yao}}, \bibinfo {author} {\bibfnamefont {M.-Y.}\ \bibnamefont {Li}}, \bibinfo {author} {\bibfnamefont {C.-Y.}\ \bibnamefont {Zhao}}, \bibinfo {author} {\bibfnamefont {W.}~\bibnamefont {Liu}},\ and\ \bibinfo {author} {\bibfnamefont {S.-P.}\ \bibnamefont {Guo}},\ }\bibfield  {title} {\bibinfo {title} {Introduction of diverse divalent main-group metal cations to tune the physical properties of rare-earth thiophosphates},\ }\href@noop {} {\bibfield  {journal} {\bibinfo  {journal} {Inorganic Chemistry}\ }\textbf {\bibinfo {volume} {64}},\ \bibinfo {pages} {15353} (\bibinfo {year} {2025})}\BibitemShut {NoStop}%
\bibitem [{\citenamefont {Ding}\ \emph {et~al.}(2009)\citenamefont {Ding}, \citenamefont {Zhu},\ and\ \citenamefont {Berakdar}}]{ding2009localized}%
  \BibitemOpen
  \bibfield  {author} {\bibinfo {author} {\bibfnamefont {K.-H.}\ \bibnamefont {Ding}}, \bibinfo {author} {\bibfnamefont {Z.-G.}\ \bibnamefont {Zhu}},\ and\ \bibinfo {author} {\bibfnamefont {J.}~\bibnamefont {Berakdar}},\ }\bibfield  {title} {\bibinfo {title} {Localized magnetic states in biased bilayer and trilayer graphene},\ }\href@noop {} {\bibfield  {journal} {\bibinfo  {journal} {Journal of Physics: Condensed Matter}\ }\textbf {\bibinfo {volume} {21}},\ \bibinfo {pages} {182002} (\bibinfo {year} {2009})}\BibitemShut {NoStop}%
\bibitem [{\citenamefont {Yu}\ \emph {et~al.}(2023)\citenamefont {Yu}, \citenamefont {Pistunova}, \citenamefont {Hu}, \citenamefont {Watanabe}, \citenamefont {Taniguchi},\ and\ \citenamefont {Heinz}}]{yu2023observation}%
  \BibitemOpen
  \bibfield  {author} {\bibinfo {author} {\bibfnamefont {L.}~\bibnamefont {Yu}}, \bibinfo {author} {\bibfnamefont {K.}~\bibnamefont {Pistunova}}, \bibinfo {author} {\bibfnamefont {J.}~\bibnamefont {Hu}}, \bibinfo {author} {\bibfnamefont {K.}~\bibnamefont {Watanabe}}, \bibinfo {author} {\bibfnamefont {T.}~\bibnamefont {Taniguchi}},\ and\ \bibinfo {author} {\bibfnamefont {T.~F.}\ \bibnamefont {Heinz}},\ }\bibfield  {title} {\bibinfo {title} {Observation of quadrupolar and dipolar excitons in a semiconductor heterotrilayer},\ }\href@noop {} {\bibfield  {journal} {\bibinfo  {journal} {Nature materials}\ }\textbf {\bibinfo {volume} {22}},\ \bibinfo {pages} {1485} (\bibinfo {year} {2023})}\BibitemShut {NoStop}%
\bibitem [{\citenamefont {Xu}\ \emph {et~al.}(2025)\citenamefont {Xu}, \citenamefont {Fu}, \citenamefont {Huang}, \citenamefont {Ge}, \citenamefont {Xu}, \citenamefont {Zheng}, \citenamefont {Deng}, \citenamefont {Xie}, \citenamefont {Tong}, \citenamefont {Li} \emph {et~al.}}]{xu2025correlated}%
  \BibitemOpen
  \bibfield  {author} {\bibinfo {author} {\bibfnamefont {B.}~\bibnamefont {Xu}}, \bibinfo {author} {\bibfnamefont {J.}~\bibnamefont {Fu}}, \bibinfo {author} {\bibfnamefont {L.}~\bibnamefont {Huang}}, \bibinfo {author} {\bibfnamefont {C.}~\bibnamefont {Ge}}, \bibinfo {author} {\bibfnamefont {Z.}~\bibnamefont {Xu}}, \bibinfo {author} {\bibfnamefont {W.}~\bibnamefont {Zheng}}, \bibinfo {author} {\bibfnamefont {Q.}~\bibnamefont {Deng}}, \bibinfo {author} {\bibfnamefont {S.}~\bibnamefont {Xie}}, \bibinfo {author} {\bibfnamefont {Q.}~\bibnamefont {Tong}}, \bibinfo {author} {\bibfnamefont {D.}~\bibnamefont {Li}}, \emph {et~al.},\ }\bibfield  {title} {\bibinfo {title} {Correlated fermionic-bosonic insulating states in twisted hetero-trilayer semiconductors},\ }\href@noop {} {\bibfield  {journal} {\bibinfo  {journal} {Nature Communications}\ }\textbf {\bibinfo {volume} {16}},\ \bibinfo {pages} {3938} (\bibinfo {year} {2025})}\BibitemShut {NoStop}%
\bibitem [{\citenamefont {Pantale{\'o}n}\ \emph {et~al.}(2023)\citenamefont {Pantale{\'o}n}, \citenamefont {Jimeno-Pozo}, \citenamefont {Sainz-Cruz}, \citenamefont {Phong}, \citenamefont {Cea},\ and\ \citenamefont {Guinea}}]{pantaleon2023superconductivity}%
  \BibitemOpen
  \bibfield  {author} {\bibinfo {author} {\bibfnamefont {P.~A.}\ \bibnamefont {Pantale{\'o}n}}, \bibinfo {author} {\bibfnamefont {A.}~\bibnamefont {Jimeno-Pozo}}, \bibinfo {author} {\bibfnamefont {H.}~\bibnamefont {Sainz-Cruz}}, \bibinfo {author} {\bibfnamefont {V.~T.}\ \bibnamefont {Phong}}, \bibinfo {author} {\bibfnamefont {T.}~\bibnamefont {Cea}},\ and\ \bibinfo {author} {\bibfnamefont {F.}~\bibnamefont {Guinea}},\ }\bibfield  {title} {\bibinfo {title} {Superconductivity and correlated phases in non-twisted bilayer and trilayer graphene},\ }\href@noop {} {\bibfield  {journal} {\bibinfo  {journal} {Nature Reviews Physics}\ }\textbf {\bibinfo {volume} {5}},\ \bibinfo {pages} {304} (\bibinfo {year} {2023})}\BibitemShut {NoStop}%
\bibitem [{\citenamefont {McGilly}\ \emph {et~al.}(2020)\citenamefont {McGilly}, \citenamefont {Kerelsky}, \citenamefont {Finney}, \citenamefont {Shapovalov}, \citenamefont {Shih}, \citenamefont {Ghiotto}, \citenamefont {Zeng}, \citenamefont {Moore}, \citenamefont {Wu}, \citenamefont {Bai} \emph {et~al.}}]{mcgilly2020visualization}%
  \BibitemOpen
  \bibfield  {author} {\bibinfo {author} {\bibfnamefont {L.~J.}\ \bibnamefont {McGilly}}, \bibinfo {author} {\bibfnamefont {A.}~\bibnamefont {Kerelsky}}, \bibinfo {author} {\bibfnamefont {N.~R.}\ \bibnamefont {Finney}}, \bibinfo {author} {\bibfnamefont {K.}~\bibnamefont {Shapovalov}}, \bibinfo {author} {\bibfnamefont {E.-M.}\ \bibnamefont {Shih}}, \bibinfo {author} {\bibfnamefont {A.}~\bibnamefont {Ghiotto}}, \bibinfo {author} {\bibfnamefont {Y.}~\bibnamefont {Zeng}}, \bibinfo {author} {\bibfnamefont {S.~L.}\ \bibnamefont {Moore}}, \bibinfo {author} {\bibfnamefont {W.}~\bibnamefont {Wu}}, \bibinfo {author} {\bibfnamefont {Y.}~\bibnamefont {Bai}}, \emph {et~al.},\ }\bibfield  {title} {\bibinfo {title} {Visualization of moir{\'e} superlattices},\ }\href@noop {} {\bibfield  {journal} {\bibinfo  {journal} {Nature Nanotechnology}\ }\textbf {\bibinfo {volume} {15}},\ \bibinfo {pages} {580} (\bibinfo {year} {2020})}\BibitemShut {NoStop}%
\bibitem [{\citenamefont {Andrei}\ \emph {et~al.}(2021)\citenamefont {Andrei}, \citenamefont {Efetov}, \citenamefont {Jarillo-Herrero}, \citenamefont {MacDonald}, \citenamefont {Mak}, \citenamefont {Senthil}, \citenamefont {Tutuc}, \citenamefont {Yazdani},\ and\ \citenamefont {Young}}]{andrei2021marvels}%
  \BibitemOpen
  \bibfield  {author} {\bibinfo {author} {\bibfnamefont {E.~Y.}\ \bibnamefont {Andrei}}, \bibinfo {author} {\bibfnamefont {D.~K.}\ \bibnamefont {Efetov}}, \bibinfo {author} {\bibfnamefont {P.}~\bibnamefont {Jarillo-Herrero}}, \bibinfo {author} {\bibfnamefont {A.~H.}\ \bibnamefont {MacDonald}}, \bibinfo {author} {\bibfnamefont {K.~F.}\ \bibnamefont {Mak}}, \bibinfo {author} {\bibfnamefont {T.}~\bibnamefont {Senthil}}, \bibinfo {author} {\bibfnamefont {E.}~\bibnamefont {Tutuc}}, \bibinfo {author} {\bibfnamefont {A.}~\bibnamefont {Yazdani}},\ and\ \bibinfo {author} {\bibfnamefont {A.~F.}\ \bibnamefont {Young}},\ }\bibfield  {title} {\bibinfo {title} {The marvels of moir{\'e} materials},\ }\href@noop {} {\bibfield  {journal} {\bibinfo  {journal} {Nature Reviews Materials}\ }\textbf {\bibinfo {volume} {6}},\ \bibinfo {pages} {201} (\bibinfo {year} {2021})}\BibitemShut {NoStop}%
\bibitem [{\citenamefont {Shi}\ \emph {et~al.}(2021)\citenamefont {Shi}, \citenamefont {Qi}, \citenamefont {Jiang}, \citenamefont {Dai}, \citenamefont {Lin}, \citenamefont {Zhang},\ and\ \citenamefont {Fang}}]{shi2021exotic}%
  \BibitemOpen
  \bibfield  {author} {\bibinfo {author} {\bibfnamefont {B.}~\bibnamefont {Shi}}, \bibinfo {author} {\bibfnamefont {P.}~\bibnamefont {Qi}}, \bibinfo {author} {\bibfnamefont {M.}~\bibnamefont {Jiang}}, \bibinfo {author} {\bibfnamefont {Y.}~\bibnamefont {Dai}}, \bibinfo {author} {\bibfnamefont {F.}~\bibnamefont {Lin}}, \bibinfo {author} {\bibfnamefont {H.}~\bibnamefont {Zhang}},\ and\ \bibinfo {author} {\bibfnamefont {Z.}~\bibnamefont {Fang}},\ }\bibfield  {title} {\bibinfo {title} {Exotic physical properties of 2d materials modulated by moir{\'e} superlattices},\ }\href@noop {} {\bibfield  {journal} {\bibinfo  {journal} {Materials Advances}\ }\textbf {\bibinfo {volume} {2}},\ \bibinfo {pages} {5542} (\bibinfo {year} {2021})}\BibitemShut {NoStop}%
\bibitem [{\citenamefont {Adak}\ \emph {et~al.}(2024)\citenamefont {Adak}, \citenamefont {Sinha}, \citenamefont {Agarwal},\ and\ \citenamefont {Deshmukh}}]{adak2024tunable}%
  \BibitemOpen
  \bibfield  {author} {\bibinfo {author} {\bibfnamefont {P.~C.}\ \bibnamefont {Adak}}, \bibinfo {author} {\bibfnamefont {S.}~\bibnamefont {Sinha}}, \bibinfo {author} {\bibfnamefont {A.}~\bibnamefont {Agarwal}},\ and\ \bibinfo {author} {\bibfnamefont {M.~M.}\ \bibnamefont {Deshmukh}},\ }\bibfield  {title} {\bibinfo {title} {Tunable moir{\'e} materials for probing berry physics and topology},\ }\href@noop {} {\bibfield  {journal} {\bibinfo  {journal} {Nature Reviews Materials}\ }\textbf {\bibinfo {volume} {9}},\ \bibinfo {pages} {481} (\bibinfo {year} {2024})}\BibitemShut {NoStop}%
\bibitem [{\citenamefont {Abbas}\ \emph {et~al.}(2020)\citenamefont {Abbas}, \citenamefont {Li}, \citenamefont {Wang}, \citenamefont {Zhang}, \citenamefont {Wang},\ and\ \citenamefont {Zhang}}]{abbas2020recent}%
  \BibitemOpen
  \bibfield  {author} {\bibinfo {author} {\bibfnamefont {G.}~\bibnamefont {Abbas}}, \bibinfo {author} {\bibfnamefont {Y.}~\bibnamefont {Li}}, \bibinfo {author} {\bibfnamefont {H.}~\bibnamefont {Wang}}, \bibinfo {author} {\bibfnamefont {W.-X.}\ \bibnamefont {Zhang}}, \bibinfo {author} {\bibfnamefont {C.}~\bibnamefont {Wang}},\ and\ \bibinfo {author} {\bibfnamefont {H.}~\bibnamefont {Zhang}},\ }\bibfield  {title} {\bibinfo {title} {Recent advances in twisted structures of flatland materials and crafting moir{\'e} superlattices},\ }\href@noop {} {\bibfield  {journal} {\bibinfo  {journal} {Advanced Functional Materials}\ }\textbf {\bibinfo {volume} {30}},\ \bibinfo {pages} {2000878} (\bibinfo {year} {2020})}\BibitemShut {NoStop}%
\bibitem [{\citenamefont {Chen}\ \emph {et~al.}(2019)\citenamefont {Chen}, \citenamefont {Zhang}, \citenamefont {Lai}, \citenamefont {Lin}, \citenamefont {Chen}, \citenamefont {Guan}, \citenamefont {Chen}, \citenamefont {Yang}, \citenamefont {Tseng}, \citenamefont {Gwo} \emph {et~al.}}]{chen2019tunable}%
  \BibitemOpen
  \bibfield  {author} {\bibinfo {author} {\bibfnamefont {P.-Y.}\ \bibnamefont {Chen}}, \bibinfo {author} {\bibfnamefont {X.-Q.}\ \bibnamefont {Zhang}}, \bibinfo {author} {\bibfnamefont {Y.-Y.}\ \bibnamefont {Lai}}, \bibinfo {author} {\bibfnamefont {E.-C.}\ \bibnamefont {Lin}}, \bibinfo {author} {\bibfnamefont {C.-A.}\ \bibnamefont {Chen}}, \bibinfo {author} {\bibfnamefont {S.-Y.}\ \bibnamefont {Guan}}, \bibinfo {author} {\bibfnamefont {J.-J.}\ \bibnamefont {Chen}}, \bibinfo {author} {\bibfnamefont {Z.-H.}\ \bibnamefont {Yang}}, \bibinfo {author} {\bibfnamefont {Y.-W.}\ \bibnamefont {Tseng}}, \bibinfo {author} {\bibfnamefont {S.}~\bibnamefont {Gwo}}, \emph {et~al.},\ }\bibfield  {title} {\bibinfo {title} {Tunable moir{\'e} superlattice of artificially twisted monolayers},\ }\href@noop {} {\bibfield  {journal} {\bibinfo  {journal} {Advanced Materials}\ }\textbf {\bibinfo {volume} {31}},\ \bibinfo {pages} {1901077} (\bibinfo {year} {2019})}\BibitemShut {NoStop}%
\bibitem [{\citenamefont {Gao}\ \emph {et~al.}(2020)\citenamefont {Gao}, \citenamefont {Lin}, \citenamefont {Smart}, \citenamefont {Ci}, \citenamefont {Watanabe}, \citenamefont {Taniguchi}, \citenamefont {Jeanloz}, \citenamefont {Ni},\ and\ \citenamefont {Wu}}]{gao2020band}%
  \BibitemOpen
  \bibfield  {author} {\bibinfo {author} {\bibfnamefont {Y.}~\bibnamefont {Gao}}, \bibinfo {author} {\bibfnamefont {X.}~\bibnamefont {Lin}}, \bibinfo {author} {\bibfnamefont {T.}~\bibnamefont {Smart}}, \bibinfo {author} {\bibfnamefont {P.}~\bibnamefont {Ci}}, \bibinfo {author} {\bibfnamefont {K.}~\bibnamefont {Watanabe}}, \bibinfo {author} {\bibfnamefont {T.}~\bibnamefont {Taniguchi}}, \bibinfo {author} {\bibfnamefont {R.}~\bibnamefont {Jeanloz}}, \bibinfo {author} {\bibfnamefont {J.}~\bibnamefont {Ni}},\ and\ \bibinfo {author} {\bibfnamefont {J.}~\bibnamefont {Wu}},\ }\bibfield  {title} {\bibinfo {title} {Band engineering of large-twist-angle graphene/h-bn moir{\'e} superlattices with pressure},\ }\href@noop {} {\bibfield  {journal} {\bibinfo  {journal} {Physical review letters}\ }\textbf {\bibinfo {volume} {125}},\ \bibinfo {pages} {226403} (\bibinfo {year} {2020})}\BibitemShut {NoStop}%
\bibitem [{\citenamefont {Escudero}\ \emph {et~al.}(2024)\citenamefont {Escudero}, \citenamefont {Sinner}, \citenamefont {Zhan}, \citenamefont {Pantale{\'o}n},\ and\ \citenamefont {Guinea}}]{escudero2024designing}%
  \BibitemOpen
  \bibfield  {author} {\bibinfo {author} {\bibfnamefont {F.}~\bibnamefont {Escudero}}, \bibinfo {author} {\bibfnamefont {A.}~\bibnamefont {Sinner}}, \bibinfo {author} {\bibfnamefont {Z.}~\bibnamefont {Zhan}}, \bibinfo {author} {\bibfnamefont {P.~A.}\ \bibnamefont {Pantale{\'o}n}},\ and\ \bibinfo {author} {\bibfnamefont {F.}~\bibnamefont {Guinea}},\ }\bibfield  {title} {\bibinfo {title} {Designing moir{\'e} patterns by strain},\ }\href@noop {} {\bibfield  {journal} {\bibinfo  {journal} {Physical Review Research}\ }\textbf {\bibinfo {volume} {6}},\ \bibinfo {pages} {023203} (\bibinfo {year} {2024})}\BibitemShut {NoStop}%
\bibitem [{\citenamefont {Rakib}\ \emph {et~al.}(2022)\citenamefont {Rakib}, \citenamefont {Pochet}, \citenamefont {Ertekin},\ and\ \citenamefont {Johnson}}]{rakib2022moire}%
  \BibitemOpen
  \bibfield  {author} {\bibinfo {author} {\bibfnamefont {T.}~\bibnamefont {Rakib}}, \bibinfo {author} {\bibfnamefont {P.}~\bibnamefont {Pochet}}, \bibinfo {author} {\bibfnamefont {E.}~\bibnamefont {Ertekin}},\ and\ \bibinfo {author} {\bibfnamefont {H.~T.}\ \bibnamefont {Johnson}},\ }\bibfield  {title} {\bibinfo {title} {Moir{\'e} engineering in van der waals heterostructures},\ }\href@noop {} {\bibfield  {journal} {\bibinfo  {journal} {Journal of Applied Physics}\ }\textbf {\bibinfo {volume} {132}} (\bibinfo {year} {2022})}\BibitemShut {NoStop}%
\bibitem [{\citenamefont {Gao}\ and\ \citenamefont {Khalaf}(2022)}]{gao2022symmetry}%
  \BibitemOpen
  \bibfield  {author} {\bibinfo {author} {\bibfnamefont {Q.}~\bibnamefont {Gao}}\ and\ \bibinfo {author} {\bibfnamefont {E.}~\bibnamefont {Khalaf}},\ }\bibfield  {title} {\bibinfo {title} {Symmetry origin of lattice vibration modes in twisted multilayer graphene: Phasons versus moir{\'e} phonons},\ }\href@noop {} {\bibfield  {journal} {\bibinfo  {journal} {Physical Review B}\ }\textbf {\bibinfo {volume} {106}},\ \bibinfo {pages} {075420} (\bibinfo {year} {2022})}\BibitemShut {NoStop}%
\bibitem [{\citenamefont {Yuan}\ and\ \citenamefont {Zhu}(2024)}]{yuan2024flat}%
  \BibitemOpen
  \bibfield  {author} {\bibinfo {author} {\bibfnamefont {X.}~\bibnamefont {Yuan}}\ and\ \bibinfo {author} {\bibfnamefont {S.}~\bibnamefont {Zhu}},\ }\bibfield  {title} {\bibinfo {title} {Flat bands and extreme pseudomagnetic fields in monolayer graphene by topography strain engineering},\ }\href@noop {} {\bibfield  {journal} {\bibinfo  {journal} {Physical Review B}\ }\textbf {\bibinfo {volume} {109}},\ \bibinfo {pages} {245408} (\bibinfo {year} {2024})}\BibitemShut {NoStop}%
\bibitem [{\citenamefont {McMillan}(1975)}]{mcmillan1975landau}%
  \BibitemOpen
  \bibfield  {author} {\bibinfo {author} {\bibfnamefont {W.}~\bibnamefont {McMillan}},\ }\bibfield  {title} {\bibinfo {title} {Landau theory of charge-density waves in transition-metal dichalcogenides},\ }\href@noop {} {\bibfield  {journal} {\bibinfo  {journal} {Physical Review B}\ }\textbf {\bibinfo {volume} {12}},\ \bibinfo {pages} {1187} (\bibinfo {year} {1975})}\BibitemShut {NoStop}%
\bibitem [{\citenamefont {Takada}\ \emph {et~al.}(1985)\citenamefont {Takada}, \citenamefont {Wong},\ and\ \citenamefont {Holstein}}]{takada1985damping}%
  \BibitemOpen
  \bibfield  {author} {\bibinfo {author} {\bibfnamefont {S.}~\bibnamefont {Takada}}, \bibinfo {author} {\bibfnamefont {K.}~\bibnamefont {Wong}},\ and\ \bibinfo {author} {\bibfnamefont {T.}~\bibnamefont {Holstein}},\ }\bibfield  {title} {\bibinfo {title} {Damping of charge-density-wave motion},\ }\href@noop {} {\bibfield  {journal} {\bibinfo  {journal} {Physical Review B}\ }\textbf {\bibinfo {volume} {32}},\ \bibinfo {pages} {4639} (\bibinfo {year} {1985})}\BibitemShut {NoStop}%
\bibitem [{\citenamefont {Gr{\"u}ner}(1988)}]{gruner1988dynamics}%
  \BibitemOpen
  \bibfield  {author} {\bibinfo {author} {\bibfnamefont {G.}~\bibnamefont {Gr{\"u}ner}},\ }\bibfield  {title} {\bibinfo {title} {The dynamics of charge-density waves},\ }\href@noop {} {\bibfield  {journal} {\bibinfo  {journal} {Reviews of modern physics}\ }\textbf {\bibinfo {volume} {60}},\ \bibinfo {pages} {1129} (\bibinfo {year} {1988})}\BibitemShut {NoStop}%
\bibitem [{\citenamefont {Tucker}\ \emph {et~al.}(1988)\citenamefont {Tucker}, \citenamefont {Lyons},\ and\ \citenamefont {Gammie}}]{tucker1988theory}%
  \BibitemOpen
  \bibfield  {author} {\bibinfo {author} {\bibfnamefont {J.}~\bibnamefont {Tucker}}, \bibinfo {author} {\bibfnamefont {W.}~\bibnamefont {Lyons}},\ and\ \bibinfo {author} {\bibfnamefont {G.}~\bibnamefont {Gammie}},\ }\bibfield  {title} {\bibinfo {title} {Theory of charge-density-wave dynamics},\ }\href@noop {} {\bibfield  {journal} {\bibinfo  {journal} {Physical Review B}\ }\textbf {\bibinfo {volume} {38}},\ \bibinfo {pages} {1148} (\bibinfo {year} {1988})}\BibitemShut {NoStop}%
\bibitem [{\citenamefont {Gor'kov}\ and\ \citenamefont {Gr{\"u}ner}(2012)}]{gor2012charge}%
  \BibitemOpen
  \bibfield  {author} {\bibinfo {author} {\bibfnamefont {L.~P.}\ \bibnamefont {Gor'kov}}\ and\ \bibinfo {author} {\bibfnamefont {G.}~\bibnamefont {Gr{\"u}ner}},\ }\href@noop {} {\emph {\bibinfo {title} {Charge density waves in solids}}},\ Vol.~\bibinfo {volume} {25}\ (\bibinfo  {publisher} {Elsevier},\ \bibinfo {year} {2012})\BibitemShut {NoStop}%
\bibitem [{\citenamefont {Rossnagel}(2011)}]{rossnagel2011origin}%
  \BibitemOpen
  \bibfield  {author} {\bibinfo {author} {\bibfnamefont {K.}~\bibnamefont {Rossnagel}},\ }\bibfield  {title} {\bibinfo {title} {On the origin of charge-density waves in select layered transition-metal dichalcogenides},\ }\href@noop {} {\bibfield  {journal} {\bibinfo  {journal} {Journal of Physics: Condensed Matter}\ }\textbf {\bibinfo {volume} {23}},\ \bibinfo {pages} {213001} (\bibinfo {year} {2011})}\BibitemShut {NoStop}%
\bibitem [{\citenamefont {Zhu}\ \emph {et~al.}(2015)\citenamefont {Zhu}, \citenamefont {Cao}, \citenamefont {Zhang}, \citenamefont {Plummer},\ and\ \citenamefont {Guo}}]{zhu2015classification}%
  \BibitemOpen
  \bibfield  {author} {\bibinfo {author} {\bibfnamefont {X.}~\bibnamefont {Zhu}}, \bibinfo {author} {\bibfnamefont {Y.}~\bibnamefont {Cao}}, \bibinfo {author} {\bibfnamefont {J.}~\bibnamefont {Zhang}}, \bibinfo {author} {\bibfnamefont {E.}~\bibnamefont {Plummer}},\ and\ \bibinfo {author} {\bibfnamefont {J.}~\bibnamefont {Guo}},\ }\bibfield  {title} {\bibinfo {title} {Classification of charge density waves based on their nature},\ }\href@noop {} {\bibfield  {journal} {\bibinfo  {journal} {Proceedings of the National Academy of Sciences}\ }\textbf {\bibinfo {volume} {112}},\ \bibinfo {pages} {2367} (\bibinfo {year} {2015})}\BibitemShut {NoStop}%
\bibitem [{\citenamefont {Imada}\ and\ \citenamefont {Scalapino}(1986)}]{imada1986competition}%
  \BibitemOpen
  \bibfield  {author} {\bibinfo {author} {\bibfnamefont {M.}~\bibnamefont {Imada}}\ and\ \bibinfo {author} {\bibfnamefont {D.}~\bibnamefont {Scalapino}},\ }\bibfield  {title} {\bibinfo {title} {Competition between the charge-density-wave and singlet-superconducting phases in a quasi-one-dimensional system},\ }\href@noop {} {\bibfield  {journal} {\bibinfo  {journal} {Physical Review B}\ }\textbf {\bibinfo {volume} {34}},\ \bibinfo {pages} {3480} (\bibinfo {year} {1986})}\BibitemShut {NoStop}%
\bibitem [{\citenamefont {Banerjee}\ \emph {et~al.}(2013)\citenamefont {Banerjee}, \citenamefont {Feng}, \citenamefont {Silevitch}, \citenamefont {Wang}, \citenamefont {Lang}, \citenamefont {Kuo}, \citenamefont {Fisher},\ and\ \citenamefont {Rosenbaum}}]{banerjee2013charge}%
  \BibitemOpen
  \bibfield  {author} {\bibinfo {author} {\bibfnamefont {A.}~\bibnamefont {Banerjee}}, \bibinfo {author} {\bibfnamefont {Y.}~\bibnamefont {Feng}}, \bibinfo {author} {\bibfnamefont {D.}~\bibnamefont {Silevitch}}, \bibinfo {author} {\bibfnamefont {J.}~\bibnamefont {Wang}}, \bibinfo {author} {\bibfnamefont {J.}~\bibnamefont {Lang}}, \bibinfo {author} {\bibfnamefont {H.-H.}\ \bibnamefont {Kuo}}, \bibinfo {author} {\bibfnamefont {I.}~\bibnamefont {Fisher}},\ and\ \bibinfo {author} {\bibfnamefont {T.}~\bibnamefont {Rosenbaum}},\ }\bibfield  {title} {\bibinfo {title} {Charge transfer and multiple density waves in the rare earth tellurides},\ }\href@noop {} {\bibfield  {journal} {\bibinfo  {journal} {Physical Review B—Condensed Matter and Materials Physics}\ }\textbf {\bibinfo {volume} {87}},\ \bibinfo {pages} {155131} (\bibinfo {year} {2013})}\BibitemShut {NoStop}%
\bibitem [{\citenamefont {Melikyan}\ and\ \citenamefont {Te{\v{s}}anovi{\'c}}(2005)}]{melikyan2005model}%
  \BibitemOpen
  \bibfield  {author} {\bibinfo {author} {\bibfnamefont {A.}~\bibnamefont {Melikyan}}\ and\ \bibinfo {author} {\bibfnamefont {Z.}~\bibnamefont {Te{\v{s}}anovi{\'c}}},\ }\bibfield  {title} {\bibinfo {title} {Model of phase fluctuations in a lattice d-wave superconductor: Application to the cooper-pair charge-density wave in underdoped cuprates},\ }\href@noop {} {\bibfield  {journal} {\bibinfo  {journal} {Physical Review B—Condensed Matter and Materials Physics}\ }\textbf {\bibinfo {volume} {71}},\ \bibinfo {pages} {214511} (\bibinfo {year} {2005})}\BibitemShut {NoStop}%
\bibitem [{\citenamefont {Urban}\ \emph {et~al.}(2007)\citenamefont {Urban}, \citenamefont {Stafford},\ and\ \citenamefont {Grabert}}]{urban2007scaling}%
  \BibitemOpen
  \bibfield  {author} {\bibinfo {author} {\bibfnamefont {D.}~\bibnamefont {Urban}}, \bibinfo {author} {\bibfnamefont {C.}~\bibnamefont {Stafford}},\ and\ \bibinfo {author} {\bibfnamefont {H.}~\bibnamefont {Grabert}},\ }\bibfield  {title} {\bibinfo {title} {Scaling theory of the peierls charge density wave in metal nanowires},\ }\href@noop {} {\bibfield  {journal} {\bibinfo  {journal} {Physical Review B—Condensed Matter and Materials Physics}\ }\textbf {\bibinfo {volume} {75}},\ \bibinfo {pages} {205428} (\bibinfo {year} {2007})}\BibitemShut {NoStop}%
\bibitem [{\citenamefont {Hirsch}(1983)}]{hirsch1983effect}%
  \BibitemOpen
  \bibfield  {author} {\bibinfo {author} {\bibfnamefont {J.}~\bibnamefont {Hirsch}},\ }\bibfield  {title} {\bibinfo {title} {Effect of coulomb interactions on the peierls instability},\ }\href@noop {} {\bibfield  {journal} {\bibinfo  {journal} {Physical Review Letters}\ }\textbf {\bibinfo {volume} {51}},\ \bibinfo {pages} {296} (\bibinfo {year} {1983})}\BibitemShut {NoStop}%
\bibitem [{\citenamefont {Boriack}\ and\ \citenamefont {Overhauser}(1977)}]{boriack1977dynamic}%
  \BibitemOpen
  \bibfield  {author} {\bibinfo {author} {\bibfnamefont {M.}~\bibnamefont {Boriack}}\ and\ \bibinfo {author} {\bibfnamefont {A.}~\bibnamefont {Overhauser}},\ }\bibfield  {title} {\bibinfo {title} {The dynamic peierls instability},\ }\href@noop {} {\bibfield  {journal} {\bibinfo  {journal} {Physical Review B}\ }\textbf {\bibinfo {volume} {15}},\ \bibinfo {pages} {2847} (\bibinfo {year} {1977})}\BibitemShut {NoStop}%
\bibitem [{\citenamefont {Johannes}\ and\ \citenamefont {Mazin}(2008)}]{johannes2008fermi}%
  \BibitemOpen
  \bibfield  {author} {\bibinfo {author} {\bibfnamefont {M.}~\bibnamefont {Johannes}}\ and\ \bibinfo {author} {\bibfnamefont {I.}~\bibnamefont {Mazin}},\ }\bibfield  {title} {\bibinfo {title} {Fermi surface nesting and the origin of charge density waves in metals},\ }\href@noop {} {\bibfield  {journal} {\bibinfo  {journal} {Physical Review B—Condensed Matter and Materials Physics}\ }\textbf {\bibinfo {volume} {77}},\ \bibinfo {pages} {165135} (\bibinfo {year} {2008})}\BibitemShut {NoStop}%
\bibitem [{\citenamefont {Hofmann}\ \emph {et~al.}(2019)\citenamefont {Hofmann}, \citenamefont {Ugeda}, \citenamefont {Tamt{\"o}gl}, \citenamefont {Ruckhofer}, \citenamefont {Ernst}, \citenamefont {Benedek}, \citenamefont {Mart{\'\i}nez-Galera}, \citenamefont {Str{\'o}{\.z}ecka}, \citenamefont {G{\'o}mez-Rodr{\'\i}guez}, \citenamefont {Rienks} \emph {et~al.}}]{hofmann2019strong}%
  \BibitemOpen
  \bibfield  {author} {\bibinfo {author} {\bibfnamefont {P.}~\bibnamefont {Hofmann}}, \bibinfo {author} {\bibfnamefont {M.~M.}\ \bibnamefont {Ugeda}}, \bibinfo {author} {\bibfnamefont {A.}~\bibnamefont {Tamt{\"o}gl}}, \bibinfo {author} {\bibfnamefont {A.}~\bibnamefont {Ruckhofer}}, \bibinfo {author} {\bibfnamefont {W.~E.}\ \bibnamefont {Ernst}}, \bibinfo {author} {\bibfnamefont {G.}~\bibnamefont {Benedek}}, \bibinfo {author} {\bibfnamefont {A.~J.}\ \bibnamefont {Mart{\'\i}nez-Galera}}, \bibinfo {author} {\bibfnamefont {A.}~\bibnamefont {Str{\'o}{\.z}ecka}}, \bibinfo {author} {\bibfnamefont {J.~M.}\ \bibnamefont {G{\'o}mez-Rodr{\'\i}guez}}, \bibinfo {author} {\bibfnamefont {E.}~\bibnamefont {Rienks}}, \emph {et~al.},\ }\bibfield  {title} {\bibinfo {title} {Strong-coupling charge density wave in a one-dimensional topological metal},\ }\href@noop {} {\bibfield  {journal} {\bibinfo  {journal} {Physical Review B}\ }\textbf {\bibinfo {volume} {99}},\ \bibinfo {pages} {035438} (\bibinfo {year} {2019})}\BibitemShut
  {NoStop}%
\bibitem [{\citenamefont {Ryu}\ \emph {et~al.}(2018)\citenamefont {Ryu}, \citenamefont {Chen}, \citenamefont {Kim}, \citenamefont {Tsai}, \citenamefont {Tang}, \citenamefont {Jiang}, \citenamefont {Liou}, \citenamefont {Kahn}, \citenamefont {Jia}, \citenamefont {Omrani} \emph {et~al.}}]{ryu2018persistent}%
  \BibitemOpen
  \bibfield  {author} {\bibinfo {author} {\bibfnamefont {H.}~\bibnamefont {Ryu}}, \bibinfo {author} {\bibfnamefont {Y.}~\bibnamefont {Chen}}, \bibinfo {author} {\bibfnamefont {H.}~\bibnamefont {Kim}}, \bibinfo {author} {\bibfnamefont {H.-Z.}\ \bibnamefont {Tsai}}, \bibinfo {author} {\bibfnamefont {S.}~\bibnamefont {Tang}}, \bibinfo {author} {\bibfnamefont {J.}~\bibnamefont {Jiang}}, \bibinfo {author} {\bibfnamefont {F.}~\bibnamefont {Liou}}, \bibinfo {author} {\bibfnamefont {S.}~\bibnamefont {Kahn}}, \bibinfo {author} {\bibfnamefont {C.}~\bibnamefont {Jia}}, \bibinfo {author} {\bibfnamefont {A.~A.}\ \bibnamefont {Omrani}}, \emph {et~al.},\ }\bibfield  {title} {\bibinfo {title} {Persistent charge-density-wave order in single-layer tase2},\ }\href@noop {} {\bibfield  {journal} {\bibinfo  {journal} {Nano letters}\ }\textbf {\bibinfo {volume} {18}},\ \bibinfo {pages} {689} (\bibinfo {year} {2018})}\BibitemShut {NoStop}%
\bibitem [{\citenamefont {Ralevi{\'c}}\ \emph {et~al.}(2016)\citenamefont {Ralevi{\'c}}, \citenamefont {Lazarevi{\'c}}, \citenamefont {Baum}, \citenamefont {Eiter}, \citenamefont {Hackl}, \citenamefont {Giraldo-Gallo}, \citenamefont {Fisher}, \citenamefont {Petrovic}, \citenamefont {Gaji{\'c}},\ and\ \citenamefont {Popovi{\'c}}}]{ralevic2016charge}%
  \BibitemOpen
  \bibfield  {author} {\bibinfo {author} {\bibfnamefont {U.}~\bibnamefont {Ralevi{\'c}}}, \bibinfo {author} {\bibfnamefont {N.}~\bibnamefont {Lazarevi{\'c}}}, \bibinfo {author} {\bibfnamefont {A.}~\bibnamefont {Baum}}, \bibinfo {author} {\bibfnamefont {H.-M.}\ \bibnamefont {Eiter}}, \bibinfo {author} {\bibfnamefont {R.}~\bibnamefont {Hackl}}, \bibinfo {author} {\bibfnamefont {P.}~\bibnamefont {Giraldo-Gallo}}, \bibinfo {author} {\bibfnamefont {I.~R.}\ \bibnamefont {Fisher}}, \bibinfo {author} {\bibfnamefont {C.}~\bibnamefont {Petrovic}}, \bibinfo {author} {\bibfnamefont {R.}~\bibnamefont {Gaji{\'c}}},\ and\ \bibinfo {author} {\bibfnamefont {Z.}~\bibnamefont {Popovi{\'c}}},\ }\bibfield  {title} {\bibinfo {title} {Charge density wave modulation and gap measurements in cete 3},\ }\href@noop {} {\bibfield  {journal} {\bibinfo  {journal} {Physical Review B}\ }\textbf {\bibinfo {volume} {94}},\ \bibinfo {pages} {165132} (\bibinfo {year} {2016})}\BibitemShut {NoStop}%
\bibitem [{\citenamefont {Lee}\ and\ \citenamefont {Fukuyama}(1978)}]{lee1978dynamics}%
  \BibitemOpen
  \bibfield  {author} {\bibinfo {author} {\bibfnamefont {P.}~\bibnamefont {Lee}}\ and\ \bibinfo {author} {\bibfnamefont {H.}~\bibnamefont {Fukuyama}},\ }\bibfield  {title} {\bibinfo {title} {Dynamics of the charge-density wave. ii. long-range coulomb effects in an array of chains},\ }\href@noop {} {\bibfield  {journal} {\bibinfo  {journal} {Physical Review B}\ }\textbf {\bibinfo {volume} {17}},\ \bibinfo {pages} {542} (\bibinfo {year} {1978})}\BibitemShut {NoStop}%
\bibitem [{\citenamefont {Efetov}\ and\ \citenamefont {Larkin}(1977)}]{efetov1977charge}%
  \BibitemOpen
  \bibfield  {author} {\bibinfo {author} {\bibfnamefont {K.}~\bibnamefont {Efetov}}\ and\ \bibinfo {author} {\bibfnamefont {A.}~\bibnamefont {Larkin}},\ }\bibfield  {title} {\bibinfo {title} {Charge-density wave in a random potential},\ }\href@noop {} {\bibfield  {journal} {\bibinfo  {journal} {Sov. Phys. JETP}\ }\textbf {\bibinfo {volume} {45}},\ \bibinfo {pages} {2350} (\bibinfo {year} {1977})}\BibitemShut {NoStop}%
\bibitem [{\citenamefont {Littlewood}\ and\ \citenamefont {Varma}(1982)}]{littlewood1982amplitude}%
  \BibitemOpen
  \bibfield  {author} {\bibinfo {author} {\bibfnamefont {P.}~\bibnamefont {Littlewood}}\ and\ \bibinfo {author} {\bibfnamefont {C.}~\bibnamefont {Varma}},\ }\bibfield  {title} {\bibinfo {title} {Amplitude collective modes in superconductors and their coupling to charge-density waves},\ }\href@noop {} {\bibfield  {journal} {\bibinfo  {journal} {Physical Review B}\ }\textbf {\bibinfo {volume} {26}},\ \bibinfo {pages} {4883} (\bibinfo {year} {1982})}\BibitemShut {NoStop}%
\bibitem [{\citenamefont {Wegner}\ \emph {et~al.}(2020)\citenamefont {Wegner}, \citenamefont {Zhao}, \citenamefont {Li}, \citenamefont {Yang}, \citenamefont {Anikin}, \citenamefont {Karapetrov}, \citenamefont {Esfarjani}, \citenamefont {Louca},\ and\ \citenamefont {Chatterjee}}]{wegner2020evidence}%
  \BibitemOpen
  \bibfield  {author} {\bibinfo {author} {\bibfnamefont {A.}~\bibnamefont {Wegner}}, \bibinfo {author} {\bibfnamefont {J.}~\bibnamefont {Zhao}}, \bibinfo {author} {\bibfnamefont {J.}~\bibnamefont {Li}}, \bibinfo {author} {\bibfnamefont {J.}~\bibnamefont {Yang}}, \bibinfo {author} {\bibfnamefont {A.}~\bibnamefont {Anikin}}, \bibinfo {author} {\bibfnamefont {G.}~\bibnamefont {Karapetrov}}, \bibinfo {author} {\bibfnamefont {K.}~\bibnamefont {Esfarjani}}, \bibinfo {author} {\bibfnamefont {D.}~\bibnamefont {Louca}},\ and\ \bibinfo {author} {\bibfnamefont {U.}~\bibnamefont {Chatterjee}},\ }\bibfield  {title} {\bibinfo {title} {Evidence for pseudo--jahn-teller distortions in the charge density wave phase of 1 t-tise 2},\ }\href@noop {} {\bibfield  {journal} {\bibinfo  {journal} {Physical Review B}\ }\textbf {\bibinfo {volume} {101}},\ \bibinfo {pages} {195145} (\bibinfo {year} {2020})}\BibitemShut {NoStop}%
\bibitem [{\citenamefont {Guo}\ \emph {et~al.}(2024)\citenamefont {Guo}, \citenamefont {Puppin}, \citenamefont {Hellbr{\"u}ck}, \citenamefont {Magrez}, \citenamefont {Guedes}, \citenamefont {Sokolovi{\'c}},\ and\ \citenamefont {Dil}}]{guo2024real}%
  \BibitemOpen
  \bibfield  {author} {\bibinfo {author} {\bibfnamefont {F.}~\bibnamefont {Guo}}, \bibinfo {author} {\bibfnamefont {M.}~\bibnamefont {Puppin}}, \bibinfo {author} {\bibfnamefont {L.}~\bibnamefont {Hellbr{\"u}ck}}, \bibinfo {author} {\bibfnamefont {A.}~\bibnamefont {Magrez}}, \bibinfo {author} {\bibfnamefont {E.~B.}\ \bibnamefont {Guedes}}, \bibinfo {author} {\bibfnamefont {I.}~\bibnamefont {Sokolovi{\'c}}},\ and\ \bibinfo {author} {\bibfnamefont {J.~H.}\ \bibnamefont {Dil}},\ }\bibfield  {title} {\bibinfo {title} {Real-and reciprocal space characterization of the three-dimensional charge density wave in quasi-one-dimensional cute},\ }\href@noop {} {\bibfield  {journal} {\bibinfo  {journal} {Physical Review B}\ }\textbf {\bibinfo {volume} {110}},\ \bibinfo {pages} {115112} (\bibinfo {year} {2024})}\BibitemShut {NoStop}%
\bibitem [{\citenamefont {Gao}\ \emph {et~al.}(2024)\citenamefont {Gao}, \citenamefont {Chan}, \citenamefont {Jiao}, \citenamefont {Chen}, \citenamefont {Yin}, \citenamefont {Tangprapha}, \citenamefont {Yang}, \citenamefont {Li}, \citenamefont {Liu}, \citenamefont {Shen} \emph {et~al.}}]{gao2024observation}%
  \BibitemOpen
  \bibfield  {author} {\bibinfo {author} {\bibfnamefont {Q.}~\bibnamefont {Gao}}, \bibinfo {author} {\bibfnamefont {Y.-h.}\ \bibnamefont {Chan}}, \bibinfo {author} {\bibfnamefont {P.}~\bibnamefont {Jiao}}, \bibinfo {author} {\bibfnamefont {H.}~\bibnamefont {Chen}}, \bibinfo {author} {\bibfnamefont {S.}~\bibnamefont {Yin}}, \bibinfo {author} {\bibfnamefont {K.}~\bibnamefont {Tangprapha}}, \bibinfo {author} {\bibfnamefont {Y.}~\bibnamefont {Yang}}, \bibinfo {author} {\bibfnamefont {X.}~\bibnamefont {Li}}, \bibinfo {author} {\bibfnamefont {Z.}~\bibnamefont {Liu}}, \bibinfo {author} {\bibfnamefont {D.}~\bibnamefont {Shen}}, \emph {et~al.},\ }\bibfield  {title} {\bibinfo {title} {Observation of possible excitonic charge density waves and metal--insulator transitions in atomically thin semimetals},\ }\href@noop {} {\bibfield  {journal} {\bibinfo  {journal} {Nature Physics}\ }\textbf {\bibinfo {volume} {20}},\ \bibinfo {pages} {597} (\bibinfo {year} {2024})}\BibitemShut {NoStop}%
\bibitem [{\citenamefont {Kwon}\ \emph {et~al.}(2024)\citenamefont {Kwon}, \citenamefont {Jung}, \citenamefont {Lee}, \citenamefont {Cho}, \citenamefont {Kong}, \citenamefont {Won}, \citenamefont {Cheong},\ and\ \citenamefont {Yeom}}]{kwon2024dual}%
  \BibitemOpen
  \bibfield  {author} {\bibinfo {author} {\bibfnamefont {S.}~\bibnamefont {Kwon}}, \bibinfo {author} {\bibfnamefont {H.}~\bibnamefont {Jung}}, \bibinfo {author} {\bibfnamefont {S.}~\bibnamefont {Lee}}, \bibinfo {author} {\bibfnamefont {G.~Y.}\ \bibnamefont {Cho}}, \bibinfo {author} {\bibfnamefont {K.}~\bibnamefont {Kong}}, \bibinfo {author} {\bibfnamefont {C.}~\bibnamefont {Won}}, \bibinfo {author} {\bibfnamefont {S.-W.}\ \bibnamefont {Cheong}},\ and\ \bibinfo {author} {\bibfnamefont {H.~W.}\ \bibnamefont {Yeom}},\ }\bibfield  {title} {\bibinfo {title} {Dual higgs modes entangled into a soliton lattice in cute},\ }\href@noop {} {\bibfield  {journal} {\bibinfo  {journal} {Nature communications}\ }\textbf {\bibinfo {volume} {15}},\ \bibinfo {pages} {984} (\bibinfo {year} {2024})}\BibitemShut {NoStop}%
\bibitem [{\citenamefont {Fisher}(1985)}]{fisher1985sliding}%
  \BibitemOpen
  \bibfield  {author} {\bibinfo {author} {\bibfnamefont {D.~S.}\ \bibnamefont {Fisher}},\ }\bibfield  {title} {\bibinfo {title} {Sliding charge-density waves as a dynamic critical phenomenon},\ }\href@noop {} {\bibfield  {journal} {\bibinfo  {journal} {Physical Review B}\ }\textbf {\bibinfo {volume} {31}},\ \bibinfo {pages} {1396} (\bibinfo {year} {1985})}\BibitemShut {NoStop}%
\bibitem [{\citenamefont {Brazovskii}\ and\ \citenamefont {Nattermann}(2004)}]{brazovskii2004pinning}%
  \BibitemOpen
  \bibfield  {author} {\bibinfo {author} {\bibfnamefont {S.}~\bibnamefont {Brazovskii}}\ and\ \bibinfo {author} {\bibfnamefont {T.}~\bibnamefont {Nattermann}},\ }\bibfield  {title} {\bibinfo {title} {Pinning and sliding of driven elastic systems: from domain walls to charge density waves},\ }\href@noop {} {\bibfield  {journal} {\bibinfo  {journal} {Advances in Physics}\ }\textbf {\bibinfo {volume} {53}},\ \bibinfo {pages} {177} (\bibinfo {year} {2004})}\BibitemShut {NoStop}%
\bibitem [{\citenamefont {Monceau}(1985)}]{monceau1985charge}%
  \BibitemOpen
  \bibfield  {author} {\bibinfo {author} {\bibfnamefont {P.}~\bibnamefont {Monceau}},\ }\bibfield  {title} {\bibinfo {title} {Charge-density wave transport in transition metal tri-and tetrachalcogenides},\ }\href@noop {} {\bibfield  {journal} {\bibinfo  {journal} {Electronic Properties of Inorganic Quasi-One-Dimensional Compounds}\ ,\ \bibinfo {pages} {139}} (\bibinfo {year} {1985})}\BibitemShut {NoStop}%
\bibitem [{\citenamefont {Wilson}(1985)}]{wilson1985charge}%
  \BibitemOpen
  \bibfield  {author} {\bibinfo {author} {\bibfnamefont {J.~A.}\ \bibnamefont {Wilson}},\ }\bibfield  {title} {\bibinfo {title} {Charge density waves: sliding and related phenomena in nbse3 and other transition-metal chalcogenides},\ }\href@noop {} {\bibfield  {journal} {\bibinfo  {journal} {Philosophical Transactions of the Royal Society of London. Series A, Mathematical and Physical Sciences}\ }\textbf {\bibinfo {volume} {314}},\ \bibinfo {pages} {159} (\bibinfo {year} {1985})}\BibitemShut {NoStop}%
\bibitem [{\citenamefont {Fleming}\ and\ \citenamefont {Grimes}(1979)}]{PhysRevLett.42.1423}%
  \BibitemOpen
  \bibfield  {author} {\bibinfo {author} {\bibfnamefont {R.~M.}\ \bibnamefont {Fleming}}\ and\ \bibinfo {author} {\bibfnamefont {C.~C.}\ \bibnamefont {Grimes}},\ }\bibfield  {title} {\bibinfo {title} {Sliding-mode conductivity in nb${\mathrm{se}}_{3}$: Observation of a threshold electric field and conduction noise},\ }\href {https://doi.org/10.1103/PhysRevLett.42.1423} {\bibfield  {journal} {\bibinfo  {journal} {Phys. Rev. Lett.}\ }\textbf {\bibinfo {volume} {42}},\ \bibinfo {pages} {1423} (\bibinfo {year} {1979})}\BibitemShut {NoStop}%
\bibitem [{\citenamefont {Ravy}\ \emph {et~al.}(2006)\citenamefont {Ravy}, \citenamefont {Rouzi{\`e}re}, \citenamefont {Pouget}, \citenamefont {Brazovskii}, \citenamefont {Marcus}, \citenamefont {B{\'e}rar},\ and\ \citenamefont {Elkaim}}]{ravy2006disorder}%
  \BibitemOpen
  \bibfield  {author} {\bibinfo {author} {\bibfnamefont {S.}~\bibnamefont {Ravy}}, \bibinfo {author} {\bibfnamefont {S.}~\bibnamefont {Rouzi{\`e}re}}, \bibinfo {author} {\bibfnamefont {J.-P.}\ \bibnamefont {Pouget}}, \bibinfo {author} {\bibfnamefont {S.}~\bibnamefont {Brazovskii}}, \bibinfo {author} {\bibfnamefont {J.}~\bibnamefont {Marcus}}, \bibinfo {author} {\bibfnamefont {J.-F.}\ \bibnamefont {B{\'e}rar}},\ and\ \bibinfo {author} {\bibfnamefont {E.}~\bibnamefont {Elkaim}},\ }\bibfield  {title} {\bibinfo {title} {Disorder effects on the charge-density waves structure in v-and w-doped blue bronzes: Friedel oscillations and charge-density wave pinning},\ }\href@noop {} {\bibfield  {journal} {\bibinfo  {journal} {Physical Review B—Condensed Matter and Materials Physics}\ }\textbf {\bibinfo {volume} {74}},\ \bibinfo {pages} {174102} (\bibinfo {year} {2006})}\BibitemShut {NoStop}%
\bibitem [{\citenamefont {Kvashnin}\ \emph {et~al.}(2020)\citenamefont {Kvashnin}, \citenamefont {VanGennep}, \citenamefont {Mito}, \citenamefont {Medvedev}, \citenamefont {Thiyagarajan}, \citenamefont {Karis}, \citenamefont {Vasiliev}, \citenamefont {Eriksson},\ and\ \citenamefont {Abdel-Hafiez}}]{kvashnin2020coexistence}%
  \BibitemOpen
  \bibfield  {author} {\bibinfo {author} {\bibfnamefont {Y.}~\bibnamefont {Kvashnin}}, \bibinfo {author} {\bibfnamefont {D.}~\bibnamefont {VanGennep}}, \bibinfo {author} {\bibfnamefont {M.}~\bibnamefont {Mito}}, \bibinfo {author} {\bibfnamefont {S.}~\bibnamefont {Medvedev}}, \bibinfo {author} {\bibfnamefont {R.}~\bibnamefont {Thiyagarajan}}, \bibinfo {author} {\bibfnamefont {O.}~\bibnamefont {Karis}}, \bibinfo {author} {\bibfnamefont {A.}~\bibnamefont {Vasiliev}}, \bibinfo {author} {\bibfnamefont {O.}~\bibnamefont {Eriksson}},\ and\ \bibinfo {author} {\bibfnamefont {M.}~\bibnamefont {Abdel-Hafiez}},\ }\bibfield  {title} {\bibinfo {title} {Coexistence of superconductivity and charge density waves in tantalum disulfide: Experiment and theory},\ }\href@noop {} {\bibfield  {journal} {\bibinfo  {journal} {Physical Review Letters}\ }\textbf {\bibinfo {volume} {125}},\ \bibinfo {pages} {186401} (\bibinfo {year} {2020})}\BibitemShut {NoStop}%
\bibitem [{\citenamefont {Cheng}\ \emph {et~al.}(2024)\citenamefont {Cheng}, \citenamefont {Niu}, \citenamefont {Meng}, \citenamefont {Han}, \citenamefont {Qiao}, \citenamefont {Zhang},\ and\ \citenamefont {Zhao}}]{cheng2024engineering}%
  \BibitemOpen
  \bibfield  {author} {\bibinfo {author} {\bibfnamefont {W.~N.}\ \bibnamefont {Cheng}}, \bibinfo {author} {\bibfnamefont {M.}~\bibnamefont {Niu}}, \bibinfo {author} {\bibfnamefont {Y.}~\bibnamefont {Meng}}, \bibinfo {author} {\bibfnamefont {X.}~\bibnamefont {Han}}, \bibinfo {author} {\bibfnamefont {J.}~\bibnamefont {Qiao}}, \bibinfo {author} {\bibfnamefont {J.}~\bibnamefont {Zhang}},\ and\ \bibinfo {author} {\bibfnamefont {X.}~\bibnamefont {Zhao}},\ }\bibfield  {title} {\bibinfo {title} {Engineering charge density waves by stackingtronics in tantalum disulfide},\ }\href@noop {} {\bibfield  {journal} {\bibinfo  {journal} {Nano Letters}\ }\textbf {\bibinfo {volume} {24}},\ \bibinfo {pages} {6441} (\bibinfo {year} {2024})}\BibitemShut {NoStop}%
\bibitem [{\citenamefont {Yao}\ \emph {et~al.}(2024)\citenamefont {Yao}, \citenamefont {Wen}, \citenamefont {Zhao}, \citenamefont {Zhu}, \citenamefont {Wang}, \citenamefont {Tang}, \citenamefont {Yan}, \citenamefont {Zhang}, \citenamefont {Huang}, \citenamefont {Sun} \emph {et~al.}}]{yao2024electric}%
  \BibitemOpen
  \bibfield  {author} {\bibinfo {author} {\bibfnamefont {C.}~\bibnamefont {Yao}}, \bibinfo {author} {\bibfnamefont {L.}~\bibnamefont {Wen}}, \bibinfo {author} {\bibfnamefont {Y.}~\bibnamefont {Zhao}}, \bibinfo {author} {\bibfnamefont {Y.}~\bibnamefont {Zhu}}, \bibinfo {author} {\bibfnamefont {Y.}~\bibnamefont {Wang}}, \bibinfo {author} {\bibfnamefont {G.}~\bibnamefont {Tang}}, \bibinfo {author} {\bibfnamefont {X.}~\bibnamefont {Yan}}, \bibinfo {author} {\bibfnamefont {J.}~\bibnamefont {Zhang}}, \bibinfo {author} {\bibfnamefont {Y.}~\bibnamefont {Huang}}, \bibinfo {author} {\bibfnamefont {H.}~\bibnamefont {Sun}}, \emph {et~al.},\ }\bibfield  {title} {\bibinfo {title} {Electric field effect on the charge density wave in the quasi-one-dimensional semimetal ta 2 ni se 7},\ }\href@noop {} {\bibfield  {journal} {\bibinfo  {journal} {Physical Review B}\ }\textbf {\bibinfo {volume} {110}},\ \bibinfo {pages} {165136} (\bibinfo {year} {2024})}\BibitemShut {NoStop}%
\bibitem [{\citenamefont {He}\ \emph {et~al.}(2017)\citenamefont {He}, \citenamefont {Zhang}, \citenamefont {Wen}, \citenamefont {Yang}, \citenamefont {Liu}, \citenamefont {Wu}, \citenamefont {Lian}, \citenamefont {Xing}, \citenamefont {Wang}, \citenamefont {Mao} \emph {et~al.}}]{he2017band}%
  \BibitemOpen
  \bibfield  {author} {\bibinfo {author} {\bibfnamefont {J.}~\bibnamefont {He}}, \bibinfo {author} {\bibfnamefont {Y.}~\bibnamefont {Zhang}}, \bibinfo {author} {\bibfnamefont {L.}~\bibnamefont {Wen}}, \bibinfo {author} {\bibfnamefont {Y.}~\bibnamefont {Yang}}, \bibinfo {author} {\bibfnamefont {J.}~\bibnamefont {Liu}}, \bibinfo {author} {\bibfnamefont {Y.}~\bibnamefont {Wu}}, \bibinfo {author} {\bibfnamefont {H.}~\bibnamefont {Lian}}, \bibinfo {author} {\bibfnamefont {H.}~\bibnamefont {Xing}}, \bibinfo {author} {\bibfnamefont {S.}~\bibnamefont {Wang}}, \bibinfo {author} {\bibfnamefont {Z.}~\bibnamefont {Mao}}, \emph {et~al.},\ }\bibfield  {title} {\bibinfo {title} {Band dependence of charge density wave in quasi-one-dimensional ta2nise7 probed by orbital magnetoresistance},\ }\href@noop {} {\bibfield  {journal} {\bibinfo  {journal} {Applied Physics Letters}\ }\textbf {\bibinfo {volume} {111}} (\bibinfo {year} {2017})}\BibitemShut {NoStop}%
\bibitem [{\citenamefont {Watson}\ \emph {et~al.}(2023)\citenamefont {Watson}, \citenamefont {Louat}, \citenamefont {Cacho}, \citenamefont {Choi}, \citenamefont {Lee}, \citenamefont {Neumann},\ and\ \citenamefont {Kim}}]{watson2023spectral}%
  \BibitemOpen
  \bibfield  {author} {\bibinfo {author} {\bibfnamefont {M.~D.}\ \bibnamefont {Watson}}, \bibinfo {author} {\bibfnamefont {A.}~\bibnamefont {Louat}}, \bibinfo {author} {\bibfnamefont {C.}~\bibnamefont {Cacho}}, \bibinfo {author} {\bibfnamefont {S.}~\bibnamefont {Choi}}, \bibinfo {author} {\bibfnamefont {Y.~H.}\ \bibnamefont {Lee}}, \bibinfo {author} {\bibfnamefont {M.}~\bibnamefont {Neumann}},\ and\ \bibinfo {author} {\bibfnamefont {G.}~\bibnamefont {Kim}},\ }\bibfield  {title} {\bibinfo {title} {Spectral signatures of a unique charge density wave in ta2nise7},\ }\href@noop {} {\bibfield  {journal} {\bibinfo  {journal} {Nature Communications}\ }\textbf {\bibinfo {volume} {14}},\ \bibinfo {pages} {3388} (\bibinfo {year} {2023})}\BibitemShut {NoStop}%
\bibitem [{\citenamefont {Xiao}\ \emph {et~al.}(2024)\citenamefont {Xiao}, \citenamefont {Sun}, \citenamefont {Chen}, \citenamefont {Han}, \citenamefont {Yu}, \citenamefont {Liu}, \citenamefont {Li}, \citenamefont {Xiao},\ and\ \citenamefont {Yao}}]{xiao2024robust}%
  \BibitemOpen
  \bibfield  {author} {\bibinfo {author} {\bibfnamefont {P.}~\bibnamefont {Xiao}}, \bibinfo {author} {\bibfnamefont {X.}~\bibnamefont {Sun}}, \bibinfo {author} {\bibfnamefont {Y.}~\bibnamefont {Chen}}, \bibinfo {author} {\bibfnamefont {Y.}~\bibnamefont {Han}}, \bibinfo {author} {\bibfnamefont {Z.-M.}\ \bibnamefont {Yu}}, \bibinfo {author} {\bibfnamefont {W.}~\bibnamefont {Liu}}, \bibinfo {author} {\bibfnamefont {X.}~\bibnamefont {Li}}, \bibinfo {author} {\bibfnamefont {W.}~\bibnamefont {Xiao}},\ and\ \bibinfo {author} {\bibfnamefont {Y.}~\bibnamefont {Yao}},\ }\bibfield  {title} {\bibinfo {title} {Robust edge states of quasi-1d material ta2nise7 and applications in saturable absorbers},\ }\href@noop {} {\bibfield  {journal} {\bibinfo  {journal} {Nano Letters}\ }\textbf {\bibinfo {volume} {24}},\ \bibinfo {pages} {10402} (\bibinfo {year} {2024})}\BibitemShut {NoStop}%
\bibitem [{\citenamefont {Birkbeck}\ \emph {et~al.}(2024)\citenamefont {Birkbeck}, \citenamefont {Xiao}, \citenamefont {Inbar}, \citenamefont {Taniguchi}, \citenamefont {Watanabe}, \citenamefont {Berg}, \citenamefont {Glazman}, \citenamefont {Guinea}, \citenamefont {von Oppen},\ and\ \citenamefont {Ilani}}]{birkbeck2024measuring}%
  \BibitemOpen
  \bibfield  {author} {\bibinfo {author} {\bibfnamefont {J.}~\bibnamefont {Birkbeck}}, \bibinfo {author} {\bibfnamefont {J.}~\bibnamefont {Xiao}}, \bibinfo {author} {\bibfnamefont {A.}~\bibnamefont {Inbar}}, \bibinfo {author} {\bibfnamefont {T.}~\bibnamefont {Taniguchi}}, \bibinfo {author} {\bibfnamefont {K.}~\bibnamefont {Watanabe}}, \bibinfo {author} {\bibfnamefont {E.}~\bibnamefont {Berg}}, \bibinfo {author} {\bibfnamefont {L.}~\bibnamefont {Glazman}}, \bibinfo {author} {\bibfnamefont {F.}~\bibnamefont {Guinea}}, \bibinfo {author} {\bibfnamefont {F.}~\bibnamefont {von Oppen}},\ and\ \bibinfo {author} {\bibfnamefont {S.}~\bibnamefont {Ilani}},\ }\bibfield  {title} {\bibinfo {title} {Measuring phonon dispersion and electron-phason coupling in twisted bilayer graphene with a cryogenic quantum twisting microscope},\ }\href@noop {} {\bibfield  {journal} {\bibinfo  {journal} {arXiv preprint arXiv:2407.13404}\ } (\bibinfo {year} {2024})}\BibitemShut {NoStop}%
\bibitem [{\citenamefont {Ochoa}(2019)}]{ochoa2019moire}%
  \BibitemOpen
  \bibfield  {author} {\bibinfo {author} {\bibfnamefont {H.}~\bibnamefont {Ochoa}},\ }\bibfield  {title} {\bibinfo {title} {Moir{\'e}-pattern fluctuations and electron-phason coupling in twisted bilayer graphene},\ }\href@noop {} {\bibfield  {journal} {\bibinfo  {journal} {Physical Review B}\ }\textbf {\bibinfo {volume} {100}},\ \bibinfo {pages} {155426} (\bibinfo {year} {2019})}\BibitemShut {NoStop}%
\bibitem [{\citenamefont {Imambekov}\ \emph {et~al.}(2012)\citenamefont {Imambekov}, \citenamefont {Schmidt},\ and\ \citenamefont {Glazman}}]{imambekov2012one}%
  \BibitemOpen
  \bibfield  {author} {\bibinfo {author} {\bibfnamefont {A.}~\bibnamefont {Imambekov}}, \bibinfo {author} {\bibfnamefont {T.~L.}\ \bibnamefont {Schmidt}},\ and\ \bibinfo {author} {\bibfnamefont {L.~I.}\ \bibnamefont {Glazman}},\ }\bibfield  {title} {\bibinfo {title} {One-dimensional quantum liquids: Beyond the luttinger liquid paradigm},\ }\href@noop {} {\bibfield  {journal} {\bibinfo  {journal} {Reviews of Modern Physics}\ }\textbf {\bibinfo {volume} {84}},\ \bibinfo {pages} {1253} (\bibinfo {year} {2012})}\BibitemShut {NoStop}%
\bibitem [{\citenamefont {Miao}\ \emph {et~al.}(2019)\citenamefont {Miao}, \citenamefont {Fumagalli}, \citenamefont {Rossi}, \citenamefont {Lorenzana}, \citenamefont {Seibold}, \citenamefont {Yakhou-Harris}, \citenamefont {Kummer}, \citenamefont {Brookes}, \citenamefont {Gu}, \citenamefont {Braicovich} \emph {et~al.}}]{miao2019formation}%
  \BibitemOpen
  \bibfield  {author} {\bibinfo {author} {\bibfnamefont {H.}~\bibnamefont {Miao}}, \bibinfo {author} {\bibfnamefont {R.}~\bibnamefont {Fumagalli}}, \bibinfo {author} {\bibfnamefont {M.}~\bibnamefont {Rossi}}, \bibinfo {author} {\bibfnamefont {J.}~\bibnamefont {Lorenzana}}, \bibinfo {author} {\bibfnamefont {G.}~\bibnamefont {Seibold}}, \bibinfo {author} {\bibfnamefont {F.}~\bibnamefont {Yakhou-Harris}}, \bibinfo {author} {\bibfnamefont {K.}~\bibnamefont {Kummer}}, \bibinfo {author} {\bibfnamefont {N.~B.}\ \bibnamefont {Brookes}}, \bibinfo {author} {\bibfnamefont {G.}~\bibnamefont {Gu}}, \bibinfo {author} {\bibfnamefont {L.}~\bibnamefont {Braicovich}}, \emph {et~al.},\ }\bibfield  {title} {\bibinfo {title} {Formation of incommensurate charge density waves in cuprates},\ }\href@noop {} {\bibfield  {journal} {\bibinfo  {journal} {Physical Review X}\ }\textbf {\bibinfo {volume} {9}},\ \bibinfo {pages} {031042} (\bibinfo {year} {2019})}\BibitemShut {NoStop}%
\bibitem [{\citenamefont {Shapiro}\ \emph {et~al.}(2007)\citenamefont {Shapiro}, \citenamefont {Vorderwisch}, \citenamefont {Habicht}, \citenamefont {Hradil},\ and\ \citenamefont {Schneider}}]{shapiro2007observation}%
  \BibitemOpen
  \bibfield  {author} {\bibinfo {author} {\bibfnamefont {S.}~\bibnamefont {Shapiro}}, \bibinfo {author} {\bibfnamefont {P.}~\bibnamefont {Vorderwisch}}, \bibinfo {author} {\bibfnamefont {K.}~\bibnamefont {Habicht}}, \bibinfo {author} {\bibfnamefont {K.}~\bibnamefont {Hradil}},\ and\ \bibinfo {author} {\bibfnamefont {H.}~\bibnamefont {Schneider}},\ }\bibfield  {title} {\bibinfo {title} {Observation of phasons in the magnetic shape memory alloy ni2mnga},\ }\href@noop {} {\bibfield  {journal} {\bibinfo  {journal} {Europhysics Letters}\ }\textbf {\bibinfo {volume} {77}},\ \bibinfo {pages} {56004} (\bibinfo {year} {2007})}\BibitemShut {NoStop}%
\bibitem [{\citenamefont {Fukuyama}\ and\ \citenamefont {Ogata}(2020)}]{fukuyama2020theory}%
  \BibitemOpen
  \bibfield  {author} {\bibinfo {author} {\bibfnamefont {H.}~\bibnamefont {Fukuyama}}\ and\ \bibinfo {author} {\bibfnamefont {M.}~\bibnamefont {Ogata}},\ }\bibfield  {title} {\bibinfo {title} {Theory of phason drag effect on thermoelectricity},\ }\href@noop {} {\bibfield  {journal} {\bibinfo  {journal} {Physical Review B}\ }\textbf {\bibinfo {volume} {102}},\ \bibinfo {pages} {205136} (\bibinfo {year} {2020})}\BibitemShut {NoStop}%
\bibitem [{\citenamefont {Abdulsalam}\ and\ \citenamefont {Joubert}(2015)}]{abdulsalam2015structural}%
  \BibitemOpen
  \bibfield  {author} {\bibinfo {author} {\bibfnamefont {M.}~\bibnamefont {Abdulsalam}}\ and\ \bibinfo {author} {\bibfnamefont {D.~P.}\ \bibnamefont {Joubert}},\ }\bibfield  {title} {\bibinfo {title} {Structural and electronic properties of mx3 (m= ti, zr and hf; x= s, se, te) from first principles calculations},\ }\href@noop {} {\bibfield  {journal} {\bibinfo  {journal} {The European Physical Journal B}\ }\textbf {\bibinfo {volume} {88}},\ \bibinfo {pages} {177} (\bibinfo {year} {2015})}\BibitemShut {NoStop}%
\bibitem [{\citenamefont {Liu}\ \emph {et~al.}(2021)\citenamefont {Liu}, \citenamefont {Li}, \citenamefont {Zhang}, \citenamefont {Li}, \citenamefont {Yang}, \citenamefont {Zhang}, \citenamefont {Chen}, \citenamefont {Uwatoko}, \citenamefont {Yang}, \citenamefont {Sui} \emph {et~al.}}]{liu2021quasi}%
  \BibitemOpen
  \bibfield  {author} {\bibinfo {author} {\bibfnamefont {Z.}~\bibnamefont {Liu}}, \bibinfo {author} {\bibfnamefont {J.}~\bibnamefont {Li}}, \bibinfo {author} {\bibfnamefont {J.}~\bibnamefont {Zhang}}, \bibinfo {author} {\bibfnamefont {J.}~\bibnamefont {Li}}, \bibinfo {author} {\bibfnamefont {P.}~\bibnamefont {Yang}}, \bibinfo {author} {\bibfnamefont {S.}~\bibnamefont {Zhang}}, \bibinfo {author} {\bibfnamefont {G.}~\bibnamefont {Chen}}, \bibinfo {author} {\bibfnamefont {Y.}~\bibnamefont {Uwatoko}}, \bibinfo {author} {\bibfnamefont {H.}~\bibnamefont {Yang}}, \bibinfo {author} {\bibfnamefont {Y.}~\bibnamefont {Sui}}, \emph {et~al.},\ }\bibfield  {title} {\bibinfo {title} {Quasi-one-dimensional superconductivity in the pressurized charge-density-wave conductor hfte3},\ }\href@noop {} {\bibfield  {journal} {\bibinfo  {journal} {npj Quantum Materials}\ }\textbf {\bibinfo {volume} {6}},\ \bibinfo {pages} {90} (\bibinfo {year} {2021})}\BibitemShut {NoStop}%
\bibitem [{\citenamefont {Schrieffer}\ and\ \citenamefont {Wolff}(1966)}]{schrieffer1966relation}%
  \BibitemOpen
  \bibfield  {author} {\bibinfo {author} {\bibfnamefont {J.~R.}\ \bibnamefont {Schrieffer}}\ and\ \bibinfo {author} {\bibfnamefont {P.~A.}\ \bibnamefont {Wolff}},\ }\bibfield  {title} {\bibinfo {title} {Relation between the anderson and kondo hamiltonians},\ }\href@noop {} {\bibfield  {journal} {\bibinfo  {journal} {Physical Review}\ }\textbf {\bibinfo {volume} {149}},\ \bibinfo {pages} {491} (\bibinfo {year} {1966})}\BibitemShut {NoStop}%
\bibitem [{\citenamefont {Mermin}(1970)}]{mermin1970lindhard}%
  \BibitemOpen
  \bibfield  {author} {\bibinfo {author} {\bibfnamefont {N.~D.}\ \bibnamefont {Mermin}},\ }\bibfield  {title} {\bibinfo {title} {Lindhard dielectric function in the relaxation-time approximation},\ }\href@noop {} {\bibfield  {journal} {\bibinfo  {journal} {Physical Review B}\ }\textbf {\bibinfo {volume} {1}},\ \bibinfo {pages} {2362} (\bibinfo {year} {1970})}\BibitemShut {NoStop}%
\bibitem [{\citenamefont {Van~Loon}\ \emph {et~al.}(2021)\citenamefont {Van~Loon}, \citenamefont {R{\"o}sner}, \citenamefont {Katsnelson},\ and\ \citenamefont {Wehling}}]{van2021random}%
  \BibitemOpen
  \bibfield  {author} {\bibinfo {author} {\bibfnamefont {E.~G.}\ \bibnamefont {Van~Loon}}, \bibinfo {author} {\bibfnamefont {M.}~\bibnamefont {R{\"o}sner}}, \bibinfo {author} {\bibfnamefont {M.~I.}\ \bibnamefont {Katsnelson}},\ and\ \bibinfo {author} {\bibfnamefont {T.~O.}\ \bibnamefont {Wehling}},\ }\bibfield  {title} {\bibinfo {title} {Random phase approximation for gapped systems: Role of vertex corrections and applicability of the constrained random phase approximation},\ }\href@noop {} {\bibfield  {journal} {\bibinfo  {journal} {Physical Review B}\ }\textbf {\bibinfo {volume} {104}},\ \bibinfo {pages} {045134} (\bibinfo {year} {2021})}\BibitemShut {NoStop}%
\bibitem [{\citenamefont {Mahan}(2013)}]{mahan2013many}%
  \BibitemOpen
  \bibfield  {author} {\bibinfo {author} {\bibfnamefont {G.~D.}\ \bibnamefont {Mahan}},\ }\href@noop {} {\emph {\bibinfo {title} {Many-particle physics}}}\ (\bibinfo  {publisher} {Springer Science \& Business Media},\ \bibinfo {year} {2013})\BibitemShut {NoStop}%
\bibitem [{\citenamefont {Altland}\ and\ \citenamefont {Simons}(2010)}]{altland2010condensed}%
  \BibitemOpen
  \bibfield  {author} {\bibinfo {author} {\bibfnamefont {A.}~\bibnamefont {Altland}}\ and\ \bibinfo {author} {\bibfnamefont {B.~D.}\ \bibnamefont {Simons}},\ }\href@noop {} {\emph {\bibinfo {title} {Condensed matter field theory}}}\ (\bibinfo  {publisher} {Cambridge university press},\ \bibinfo {year} {2010})\BibitemShut {NoStop}%
\bibitem [{\citenamefont {Tsvelik}(2007)}]{tsvelik2007quantum}%
  \BibitemOpen
  \bibfield  {author} {\bibinfo {author} {\bibfnamefont {A.~M.}\ \bibnamefont {Tsvelik}},\ }\href@noop {} {\emph {\bibinfo {title} {Quantum field theory in condensed matter physics}}}\ (\bibinfo  {publisher} {Cambridge university press},\ \bibinfo {year} {2007})\BibitemShut {NoStop}%
\bibitem [{\citenamefont {Giamarchi}(2003)}]{giamarchi2003quantum}%
  \BibitemOpen
  \bibfield  {author} {\bibinfo {author} {\bibfnamefont {T.}~\bibnamefont {Giamarchi}},\ }\href@noop {} {\emph {\bibinfo {title} {Quantum physics in one dimension}}},\ Vol.\ \bibinfo {volume} {121}\ (\bibinfo  {publisher} {Clarendon press},\ \bibinfo {year} {2003})\BibitemShut {NoStop}%
\bibitem [{\citenamefont {Giamarchi}(1992)}]{giamarchi1992resistivity}%
  \BibitemOpen
  \bibfield  {author} {\bibinfo {author} {\bibfnamefont {T.}~\bibnamefont {Giamarchi}},\ }\bibfield  {title} {\bibinfo {title} {Resistivity of a one-dimensional interacting quantum fluid},\ }\href@noop {} {\bibfield  {journal} {\bibinfo  {journal} {Physical Review B}\ }\textbf {\bibinfo {volume} {46}},\ \bibinfo {pages} {342} (\bibinfo {year} {1992})}\BibitemShut {NoStop}%
\bibitem [{\citenamefont {Coddens}(2006)}]{coddens2006problem}%
  \BibitemOpen
  \bibfield  {author} {\bibinfo {author} {\bibfnamefont {G.}~\bibnamefont {Coddens}},\ }\bibfield  {title} {\bibinfo {title} {On the problem of the relation between phason elasticity and phason dynamics in quasicrystals},\ }\href@noop {} {\bibfield  {journal} {\bibinfo  {journal} {The European Physical Journal B-Condensed Matter and Complex Systems}\ }\textbf {\bibinfo {volume} {54}},\ \bibinfo {pages} {37} (\bibinfo {year} {2006})}\BibitemShut {NoStop}%
\bibitem [{\citenamefont {Chaikin}\ \emph {et~al.}(1995)\citenamefont {Chaikin}, \citenamefont {Lubensky},\ and\ \citenamefont {Witten}}]{chaikin1995principles}%
  \BibitemOpen
  \bibfield  {author} {\bibinfo {author} {\bibfnamefont {P.~M.}\ \bibnamefont {Chaikin}}, \bibinfo {author} {\bibfnamefont {T.~C.}\ \bibnamefont {Lubensky}},\ and\ \bibinfo {author} {\bibfnamefont {T.~A.}\ \bibnamefont {Witten}},\ }\href@noop {} {\emph {\bibinfo {title} {Principles of condensed matter physics}}},\ Vol.~\bibinfo {volume} {10}\ (\bibinfo  {publisher} {Cambridge university press Cambridge},\ \bibinfo {year} {1995})\BibitemShut {NoStop}%
\bibitem [{\citenamefont {Zhao}\ \emph {et~al.}(2018)\citenamefont {Zhao}, \citenamefont {Guo}, \citenamefont {Zhou}, \citenamefont {Yao}, \citenamefont {Ren}, \citenamefont {Bai},\ and\ \citenamefont {Xu}}]{zhao2018band}%
  \BibitemOpen
  \bibfield  {author} {\bibinfo {author} {\bibfnamefont {Q.}~\bibnamefont {Zhao}}, \bibinfo {author} {\bibfnamefont {Y.}~\bibnamefont {Guo}}, \bibinfo {author} {\bibfnamefont {Y.}~\bibnamefont {Zhou}}, \bibinfo {author} {\bibfnamefont {Z.}~\bibnamefont {Yao}}, \bibinfo {author} {\bibfnamefont {Z.}~\bibnamefont {Ren}}, \bibinfo {author} {\bibfnamefont {J.}~\bibnamefont {Bai}},\ and\ \bibinfo {author} {\bibfnamefont {X.}~\bibnamefont {Xu}},\ }\bibfield  {title} {\bibinfo {title} {Band alignments and heterostructures of monolayer transition metal trichalcogenides mx 3 (m= zr, hf; x= s, se) and dichalcogenides mx 2 (m= tc, re; x= s, se) for solar applications},\ }\href@noop {} {\bibfield  {journal} {\bibinfo  {journal} {Nanoscale}\ }\textbf {\bibinfo {volume} {10}},\ \bibinfo {pages} {3547} (\bibinfo {year} {2018})}\BibitemShut {NoStop}%
\bibitem [{\citenamefont {Li}\ \emph {et~al.}(2017)\citenamefont {Li}, \citenamefont {Peng}, \citenamefont {Zhang},\ and\ \citenamefont {Chen}}]{li2017anisotropic}%
  \BibitemOpen
  \bibfield  {author} {\bibinfo {author} {\bibfnamefont {J.}~\bibnamefont {Li}}, \bibinfo {author} {\bibfnamefont {J.}~\bibnamefont {Peng}}, \bibinfo {author} {\bibfnamefont {S.}~\bibnamefont {Zhang}},\ and\ \bibinfo {author} {\bibfnamefont {G.}~\bibnamefont {Chen}},\ }\bibfield  {title} {\bibinfo {title} {Anisotropic multichain nature and filamentary superconductivity in the charge density wave system hfte 3},\ }\href@noop {} {\bibfield  {journal} {\bibinfo  {journal} {Physical Review B}\ }\textbf {\bibinfo {volume} {96}},\ \bibinfo {pages} {174510} (\bibinfo {year} {2017})}\BibitemShut {NoStop}%
\bibitem [{\citenamefont {Island}\ \emph {et~al.}(2016)\citenamefont {Island}, \citenamefont {Biele}, \citenamefont {Barawi}, \citenamefont {Clamagirand}, \citenamefont {Ares}, \citenamefont {S{\'a}nchez}, \citenamefont {van~der Zant}, \citenamefont {Ferrer}, \citenamefont {D’Agosta},\ and\ \citenamefont {Castellanos-Gomez}}]{island2016titanium}%
  \BibitemOpen
  \bibfield  {author} {\bibinfo {author} {\bibfnamefont {J.~O.}\ \bibnamefont {Island}}, \bibinfo {author} {\bibfnamefont {R.}~\bibnamefont {Biele}}, \bibinfo {author} {\bibfnamefont {M.}~\bibnamefont {Barawi}}, \bibinfo {author} {\bibfnamefont {J.~M.}\ \bibnamefont {Clamagirand}}, \bibinfo {author} {\bibfnamefont {J.~R.}\ \bibnamefont {Ares}}, \bibinfo {author} {\bibfnamefont {C.}~\bibnamefont {S{\'a}nchez}}, \bibinfo {author} {\bibfnamefont {H.~S.}\ \bibnamefont {van~der Zant}}, \bibinfo {author} {\bibfnamefont {I.~J.}\ \bibnamefont {Ferrer}}, \bibinfo {author} {\bibfnamefont {R.}~\bibnamefont {D’Agosta}},\ and\ \bibinfo {author} {\bibfnamefont {A.}~\bibnamefont {Castellanos-Gomez}},\ }\bibfield  {title} {\bibinfo {title} {Titanium trisulfide (tis3): a 2d semiconductor with quasi-1d optical and electronic properties},\ }\href@noop {} {\bibfield  {journal} {\bibinfo  {journal} {Scientific reports}\ }\textbf {\bibinfo {volume} {6}},\ \bibinfo {pages} {22214} (\bibinfo {year} {2016})}\BibitemShut {NoStop}%
\bibitem [{\citenamefont {Denholme}\ \emph {et~al.}(2017)\citenamefont {Denholme}, \citenamefont {Yukawa}, \citenamefont {Tsumura}, \citenamefont {Nagao}, \citenamefont {Tamura}, \citenamefont {Watauchi}, \citenamefont {Tanaka}, \citenamefont {Takayanagi},\ and\ \citenamefont {Miyakawa}}]{denholme2017coexistence}%
  \BibitemOpen
  \bibfield  {author} {\bibinfo {author} {\bibfnamefont {S.~J.}\ \bibnamefont {Denholme}}, \bibinfo {author} {\bibfnamefont {A.}~\bibnamefont {Yukawa}}, \bibinfo {author} {\bibfnamefont {K.}~\bibnamefont {Tsumura}}, \bibinfo {author} {\bibfnamefont {M.}~\bibnamefont {Nagao}}, \bibinfo {author} {\bibfnamefont {R.}~\bibnamefont {Tamura}}, \bibinfo {author} {\bibfnamefont {S.}~\bibnamefont {Watauchi}}, \bibinfo {author} {\bibfnamefont {I.}~\bibnamefont {Tanaka}}, \bibinfo {author} {\bibfnamefont {H.}~\bibnamefont {Takayanagi}},\ and\ \bibinfo {author} {\bibfnamefont {N.}~\bibnamefont {Miyakawa}},\ }\bibfield  {title} {\bibinfo {title} {Coexistence of superconductivity and charge-density wave in the quasi-one-dimensional material hfte3},\ }\href@noop {} {\bibfield  {journal} {\bibinfo  {journal} {Scientific reports}\ }\textbf {\bibinfo {volume} {7}},\ \bibinfo {pages} {45217} (\bibinfo {year} {2017})}\BibitemShut {NoStop}%
\bibitem [{\citenamefont {Meyer}\ \emph {et~al.}(2019)\citenamefont {Meyer}, \citenamefont {Pham}, \citenamefont {Oh}, \citenamefont {Ercius}, \citenamefont {Kisielowski}, \citenamefont {Cohen},\ and\ \citenamefont {Zettl}}]{meyer2019metal}%
  \BibitemOpen
  \bibfield  {author} {\bibinfo {author} {\bibfnamefont {S.}~\bibnamefont {Meyer}}, \bibinfo {author} {\bibfnamefont {T.}~\bibnamefont {Pham}}, \bibinfo {author} {\bibfnamefont {S.}~\bibnamefont {Oh}}, \bibinfo {author} {\bibfnamefont {P.}~\bibnamefont {Ercius}}, \bibinfo {author} {\bibfnamefont {C.}~\bibnamefont {Kisielowski}}, \bibinfo {author} {\bibfnamefont {M.~L.}\ \bibnamefont {Cohen}},\ and\ \bibinfo {author} {\bibfnamefont {A.}~\bibnamefont {Zettl}},\ }\bibfield  {title} {\bibinfo {title} {Metal-insulator transition in quasi-one-dimensional hfte 3 in the few-chain limit},\ }\href@noop {} {\bibfield  {journal} {\bibinfo  {journal} {Physical Review B}\ }\textbf {\bibinfo {volume} {100}},\ \bibinfo {pages} {041403} (\bibinfo {year} {2019})}\BibitemShut {NoStop}%
\bibitem [{\citenamefont {Chen}\ \emph {et~al.}(2024)\citenamefont {Chen}, \citenamefont {Zhu}, \citenamefont {Lei}, \citenamefont {Zhuo}, \citenamefont {Wang}, \citenamefont {Ma}, \citenamefont {Luo}, \citenamefont {Xiang},\ and\ \citenamefont {Chen}}]{chen2024thickness}%
  \BibitemOpen
  \bibfield  {author} {\bibinfo {author} {\bibfnamefont {X.}~\bibnamefont {Chen}}, \bibinfo {author} {\bibfnamefont {C.}~\bibnamefont {Zhu}}, \bibinfo {author} {\bibfnamefont {B.}~\bibnamefont {Lei}}, \bibinfo {author} {\bibfnamefont {W.}~\bibnamefont {Zhuo}}, \bibinfo {author} {\bibfnamefont {W.}~\bibnamefont {Wang}}, \bibinfo {author} {\bibfnamefont {J.}~\bibnamefont {Ma}}, \bibinfo {author} {\bibfnamefont {X.}~\bibnamefont {Luo}}, \bibinfo {author} {\bibfnamefont {Z.}~\bibnamefont {Xiang}},\ and\ \bibinfo {author} {\bibfnamefont {X.}~\bibnamefont {Chen}},\ }\bibfield  {title} {\bibinfo {title} {Thickness-dependent anisotropic superconductivity and charge density wave in zrte 3 down to the two-dimensional limit},\ }\href@noop {} {\bibfield  {journal} {\bibinfo  {journal} {Physical Review B}\ }\textbf {\bibinfo {volume} {109}},\ \bibinfo {pages} {144513} (\bibinfo {year} {2024})}\BibitemShut {NoStop}%
\bibitem [{\citenamefont {Gleason}\ \emph {et~al.}(2015)\citenamefont {Gleason}, \citenamefont {Gim}, \citenamefont {Byrum}, \citenamefont {Kogar}, \citenamefont {Abbamonte}, \citenamefont {Fradkin}, \citenamefont {MacDougall}, \citenamefont {Van~Harlingen}, \citenamefont {Zhu}, \citenamefont {Petrovic} \emph {et~al.}}]{gleason2015structural}%
  \BibitemOpen
  \bibfield  {author} {\bibinfo {author} {\bibfnamefont {S.}~\bibnamefont {Gleason}}, \bibinfo {author} {\bibfnamefont {Y.}~\bibnamefont {Gim}}, \bibinfo {author} {\bibfnamefont {T.}~\bibnamefont {Byrum}}, \bibinfo {author} {\bibfnamefont {A.}~\bibnamefont {Kogar}}, \bibinfo {author} {\bibfnamefont {P.}~\bibnamefont {Abbamonte}}, \bibinfo {author} {\bibfnamefont {E.}~\bibnamefont {Fradkin}}, \bibinfo {author} {\bibfnamefont {G.}~\bibnamefont {MacDougall}}, \bibinfo {author} {\bibfnamefont {D.}~\bibnamefont {Van~Harlingen}}, \bibinfo {author} {\bibfnamefont {X.}~\bibnamefont {Zhu}}, \bibinfo {author} {\bibfnamefont {C.}~\bibnamefont {Petrovic}}, \emph {et~al.},\ }\bibfield  {title} {\bibinfo {title} {Structural contributions to the pressure-tuned charge-density-wave to superconductor transition in zrte 3: Raman scattering studies},\ }\href@noop {} {\bibfield  {journal} {\bibinfo  {journal} {Physical Review B}\ }\textbf {\bibinfo {volume} {91}},\ \bibinfo {pages} {155124} (\bibinfo {year} {2015})}\BibitemShut
  {NoStop}%
\bibitem [{\citenamefont {Hu}\ \emph {et~al.}(2015)\citenamefont {Hu}, \citenamefont {Zheng}, \citenamefont {Ren}, \citenamefont {Feng},\ and\ \citenamefont {Li}}]{hu2015charge}%
  \BibitemOpen
  \bibfield  {author} {\bibinfo {author} {\bibfnamefont {Y.}~\bibnamefont {Hu}}, \bibinfo {author} {\bibfnamefont {F.}~\bibnamefont {Zheng}}, \bibinfo {author} {\bibfnamefont {X.}~\bibnamefont {Ren}}, \bibinfo {author} {\bibfnamefont {J.}~\bibnamefont {Feng}},\ and\ \bibinfo {author} {\bibfnamefont {Y.}~\bibnamefont {Li}},\ }\bibfield  {title} {\bibinfo {title} {Charge density waves and phonon-electron coupling in zrte 3},\ }\href@noop {} {\bibfield  {journal} {\bibinfo  {journal} {Physical Review B}\ }\textbf {\bibinfo {volume} {91}},\ \bibinfo {pages} {144502} (\bibinfo {year} {2015})}\BibitemShut {NoStop}%
\bibitem [{\citenamefont {Kresse}\ and\ \citenamefont {Hafner}(1993)}]{vasp1}%
  \BibitemOpen
  \bibfield  {author} {\bibinfo {author} {\bibfnamefont {G.}~\bibnamefont {Kresse}}\ and\ \bibinfo {author} {\bibfnamefont {J.}~\bibnamefont {Hafner}},\ }\bibfield  {title} {\bibinfo {title} {Ab initio molecular dynamics for liquid metals},\ }\href {https://doi.org/10.1103/PhysRevB.47.558} {\bibfield  {journal} {\bibinfo  {journal} {Phys. Rev. B}\ }\textbf {\bibinfo {volume} {47}},\ \bibinfo {pages} {558(R)} (\bibinfo {year} {1993})}\BibitemShut {NoStop}%
\bibitem [{\citenamefont {Kresse}\ and\ \citenamefont {Hafner}(1994)}]{vasp2}%
  \BibitemOpen
  \bibfield  {author} {\bibinfo {author} {\bibfnamefont {G.}~\bibnamefont {Kresse}}\ and\ \bibinfo {author} {\bibfnamefont {J.}~\bibnamefont {Hafner}},\ }\bibfield  {title} {\bibinfo {title} {Ab initio molecular-dynamics simulation of the liquid-metal--amorphous-semiconductor transition in germanium},\ }\href {https://doi.org/10.1103/PhysRevB.49.14251} {\bibfield  {journal} {\bibinfo  {journal} {Phys. Rev. B}\ }\textbf {\bibinfo {volume} {49}},\ \bibinfo {pages} {14251} (\bibinfo {year} {1994})}\BibitemShut {NoStop}%
\bibitem [{\citenamefont {Kresse}\ and\ \citenamefont {Furthmüller}(1996)}]{vasp3}%
  \BibitemOpen
  \bibfield  {author} {\bibinfo {author} {\bibfnamefont {G.}~\bibnamefont {Kresse}}\ and\ \bibinfo {author} {\bibfnamefont {J.}~\bibnamefont {Furthmüller}},\ }\bibfield  {title} {\bibinfo {title} {Efficiency of ab-initio total energy calculations for metals and semiconductors using a plane-wave basis set},\ }\href {https://doi.org/10.1016/0927-0256(96)00008-0} {\bibfield  {journal} {\bibinfo  {journal} {Comput. Mater. Sci.}\ }\textbf {\bibinfo {volume} {6}},\ \bibinfo {pages} {15} (\bibinfo {year} {1996})}\BibitemShut {NoStop}%
\bibitem [{\citenamefont {Kresse}\ and\ \citenamefont {Furthm\"uller}(1996)}]{vasp4}%
  \BibitemOpen
  \bibfield  {author} {\bibinfo {author} {\bibfnamefont {G.}~\bibnamefont {Kresse}}\ and\ \bibinfo {author} {\bibfnamefont {J.}~\bibnamefont {Furthm\"uller}},\ }\bibfield  {title} {\bibinfo {title} {Efficient iterative schemes for ab initio total-energy calculations using a plane-wave basis set},\ }\href {https://doi.org/10.1103/PhysRevB.54.11169} {\bibfield  {journal} {\bibinfo  {journal} {Phys. Rev. B}\ }\textbf {\bibinfo {volume} {54}},\ \bibinfo {pages} {11169} (\bibinfo {year} {1996})}\BibitemShut {NoStop}%
\bibitem [{\citenamefont {Kresse}\ and\ \citenamefont {Joubert}(1999)}]{paw}%
  \BibitemOpen
  \bibfield  {author} {\bibinfo {author} {\bibfnamefont {G.}~\bibnamefont {Kresse}}\ and\ \bibinfo {author} {\bibfnamefont {D.}~\bibnamefont {Joubert}},\ }\bibfield  {title} {\bibinfo {title} {From ultrasoft pseudopotentials to the projector augmented-wave method},\ }\href {https://doi.org/10.1103/PhysRevB.59.1758} {\bibfield  {journal} {\bibinfo  {journal} {Phys. Rev. B}\ }\textbf {\bibinfo {volume} {59}},\ \bibinfo {pages} {1758} (\bibinfo {year} {1999})}\BibitemShut {NoStop}%
\bibitem [{\citenamefont {Perdew}\ \emph {et~al.}(1996)\citenamefont {Perdew}, \citenamefont {Burke},\ and\ \citenamefont {Ernzerhof}}]{pbe}%
  \BibitemOpen
  \bibfield  {author} {\bibinfo {author} {\bibfnamefont {J.~P.}\ \bibnamefont {Perdew}}, \bibinfo {author} {\bibfnamefont {K.}~\bibnamefont {Burke}},\ and\ \bibinfo {author} {\bibfnamefont {M.}~\bibnamefont {Ernzerhof}},\ }\bibfield  {title} {\bibinfo {title} {Generalized gradient approximation made simple},\ }\href {https://doi.org/10.1103/PhysRevLett.77.3865} {\bibfield  {journal} {\bibinfo  {journal} {Phys. Rev. Lett.}\ }\textbf {\bibinfo {volume} {77}},\ \bibinfo {pages} {3865} (\bibinfo {year} {1996})}\BibitemShut {NoStop}%
\bibitem [{\citenamefont {Herath}\ \emph {et~al.}(2019)\citenamefont {Herath}, \citenamefont {Tavadze}, \citenamefont {He}, \citenamefont {Bousquet}, \citenamefont {Singh}, \citenamefont {Mu{\~n}oz},\ and\ \citenamefont {Romero}}]{pyprocar}%
  \BibitemOpen
  \bibfield  {author} {\bibinfo {author} {\bibfnamefont {U.}~\bibnamefont {Herath}}, \bibinfo {author} {\bibfnamefont {P.}~\bibnamefont {Tavadze}}, \bibinfo {author} {\bibfnamefont {X.}~\bibnamefont {He}}, \bibinfo {author} {\bibfnamefont {E.}~\bibnamefont {Bousquet}}, \bibinfo {author} {\bibfnamefont {S.}~\bibnamefont {Singh}}, \bibinfo {author} {\bibfnamefont {F.}~\bibnamefont {Mu{\~n}oz}},\ and\ \bibinfo {author} {\bibfnamefont {A.~H.}\ \bibnamefont {Romero}},\ }\bibfield  {title} {\bibinfo {title} {Pyprocar: A python library for electronic structure pre/post-processing},\ }\href {https://doi.org/10.1016/j.cpc.2019.107080} {\bibfield  {journal} {\bibinfo  {journal} {Comput. Phys. Commun.}\ }\textbf {\bibinfo {volume} {251}},\ \bibinfo {pages} {107080} (\bibinfo {year} {2019})}\BibitemShut {NoStop}%
\bibitem [{\citenamefont {Momma}\ and\ \citenamefont {Izumi}(2011)}]{vesta}%
  \BibitemOpen
  \bibfield  {author} {\bibinfo {author} {\bibfnamefont {K.}~\bibnamefont {Momma}}\ and\ \bibinfo {author} {\bibfnamefont {F.}~\bibnamefont {Izumi}},\ }\bibfield  {title} {\bibinfo {title} {Vesta 3 for three-dimensional visualization of crystal, volumetric and morphology data},\ }\href {https://doi.org/10.1107/S0021889811038970} {\bibfield  {journal} {\bibinfo  {journal} {Journal of applied crystallography}\ }\textbf {\bibinfo {volume} {44}},\ \bibinfo {pages} {1272} (\bibinfo {year} {2011})}\BibitemShut {NoStop}%
\bibitem [{\citenamefont {Togo}\ and\ \citenamefont {Tanaka}(2015)}]{togo2015first}%
  \BibitemOpen
  \bibfield  {author} {\bibinfo {author} {\bibfnamefont {A.}~\bibnamefont {Togo}}\ and\ \bibinfo {author} {\bibfnamefont {I.}~\bibnamefont {Tanaka}},\ }\bibfield  {title} {\bibinfo {title} {First principles phonon calculations in materials science},\ }\href@noop {} {\bibfield  {journal} {\bibinfo  {journal} {Scr. Mater.}\ }\textbf {\bibinfo {volume} {108}},\ \bibinfo {pages} {1} (\bibinfo {year} {2015})}\BibitemShut {NoStop}%
\bibitem [{\citenamefont {Togo}(2023)}]{togo2023first}%
  \BibitemOpen
  \bibfield  {author} {\bibinfo {author} {\bibfnamefont {A.}~\bibnamefont {Togo}},\ }\bibfield  {title} {\bibinfo {title} {First-principles phonon calculations with phonopy and phono3py},\ }\href@noop {} {\bibfield  {journal} {\bibinfo  {journal} {J. Phys. Soc. Jpn.}\ }\textbf {\bibinfo {volume} {92}},\ \bibinfo {pages} {012001} (\bibinfo {year} {2023})}\BibitemShut {NoStop}%
\bibitem [{\citenamefont {Togo}\ \emph {et~al.}(2023)\citenamefont {Togo}, \citenamefont {Chaput}, \citenamefont {Tadano},\ and\ \citenamefont {Tanaka}}]{togo2023implementation}%
  \BibitemOpen
  \bibfield  {author} {\bibinfo {author} {\bibfnamefont {A.}~\bibnamefont {Togo}}, \bibinfo {author} {\bibfnamefont {L.}~\bibnamefont {Chaput}}, \bibinfo {author} {\bibfnamefont {T.}~\bibnamefont {Tadano}},\ and\ \bibinfo {author} {\bibfnamefont {I.}~\bibnamefont {Tanaka}},\ }\bibfield  {title} {\bibinfo {title} {Implementation strategies in phonopy and phono3py},\ }\href@noop {} {\bibfield  {journal} {\bibinfo  {journal} {J. Phys. Condens. Matter}\ } (\bibinfo {year} {2023})}\BibitemShut {NoStop}%
\end{thebibliography}
\end{document}